\documentclass[dvipsnames,11pt]{article}
\usepackage{adjustbox}
\usepackage{tikz,pgfplots}
\usetikzlibrary{arrows.meta,positioning}
\usetikzlibrary{positioning}
\usepackage{graphicx} 
\usepackage{amssymb}
\usepackage{amsthm}
\usepackage{amsmath}
\usepackage{authblk}
\usepackage{blkarray}
\usepackage{booktabs}
\usepackage{multirow}
\usepackage{bm}
\usepackage{calc}
\usepackage{caption}
\usepackage{cases}
\usepackage{centernot}
\usepackage{cite}
\usepackage[utf8]{inputenc}
\usepackage{dcolumn}
\usepackage{float}
\usepackage{placeins}
\usepackage[margin = 1in]{geometry}
\usepackage{rotating}
\usepackage{lineno}
\usepackage{listings}
\usepackage{makecell}
\usepackage{mathtools}
\usepackage{pifont}

\usepackage{setspace}
\usepackage{soul}
\usepackage{subcaption}
\usepackage{tabularx}
\usepackage{makecell}
\usepackage{tgpagella}
\usepackage{verbatim}
\usepackage{xcolor}
\usepackage{xfrac}
\usepackage[style=english]{csquotes}
\MakeOuterQuote{"}
\pgfplotsset{compat=1.18}

\newtheorem{proposition}{Proposition}
\DeclareMathOperator{\tr}{tr}

\title{Strategies for tumor elimination and control under immune evasion and chemotherapy resistance}
\author{Nazanin Mokari \& Bryce Morsky}
\affil{Department of Mathematics, Florida State University, Tallahassee, FL, USA}
\date{\today}

\begin{document}
\maketitle

\begin{abstract}
The evolutionary and ecological dynamics of tumors under immune responses and therapeutic interventions pose major challenges to long-term treatment success. Although treatment may initially achieve short-term disease control, resistant cancer cell subpopulations often arise, leading to relapse with more aggressive and treatment-resistant forms of the disease. Here, we develop and analyze mathematical models describing the interactions among effector cells, chemo-resistant tumor cells, and immuno-resistant tumor cells under distinct immune-evasion strategies. The models incorporate competition and cooperation between resistant and sensitive tumor subpopulations. We identify threshold conditions governing tumor persistence, elimination, and phenotype dominance under varying therapeutic intensities. These findings provide a theoretical framework for designing targeted and combination therapies and offer insights into strategies for mitigating the treatment resistance.
\end{abstract}

\section{Introduction}

Cancer is a genetic disorder driven by the process of somatic evolution. This disease is characterized by the uncontrolled proliferation of abnormal cells caused by genetic mutations \cite{weinberg06}. Over time, cancer cells achieve abilities such as sustained proliferative signaling, immune evasion, resistance to apoptosis, and the ability to invade surrounding tissues \cite{hanahan11}. Despite decades of study and research in cancer therapy including chemotherapy, targeted therapies, and immunotherapy treatment, failure and relapse remain major clinical challenges. These failures are often due to the diversity of tumors and their ability to develop resistance. Factors such as tumor heterogeneity, clonal evolution, immune suppression, and the tumor microenvironment contribute to dynamic resistance to treatment, making design of therapies and long-term control complicated \cite{longley05}.

Given this biological complexity of managing resistance to treatment, traditional experimental approaches alone are insufficient to predict treatment outcomes or design optimal therapies. Mathematical modeling has emerged as a critical tool to bridge this gap by allowing researchers to simulate biological systems, generate quantitative predictions, and test hypotheses \cite{altrock15}. By translating biological interactions into meaningful mathematical equations, these models provide a structured framework for investigating tumor growth, immune dynamics, and treatment responses over time. In particular, they have been used extensively in the study of adaptive therapies. Each type of model serves different purposes, depending on the nature of the system being studied and the level of detail required \cite{depillis14,gatenby09,gallagher24,gluzman20,vilela24,strobl22}.

Various classes of mathematical models have been developed to study cancer at different scales, from molecular interactions and intracellular signaling to tissue level tumor dynamics. Models often incorporate biological factors such as cytokine signaling (e.g., IL-2), cytotoxic T cell activity, tumor proliferation rates, and drug-induced cell death \cite{depillis05, donofrio04}. Incorporating multiple tumor subpopulations, each with specific growth rates and drug sensitivities, enables the modeling of heterogeneity of tumor and the emergence of resistant clones \cite{komarova05}. By parameterizing these models with biological data, specific treatments that alter the composition and behavior of the tumor ecosystem can be simulated and analyzed.

Evolutionary game theory (EGT) offers a powerful framework to model the strategic interactions between heterogeneous cell populations within a tumor \cite{basanta13, archetti19, wolfl22}. In EGT, cancer cells are considered as players in a game, each adopting distinct strategies such as drug sensitivity, immune evasion, or rapid proliferation that confer context-dependent fitness advantages. This framework emphasizes frequency-dependent selection, capturing how the fitness of a strategy depends on its prevalence relative to others in the population. As such, EGT enables the modeling of dynamic shifts in subpopulation composition under selective pressures imposed by therapies or immune responses \cite{gatenby03}. EGT has been widely applied to various aspects of cancer biology and treatment. For example, it has been used to model tumor–stroma interactions \cite{basanta12}, public goods dynamics such as growth factor sharing \cite{gerlee17, morsky18}, and non-cooperative games like the hawk–dove game to study aggression and resistance patterns \cite{laruelle23}. Concepts such as anti-fragility, where tumors may benefit from certain stressors, have also been explored through this lens \cite{bayer23}. Overall, EGT provides a particularly suitable tool for investigating the emergence and dominance of resistant phenotypes, and for designing therapeutic strategies that alter the evolutionary landscape, effectively changing the  rules of the game  to steer tumor evolution toward more treatable or less aggressive.

Insights from both differential equation models and EGT have inspired new therapeutic strategies to manage treatment of resistance in cancer. One such approach is adaptive therapy, which seeks to control rather than eliminate the tumor by maintaining a balance between drug-sensitive and drug-resistant cells \cite{gatenby09}. Unlike traditional maximum tolerated dose (MTD) strategies, adaptive therapy adjusts treatment dynamically based on tumor response, aiming to exploit competition between cell types and delay resistance. Similarly, combination therapy that employs multiple drugs or treatments like chemotherapy and immunotherapy to target different tumor vulnerabilities simultaneously can reduce the likelihood that any single clone can survive \cite{bozic12}. However, the success of such strategies depends heavily on timing, dosing, and sequence, all of which can be systematically studied and optimized using mathematical modeling \cite{castiglione07}.

There remain many open questions in adaptive therapy \cite{west23}. Despite the promise of these approaches, several critical questions remain unanswered. For example, how does treatment timing affect the emergence of resistance? What are the optimal sequences or combinations of chemotherapy and immunotherapy that minimize tumor burden while delaying relapse? Can we predict when and how to switch or pause treatments to maintain long-term tumor control? How do different subpopulations within the tumor interact under therapy, and can these interactions be manipulated to our advantage? Addressing these questions requires models that integrate biological realism with predictive power --- models that can simulate not just how cancer grows, but how it adapts and evolves in response to therapy.

In this study, we develop an ODE-based model to investigate tumor–immune dynamics in B-cell lymphoma (BCL) with a focus on understanding the effects of combination therapy. Our model incorporates an immune system population of effector cells and two tumor subpopulations, one resistant to the immune system and the other sensitive to it. We also consider the case where the immuno-sensitive type is chemo-resistant. The structure and parameters of the tumor–immune interaction model are primarily inspired by the foundational nonlinear system in \cite{kuznetsov94}. Using insights from our analyses, we construct treatment protocols that consider different timing and sequencing of therapies so as to control or eliminate tumors.

\section{Methods}

\subsection{Model Summary} \label{methods:model_summary}

One of the foundational models in the study of cancer-immune system interactions is the system of ODEs introduced in \cite{kuznetsov94}. Their model captures the interaction between immunogenic tumor cells and cytotoxic effector cells (such as CD8$^+$ T-cells), incorporating tumor growth, immune stimulation, immune-mediated killing, and effector exhaustion. In the model, when the tumor population exceeds a critical threshold, effector cells are no longer able to control tumor growth, leading to a large tumor. Below this threshold, the tumor is effectively suppressed. This bistable behavior has since become a cornerstone in mathematical oncology, illustrating how immune surveillance can fail once tumors surpass a certain size. However, what occurs if an immune evasive subpopulation emerges? Assuming some costs to such a mutant subpopulation, how then does it interact with immune sensitive cells? And, what are the implications for treatment? A broader consideration for the therapeutic strategy is whether treatment should aim to aggressively eliminate the tumor (``fight to win") or instead focus on long-term control and resistance management (``fight to not lose"). 

To address these questions, we introduce a type of tumor cell that is resistant to the immune system. $T_1$ and $T_2$ are the amount of immune resistant and immune sensitive tumor cells, respectively. $E$ is the size of the effector population, representing the effects of cytotoxic T-cells. The resistant type $T_1$ has evolved mechanisms to resist the immune system \cite{tufail25}. We consider two types of mechanisms of immune resistance, immune checkpoint regulation and reduced antigen presentation, and study each individually. In both cases, we assume that immune evasion entails both a reduction in the growth rate by $\zeta_1 > 0$ (the cost for resistance) and a benefit represented by $\theta_1 \in [0,1]$, which measures the effectiveness of resistance. Specifically, the growth rate of $T_1$ is $\alpha_1 = \alpha - \zeta_1$, where $\alpha$ is the default growth rate of tumor cells. $\theta_1$ reduces the rate at which effectors kill $T_1$, but may also affect the dynamics of $E$, depending on the type of resistant (to be elaborated upon later).

Though $T_1$ are immune resistant, we assume that they are sensitive to chemotherapy. $T_2$ represents immune sensitive tumor cells that may or may not be chemo-resistant: we consider both scenarios. Their growth rate is $\alpha_2 = \alpha - \zeta_2$ where $\zeta_2$ is the cost of resistance. Chemo-resistance reduces the rate at which chemotherapy kills tumor cells by $\theta_2 \in [0,1]$, while $T_1$ is fully susceptible to chemotherapy ($\zeta_2=0$  and $\theta_2=1$ in the case where $T_2$ are not chemo-resistant). $\chi$ is the baseline rate at which chemotherapy kills tumor cells. For modeling convenience, this rate includes the concentration and kill rate relative to concentration level. Therefore, chemotherapy kills $T_1$ and $T_2$ cells at rates $\chi$ and $\theta_2\chi$, respectively.

In total, the model features three state variables, $T_1$, $T_2$, and $E$, and fourteen parameters summarized in Table \ref{tbl:param}. All parameter values unassociated with resistance are the same as those for the tumor and effectors in \cite{kuznetsov94}. We assume that the baseline cost of resistance $\zeta_i$ and chemotherapy kill rate $\chi$ are $30\%$ and $50\%$ of the growth rate $\alpha$, respectively. A relatively low value of $\chi$ was chosen, since we primarily consider chemotherapy as a complement to immunotherapy. However, we consider larger $\chi$ in our bifurcation analyses. The effector cell date rate from chemotherapy coefficient $\omega$ is taken from the relative kill rates of effectors and tumor cells in \cite{de09}. Specifically, $\omega = [K_L(1-e^{\delta_L M})]/[K_T(1-e^{\delta_T M})]$, where $K_L$ and $K_T$ are the rates of CD8$^+$ T-cell depletion from medicine toxicity and chemotherapy-induced tumor death, respectively, $\delta_L=\delta_T$ are medicine efficacy coefficients, and $M$ is the dose. Note that these data are derived from humans. The remaining parameters $\theta_1$, $\theta_2$, and $\phi$ are varied throughout this paper. The generalized system of equations for the model is:
\begin{subequations}
\begin{align}
    \dot{T}_1 &= \left(\alpha_1\left(1 - \frac{T}{\kappa}\right) - \theta_1 E - \chi\right)T_1, \label{eq:ODEs_T1}\\
    \dot{T}_2 &= \left(\alpha_2\left(1 - \frac{T}{\kappa}\right) - E-\theta_2\chi\right)T_2, \label{eq:ODEs_T2}\\
    \dot{E} &=\left(\frac{\rho_1 T_1 + \rho T_2}{T + \eta} - \mu_1 T_1 - \mu T_2 - \delta - \omega\chi\right)E + \sigma, \label{eq:ODEs_E}
    \end{align} \label{eq:ODEs}
\end{subequations}
where $T = T_1 + T_2$ is the total size of the tumor population. Tumors cells are killed by effectors and chemotherapy and growth is limited by a carrying capacity $\kappa$. Effector cells are activated with a nonlinear activation term with parameters $\rho_1$, $\rho$, and $\eta$. Tumor cells also exhaust and kill effector cells at rates $\mu_1$ and $\mu$ for $T_1$ and $T_2$, respectively. Though the effects of chemotherapy on effectors can be complex \cite{menard08,bracci14}, we assume that chemotherapy damages and thus inactivates effectors at rate $\omega\chi$. Effector cells have a natural death rate $\delta$ and are assumed to be generated at a rate $\sigma$. When $T_1$ exhibits immune checkpoint evasion, $\rho_1 = \rho$ and $\mu_1 = \mu + \phi(1-\theta_1)$. When it exhibits reduced antigen presentation, $\rho_1 = \theta_1 \rho$ and $\mu_1 = \theta_1 \mu$. Note that for both immune evasion strategies, $T_2$ is a free-rider. It indirectly gains a benefit from reduced effector availability due to $T_1$. The derivation of these parameter values are detailed in the following subsection.

\begin{table}[!htb]
\begin{center}
\begin{tabular}{ccl}
\toprule
\textbf{Parameter} & \textbf{Default value} & \textbf{Definition} \\
\midrule
$\alpha$ & $1.636$ & Default growth rate of tumor cells \\
$\alpha_i=\alpha-\zeta_i$ & ---- & Growth rate of $T_i$ \\
$\delta$ & $0.3743$ & Death rate of effectors \\
$\zeta_i$ & $0.4908$ & Reduction in growth rate from resistance \\
$\eta$ & $20.19$ & Activation parameter \\
$\theta_1$ & --- & Reduction in death rate from effectors for $T_1$ \\
$\theta_2$ & --- & Reduction in death rate from chemotherapy for $T_2$ \\
$\kappa$ & $500$ & Carrying capacity \\
$\chi$ & $0.818$ & Baseline chemotherapy kill rate \\
$\mu$ & $0.00311$ & Inactivation rate for tumor type $i$ \\
$\rho$ & $1.131$ & Activation parameter for tumor type $i$  \\
$\sigma$ & $0.1181$ & Birth rate of effectors \\
$\phi$ & $0.00311$ & Immune checkpoint inactivation weight \\
$\omega$ & $0.054$ & Effector cell death rate from chemotherapy coefficient \\
$T_i$ & --- & $10^6$ type $i$ tumor cells \\
$E$ & --- & $10^6$ effectors \\
\bottomrule
\end{tabular}
\caption{Summary of parameter definitions and default values of them and variables.} \label{tbl:param}
\end{center}
\end{table}

\subsection{Immune Evasion} \label{methods:immune_evasion}

\begin{figure}[ht!]
    \centering
    \begin{tikzpicture}[baseline=(C.base), >=Stealth, node distance=20mm]
    \node (ET) {$E+T_i$};
    \node (C)  [right=of ET] {$C_i$};
    \node (ETs) [right=20mm of C, yshift=7mm] {$E+T_i^\dagger$};
    \node (EsT) [right=20mm of C, yshift=-7mm] {$E^\dagger+T_i$};

    \draw[->] ([yshift=2pt]ET.east) -- node[above] {$\gamma_{i,1}$} ([yshift=2pt]C.west);
    \draw[->] ([yshift=-2pt]C.west) -- node[below] {$\gamma_{i,-1}$} ([yshift=-2pt]ET.east);

    \draw[->] (C) -- node[above] {$\gamma_{i,2}$} (ETs);
    \draw[->] (C) -- node[below] {$\gamma_{i,3}$} (EsT);
    \end{tikzpicture}
\caption{Tumor ($T_i$) and effector ($E$) cells can combine into a conglomerate ($C_i$) for $i=1,2$. $C_i$ may then dissociate leaving either one or no cells damaged with damaged cells marked by $\dagger$. Rates of these reaction are $\gamma_{i,j}$ for $T_i$ and for $j=-1,1,2,3$.}\label{fig:ET_diagram}
\end{figure}
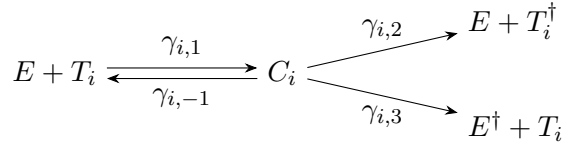

We assume that the tumor population $T_2$ consists of cancer cells identical to those considered in \cite{kuznetsov94}. To determine how immune evasion may affect the parameters for $T_1$, we consider the interactions between effectors and tumor cells. Effectors may bind to tumor cells of type $i$ to form a conglomerate $C_i$, which may then fracture with no effect to either cell or one cell being damaged (and thus inactivated or dead). Figure \ref{fig:ET_diagram} depicts these interactions.

The parameters in Table \ref{tbl:param} are partly derived from the rates $\gamma_{i,j}$ by assuming that these interactions occur on a relatively fast time scale and thus are in steady state. To see this, consider first the the system of ODEs that includes the dynamics of conglomerates $C_i$:
\begin{subequations}
    \begin{align}
        \dot{T}_i &= \alpha_i'T_i\left(1-\frac{T}{\kappa}\right) - (\gamma_{i,1}E + \theta_i''\chi')T_i + (\gamma_{i,-1}+\gamma_{i,3})C_i, \\
        \dot{C}_i &= \gamma_{i,1}ET_i - (\gamma_{i,-1}+\gamma_{i,2}+\gamma_{i,3})C_i, \\
        \dot{E} &= \frac{\rho'(C_1+C_2)}{T+\eta} + (\gamma_{1,-1}+\gamma_{1,2})C_1 + (\gamma_{2,-1}+\gamma_{2,2})C_2 - (\gamma_{1,1}T_1 + \gamma_{2,1}T_2 + \omega\chi' + \delta')E + \sigma'.
    \end{align}
\end{subequations}
Assuming that the dynamics for $C_i$ are fast and are thus at equilibrium, $C_i = \Gamma_iET_i$ where
\begin{equation}
    \Gamma_i = \frac{\gamma_{i,1}}{\gamma_{i,-1}+\gamma_{i,2}+\gamma_{i,3}}.
\end{equation}
Setting $C_i = \Gamma_iET_i$ and simplifying by dividing by $\gamma_{2,2}\Gamma_2$ we have
\begin{subequations}
    \begin{align}
        \dot{T}_i &= \alpha_iT_i\left(1-\frac{T}{\kappa}\right) - \theta_i'ET_i - \theta_i''T_i, \\
        \dot{E} &= \frac{\rho_1 ET_1+\rho ET_2}{T+\eta} - \mu_1 ET_1 - \mu ET_2 - \delta E + \sigma, \label{eq:red_Edot}
    \end{align}
\end{subequations}
where
\begin{center}
\begin{subequations}
\begin{minipage}{0.3\textwidth}
  \begin{equation}
  \theta_i' = \frac{\gamma_{i,2}\Gamma_i}{\gamma_{2,2}\Gamma_2},
  \end{equation}
\end{minipage}%
\begin{minipage}{0.3\textwidth}
  \begin{equation}
  \rho_1 = \rho\frac{\Gamma_1}{\Gamma_2},
  \end{equation}
\end{minipage}
\begin{minipage}{0.3\textwidth}
  \begin{equation}
  \mu_1 = \frac{\gamma_{1,3}\Gamma_1}{\gamma_{2,2}\Gamma_2}.
  \end{equation}
\end{minipage}
\end{subequations}
\end{center}
$\theta_2'=\theta_1''=1$ and $\theta_2''=\theta_2$, while $\theta_1'$, $\rho_1$, and $\mu_1$ are determined by the specific immune evasion strategy employed by $T_1$. The remaining parameters remain unchanged by the immune strategy and are related to the values in Table \ref{tbl:param} as follows: $\alpha_i = \alpha_i'/(\gamma_{2,2}\Gamma_2)$, $\rho = \rho'/\gamma_{2,2}$, $\mu = \gamma_{2,3}/\gamma_{2,2}$, $\chi = \chi'/(\gamma_{2,2}\Gamma_2)$, $\delta = \delta'/(\gamma_{2,2}\Gamma_2)$, and $\sigma = \sigma'/(\gamma_{2,2}\Gamma_2)$.

Under immune checkpoint evasion, the population $T_1$ cells gain a new receptor (PD-L1) by mutation. This receptor can bind to PD-1 receptors on effector cells, inactivating them. This mechanism, commonly referred to as immune checkpoint inhibition, reduces the effectiveness of immune-mediated killing. Consequently, fewer $T_1$ cells are eliminated by effector cells. Specifically, this effect reduces the rate that the $C_2$ conjugation leads to lysis of $T_1$. At rate $\gamma_{1,2}$, the effector attempts to kill the tumor cell and is successful with probability $\theta_1$. Therefore, $\gamma_{1,2} = \theta_1 \gamma_{2,2}$. However, with probability $\phi(1-\theta_1)$, the effector dies in lieu of $T_1$ and, with probability $(1-\phi)(1-\theta_1)$, the effector cells and $T_1$ cells detach without damaging cells. Here $\phi\in[0,1]$ is an inactivation weight. Since effectors may also be inactivated at the same rate as $C_2$, $\gamma_{1,3} = \gamma_{2,3} + \phi(1-\theta_1)\gamma_{2,2}$. Also, $\gamma_{1,-1} = \gamma_{2,-1}+ (1-\phi)(1-\theta_1)\gamma_{2,2}$.

The other reaction rate, binding rate of effector cells to tumor cells, is identical for both $T_1$ and $T_2$, $\gamma_{1,1} = \gamma_{2,1}$. Because, immune checkpoint evasion does not affect it. These relationships imply that 
\begin{equation}
    \Gamma_1 = \frac{\gamma_{1,1}}{\gamma_{1,-1}+\gamma_{1,2}+\gamma_{1,3}} = \frac{\gamma_{2,1}}{\gamma_{2,-1}+(1-\phi)(1-\theta_1)\gamma_{2,2} + \theta_1\gamma_{2,2}+\gamma_{2,3}+\phi(1-\theta_1)\gamma_{2,2}} = \Gamma_2.
\end{equation}
Therefore, $\theta_1'=\theta_1$, $\rho_1 = \rho$, and
\begin{equation}
    \mu_1 = \frac{(\gamma_{2,3} + \phi(1-\theta_1)\gamma_{2,2})\Gamma_2}{\gamma_{2,2}\Gamma_2} = \mu + \phi(1-\theta_1),
\end{equation}
as summarized in Section \ref{methods:model_summary}.

Next consider the scenario where $T_1$ employs the strategy of reduced antigen presentation, which decreases the rate at which effectors recognize and bind to $T_1$ cells to create $C_1$ conglomerates. Consequently, $\gamma_{1,1} = \theta_1\gamma_{2,1}$. Once bound, the dynamics are unchanged relative to those of $E$ and $T_2$ (i.e. the remaining rates are unchanged). Therefore, $\Gamma_1 = \theta_1\Gamma_2$, which implies that $\theta_1' = \theta_1$, $\rho_1 = \theta_1\rho$, and $\mu_1 = \theta_1\mu$. Note that this effect reduces both the activation and inactivation rates of effectors.

\section{Results}

We investigate the dynamics of the system under different immune-evasion strategies employed by tumor cells $T_1$ and when $T_2$ is and is not chemo-resistant. We first characterize the equilibrium structure of the tumor–immune system in Section \ref{results:general_structure_immune_evasion}. We then complement these results with bifurcation diagrams in Sections \ref{results:bifurcation_immuno_resistant} and \ref{results:bifurcation_chemo_resistant}, which provide the foundation for designing therapeutic strategies aimed at suppressing resistance and achieving long-term tumor control. These therapies alter parameters, qualitatively changing outcomes.

\subsection{Equilibrium and Stability Analysis}\label{results:general_structure_immune_evasion}

We begin by deriving the conditions for equilibria of the system and their stability without specifying the immune-evasion mechanism employed by $T_1$. Throughout, we will discuss the implications of whether or not $T_2$ is chemo-resistant. There are essentially four possible equilibria: a tumor-free equilibrium $\mathcal{E}_0=(0,0,\sigma/(\delta+\omega\chi)$, a $T_1$-only equilibrium $\mathcal{E}_1=(T^*_1,0,E^*)$, a $T_2$-only equilibrium $\mathcal{E}_2=(0,T^*_2,E^*)$, and a coexistence equilibrium $\mathcal{E}_3=(T^*_1,T^*_2,E^*)$ with $T^*_1,T^*_2>0$. The total population at equilibrium is $T^*=T_1^*+T_2^*$, and we represent the coexistence equilibria as $T_1^*=rT^*$ and $T_2^*=(1-r)T^*$, where $r\in (0,1)$ is the portion of the tumor that is $T_1$.

\paragraph{Tumor-free equilibrium: } The Jacobian evaluated at $\mathcal{E}_0$ is
\begin{equation}
    J(\mathcal{E}_0)=\begin{pmatrix}
        \alpha_1-\theta_1E^*-\chi& 0 & 0\\[8pt]
        0 & \alpha_2-E^*-\theta_2\chi & 0\\[8pt]
        E^*\left(\dfrac{\rho_1}{\eta}-\mu_1\right) & E^*\left(\dfrac{\rho}{\eta}-\mu\right) & -\dfrac{\sigma}{E^*}
        \end{pmatrix}.
\end{equation}
Its eigenvalues are $\lambda_1=\alpha_1-\theta_1\sigma/(\delta+\omega\chi) - \chi$, $\lambda_2=\alpha_2-\sigma/(\delta+\omega\chi) - \theta_2\chi$, and $\lambda_3=-\delta-\omega\chi$. Hence, we have the following proposition.
\begin{proposition}
The tumor-free equilibrium $\mathcal{E}_0 = (0,0,\sigma/(\delta+\omega\chi))$ is stable iff
\[\frac{\sigma}{\delta+\omega\chi}>\max\!\left\{\alpha_2-\theta_2\chi,\ \frac{\alpha_1-\chi}{\theta_1}\right\}.
\]
Otherwise, the model admits a nontrivial equilibrium in which at least one tumor population persists.
\end{proposition}\label{prop_1}
\noindent 
This proposition states that the effective immune strength, measured by the ratio of effector cell recruitment $\sigma$ to their total loss rate (the natural death rate plus any chemotherapy-induced suppression) must exceed the maximal effective growth rate of the tumor populations. Note that when there is no chemotherapy and presumably no resistance to it, such as during the emergence of a tumor, $\chi=\theta_2=0$ and $\alpha_2=\alpha$. Thus, the condition becomes $\sigma/\delta > \max\{\alpha,\alpha_1/\theta \}$. In the remaining analyses of this subsection, we will generally consider this case --- i.e.\ when $\mathcal{E}_0$ without chemotherapy is unstable.

For equilibria with at least one tumor population present, the steady states are determined by the equation
\begin{equation}
\left(\frac{P_j T^*}{T^*+\eta} - M_jT^* - \delta - \omega\chi \right) \left(A_j \left(1 - \frac{T^*}{\kappa}\right) - \Theta_j\chi \right) + \sigma = 0 \label{eq:eq_condition}
\end{equation}
for equilibrium $\mathcal{E}_j$. $P_1=\rho_1$, $P_2=\rho$, and $P_3 = r\rho_1 + (1-r)\rho$. Similarly, $M_1=\mu_1$, $M_2=\mu$, and $M_3=r\mu_1 + (1-r)\mu$. $A_1=\alpha_1/\theta_1$, $A_2=\alpha_2$, $\Theta_1=1/\theta_1$, and $\Theta_2=\theta_2$. In the case of coexistence, either $j=1$ or $j=2$ may be used for $A_3$ and $X_3$, since $A_1(1-T^*/\kappa)-\Theta_1\chi = A_2(1-T^*/\kappa)-\Theta_2\chi$. In the case where there is no chemotherapy, coexistence simplifies to the condition $\alpha_1/\theta_1=\alpha_2$.

Rearranging Equation \ref{eq:eq_condition} yields the cubic polynomial
\begin{equation}
    F(\mathcal{E}_j) = B_3 T^{*3}+B_2 T^{*2}+B_1 T^*+B_0 = 0,
    \label{eq:F_i_j}
\end{equation}
where
\begin{subequations}
    \begin{align}
        &B_0= \kappa\eta[(\Theta_j \chi - A_j)(\delta+\omega\chi) + \sigma],\\
        &B_1=\kappa(A_j - \Theta_j\chi)(P_j-\eta M_j-\delta-\omega\chi) + A_j\eta(\delta+\omega\chi) + \kappa\sigma,\\
        &B_2= A_jM_j(\eta - \kappa) + A_j(\delta + \omega\chi - P_j) + M_j\Theta_j \kappa\chi\\
        &B_3=A_jM_j.
    \end{align}
\end{subequations} 
In general, we may thus have up to three solutions. $B_3>0$. Note that $B_0 > 0$ implies that the tumor-free equilibrium is stable due to Proposition \ref{prop_1}. Thus, we consider the case where $B_0 < 0$. Therefore, by Descartes' Rule of Signs, we have either one or three equilibria.

In absent of chemotherapy, assume that $P_j>\delta$ and $\kappa>\eta$, following the assumptions in \cite{kuznetsov94}. Under these assumptions, it follows that $B_2<0$. If $P_j + \delta\eta > M_j \eta$, then $B_1>0$ and $F(\mathcal{E}_j)$ admits up to three real roots. This corresponds to the existence of three equilibria, leading to bistability in the system. On the other hand, if $P_j+\delta\eta < M_j\eta$, then $B_1<0$ and $F(\mathcal{E}_j)$ has only one real root. In this case, the system admits a single equilibrium, which corresponds either to a high tumor population equilibrium or to a low tumor population equilibrium. The functional form of $F(\mathcal{E}_j)$ is identical across strategies; only the parameter substitutions differ depending on the dominant phenotype.

In general, the Jacobian evaluated at $\mathcal{E}_j$ is
\begin{equation}
    J(\mathcal{E}_j)=\begin{pmatrix}
        \alpha_1\left(1-\dfrac{T^*}{\kappa}\right)-\theta_1E^*-\chi-\dfrac{\alpha_1}{\kappa}T^*_1 & -\dfrac{\alpha_1}{\kappa}T^*_1 & -\theta_1 T^*_1\\[8pt]
        -\dfrac{\alpha_2}{\kappa}T^*_2 & \alpha_2\left(1-\dfrac{T^*}{\kappa}\right)-E^*-\theta_2\chi-\dfrac{\alpha_2}{\kappa}T^*_2 & -T^*_2\\[8pt]
        E^*\left(\dfrac{(\rho_1-\rho)(1-r)T^* + \rho_1\eta}{(T^*+\eta)^2}-\mu_1\right) 
        & E^*\left(\dfrac{(\rho-\rho_1)rT^* + \rho\eta}{(T^*+\eta)^2}-\mu\right)
        & -\dfrac{\sigma}{E^*}
    \end{pmatrix}.
\end{equation}
Below we evaluate the stability for $j=1,2,3$.

\paragraph{$T_1$-only equilibrium: }
Letting $T_2^*=0$, the Jacobian matrix is
\begin{equation}
    J=\begin{pmatrix}
- \dfrac{\alpha_1}{\kappa}T^*_1 & -\dfrac{\alpha_1}{\kappa}T^*_1 & -\theta_1 T_1^*\\[8pt]
0 & \alpha_2\left(1-\dfrac{T^*_1}{\kappa}\right)-E^*-\theta_2\chi & 0\\[8pt]
E^*\left(\dfrac{\rho_1\eta}{(T^*_1+\eta)^2}-\mu_1\right) 
& E^*\left(\dfrac{(\rho-\rho_1)T^*_1 + \rho\eta}{(T^*_1+\eta)^2}-\mu\right) &-\dfrac{\sigma}{E^*}
\end{pmatrix},
\end{equation}
which has the eigenvalue $\lambda_2= \alpha_2(1-T^*_1/\kappa)-E^*-\theta_2\chi$ and the remaining matrix
\begin{equation}
    J_1=\begin{pmatrix}
        - \dfrac{\alpha_1}{\kappa}T^*_1 &-\theta_1 T^*_1\\[8pt]
        E^*\left(\dfrac{\rho_1\eta}{(T^*_1+\eta)^2}-\mu_1\right) & -\dfrac{\sigma}{E^*}
    \end{pmatrix}.
\end{equation}
Note that $\tr(J_1) < 0$. Furthermore,
\begin{equation}
\det(J_1)= \dfrac{\alpha_1 \sigma}{\kappa E^* }T^*_1+ \left(\dfrac{\rho_1\eta}{(T^*_1+\eta)^2}-\mu_1\right)\theta_1 E^* T^*_1.
\end{equation}
Thus, $\det(J_1)$ is positive if $\rho_1\eta/(T^*_1+\eta)^2>\mu_1$. Moreover, if $\det(J_1) > 0$, the other two eigenvalues are negative.
Consider the case with no chemotherapy and where $T_2$ is not chemo-resistant ($\chi=0$ and $\alpha_2=\alpha$). Since this is a monomorphic equilibrium, $\alpha_1/\theta_1 \neq \alpha$. If $\alpha_1/\theta_1 > \alpha$, then 
\begin{equation}
    \lambda_2 = \alpha\left(1-\frac{T^*_1}{\kappa}\right)-E^*<\frac{\alpha_1}{\theta_1}\left(1-\frac{T^*_1}{\kappa}\right)-E^* = \frac{1}{\theta_1}\left(\alpha_1\left(1-\frac{T^*_1}{\kappa}\right)-\theta_1E^*\right)=0.
\end{equation}
Thus, $\lambda_2 < 0$. By a similar argument, $\lambda_2>0$ if $\alpha_1/\theta_1 < \alpha$. As a result, $\mathcal{E}_1$ is stable if $\det(J_1) > 0$.

\paragraph{$T_2$-only equilibrium: } Letting $T_1^*=0$, the Jacobian matrix is
\begin{equation}
J=\begin{pmatrix}
\alpha_1\left(1-\dfrac{T_2}{\kappa}\right)-\theta_1E^*-\chi & 0 & 0\\[8pt]
-\dfrac{\alpha_2}{\kappa}T_2 &  - \dfrac{\alpha_2}{\kappa}T_2 & -T_2\\[8pt]
E^*\left(\dfrac{(\rho_1-\rho)T^*_2+\rho_1\eta}{(T^*_2+\eta)^2}-\mu_1\right) 
& E^*\left(\dfrac{\rho\eta}{(T^*_2+\eta)^2}-\mu\right)
& -\dfrac{\sigma}{E^*}
\end{pmatrix}.
\end{equation}
Therefore, we have the eigenvalue $\lambda_1=\alpha_1 (1-T_2/\kappa)-\theta_1E^*-\chi$ and the remaining matrix
\begin{equation}
J_2=\begin{pmatrix}
 -\dfrac{\alpha_2}{\kappa}T_2^*  & -T_2\\[8pt]
 E^*\left(\dfrac{\rho\eta}{(T^*_2+\eta)^2}-\mu\right) 
& -\dfrac{\sigma}{E^*}
\end{pmatrix},
\end{equation}
where $\tr(J_2)< 0$. Similarly to the $T_1$-only case, if $\rho\eta/(T^*_2+\eta)^2>\mu$, then $\det(J_1) > 0$, implying that the remaining two eigenvalues are negative.

If $\chi=0$, $\alpha_2=\alpha$, and $\alpha_1/\theta_1 < \alpha$, then
\begin{equation}
    \lambda_1= \alpha_1\left(1-\frac{T^*_2}{\kappa}\right)-\theta_1 E^*< \theta_1\alpha\left(1-\frac{T^*_2}{\kappa}\right)-\theta_1 E^* = \theta_1\left(\alpha\left(1-\frac{T^*_2}{\kappa}\right)-E^*\right)=0,
\end{equation}
and thus $\lambda_1<0$. Similarly, if $\alpha_1/\theta_1 > \alpha$, then $\lambda_1>0$ and it is unstable. Essentially, the cost of resistance is greater than its benefit.

\paragraph{Coexistence equilibria (CE): } Assume that $T_1^*,T_2^*>0$. Each positive root $T^*$ of Equation \ref{eq:F_i_j} generates a continuum of coexistence equilibria. At any such equilibrium, the Jacobian matrix simplifies to
\begin{equation}
    J=\begin{pmatrix}
    -\dfrac{\alpha_1}{\kappa}rT^* & -\dfrac{\alpha_1}{\kappa}rT^* & -\theta_1 rT^*\\[8pt]
    -\dfrac{\alpha_2}{\kappa}(1-r)T^* & -\dfrac{\alpha_2}{\kappa}(1-r)T^* & -(1-r)T^*\\[8pt]
    E^*\left(\dfrac{(\rho_1-\rho)(1-r)T^*+\rho_1\eta}{(T^*+\eta)^2}-\mu_1\right) & E^*\left(\dfrac{(\rho-\rho_1)rT^*+\rho\eta}{(T^*+\eta)^2}-\mu\right) & -\dfrac{\sigma}{E^*}
\end{pmatrix},
\end{equation}
and
\begin{equation}
    \det(J)=\frac{r(1-r)E^*(T^*)^2}{\kappa}\left(\frac{\rho - \rho_1}{T^* + \eta} + \mu_1 - \mu \right)(\theta_1\alpha_2-\alpha_1).
\end{equation}
When $T_1$ employs immune checkpoint evasion, the middle term reduces to $\phi(1-\theta_1)$ and thus $\det(J) > 0$ whenever $\theta_1 \alpha_2 > \alpha_1$. In this case, since $\tr(J) < 0$, the Routh–Hurwitz criterion ensures stability. Consequently, a sufficient condition for the existence of at least one stable equilibrium is $(1-\theta_2)\chi+\alpha_2<\alpha_1<\theta_1\alpha_2$. Note that this lower bound is given by our assumption that $B_0<0$. This inequality characterizes the parameter range under which the system admits at least one stable equilibrium point. However, when $T_1$ exhibits reduced antigen mutation, we obtain $\det(J)>0$ whenever $(\theta_1 \alpha_2-\alpha_1)(\rho-\mu(T^*+\eta)) > 0$. Hence, the existence of a stable equilibrium, additional to those determined by $\alpha_i$ and $\theta_i$, also depends on the parameters $\rho$, $\mu$, and $\eta$, and the state $T^*$. In absence of chemotherapy, $\det(J(\mathcal{E}_3)) = 0$, since $\alpha_1=\theta_1\alpha_2$. Therefore, at least one eigenvalue is zero.

\subsection{Bifurcation diagrams for Immuno-resistance Only} \label{results:bifurcation_immuno_resistant}

\begin{figure}[!ht]
\centering
\captionsetup[subfigure]{justification=centering}
\begin{subfigure}{0.45\columnwidth}
    \caption{}
    \includegraphics[width=\textwidth]{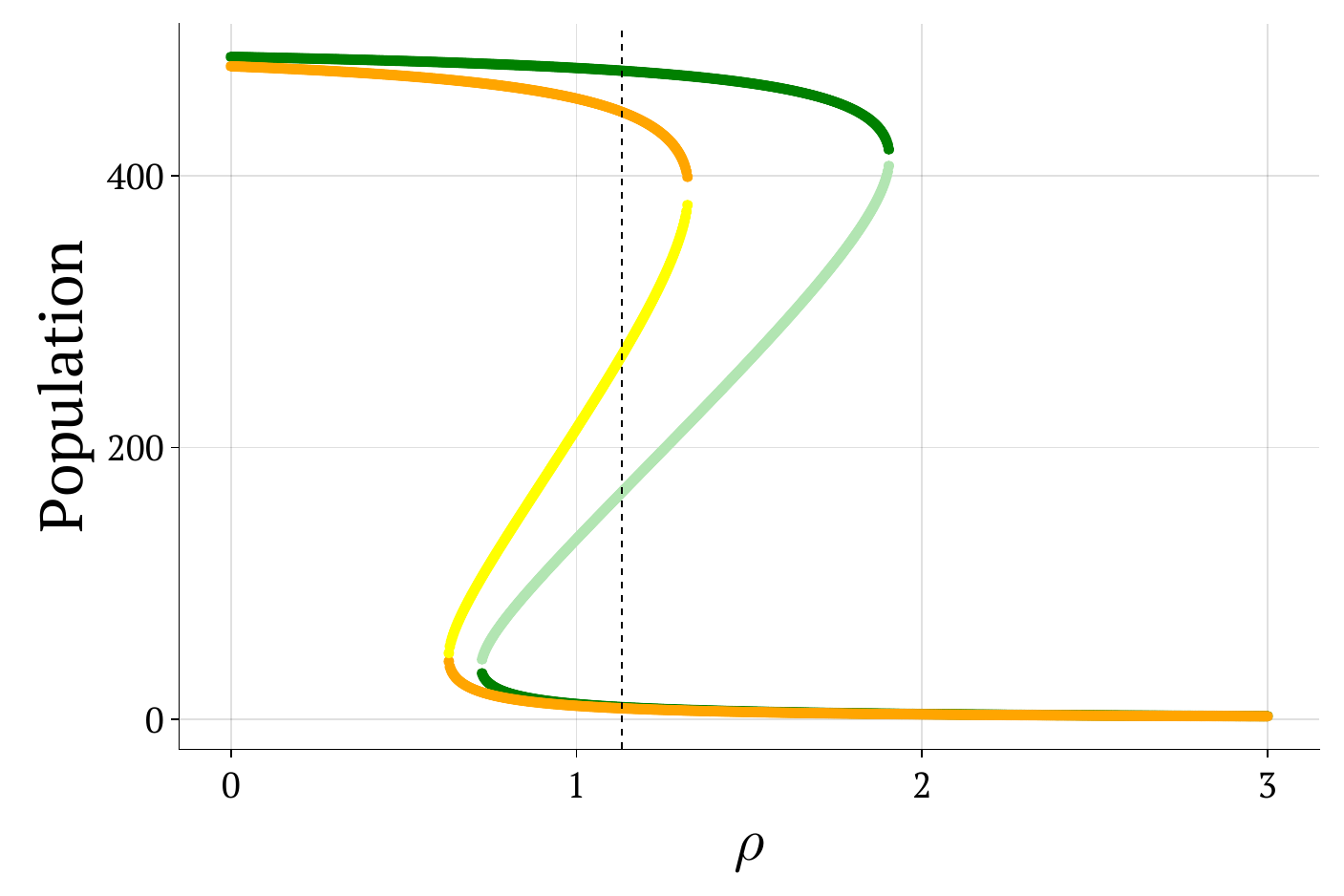}
    \label{fig:rho_1}
\end{subfigure}
\begin{subfigure}{0.45\columnwidth}
 \caption{}
    \includegraphics[width=\textwidth]{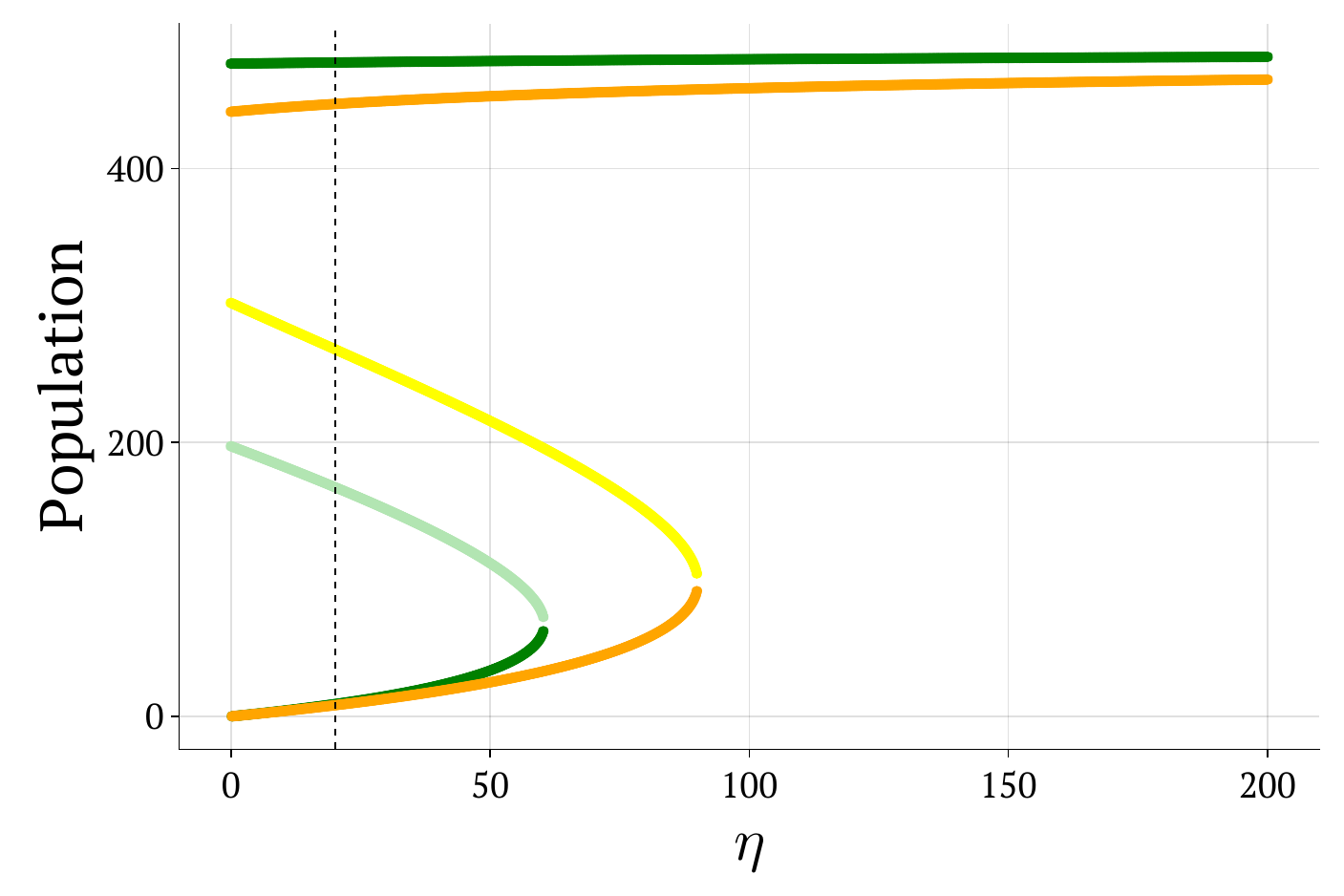}
    \label{fig:eta_1}
\end{subfigure}
\begin{subfigure}{0.45\columnwidth}
 \caption{}
    \includegraphics[width=\textwidth]{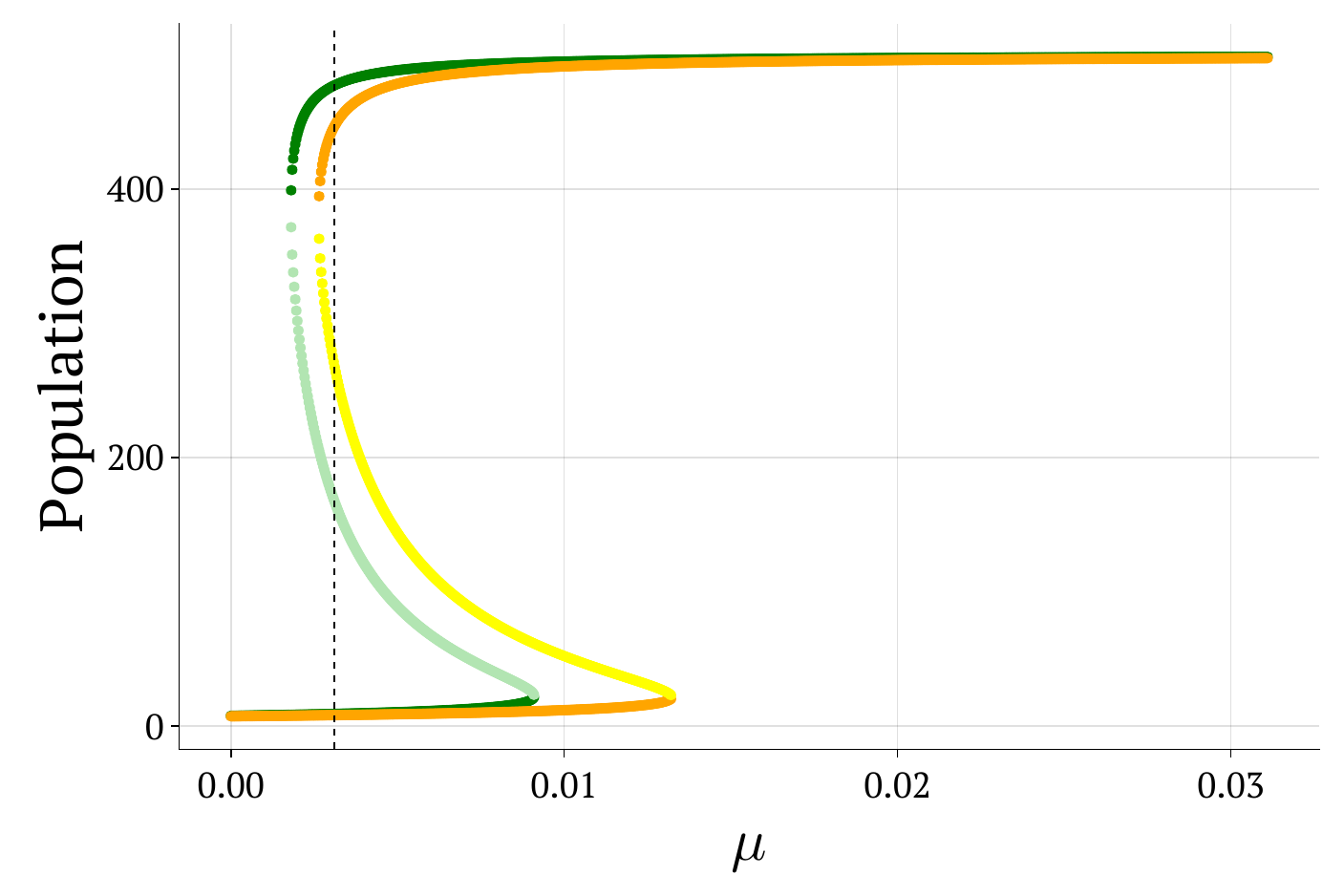}
    \label{fig:mu_1}
\end{subfigure}
\begin{subfigure}{0.45\columnwidth}
 \caption{}
    \includegraphics[width=\textwidth]{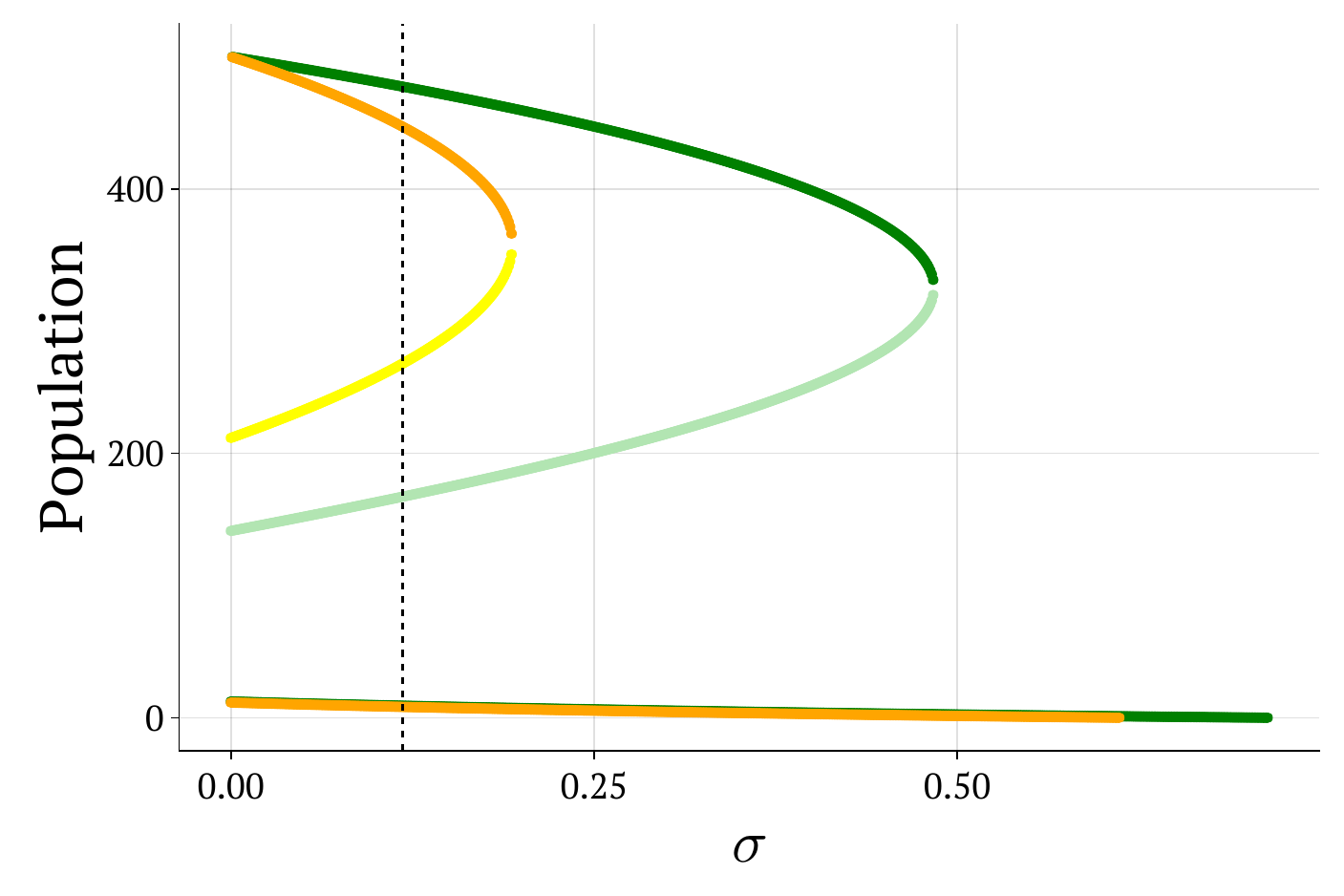}
    \label{fig:sigma_1}
\end{subfigure}
\begin{subfigure}{0.45\columnwidth}
 \caption{}
    \includegraphics[width=\textwidth]{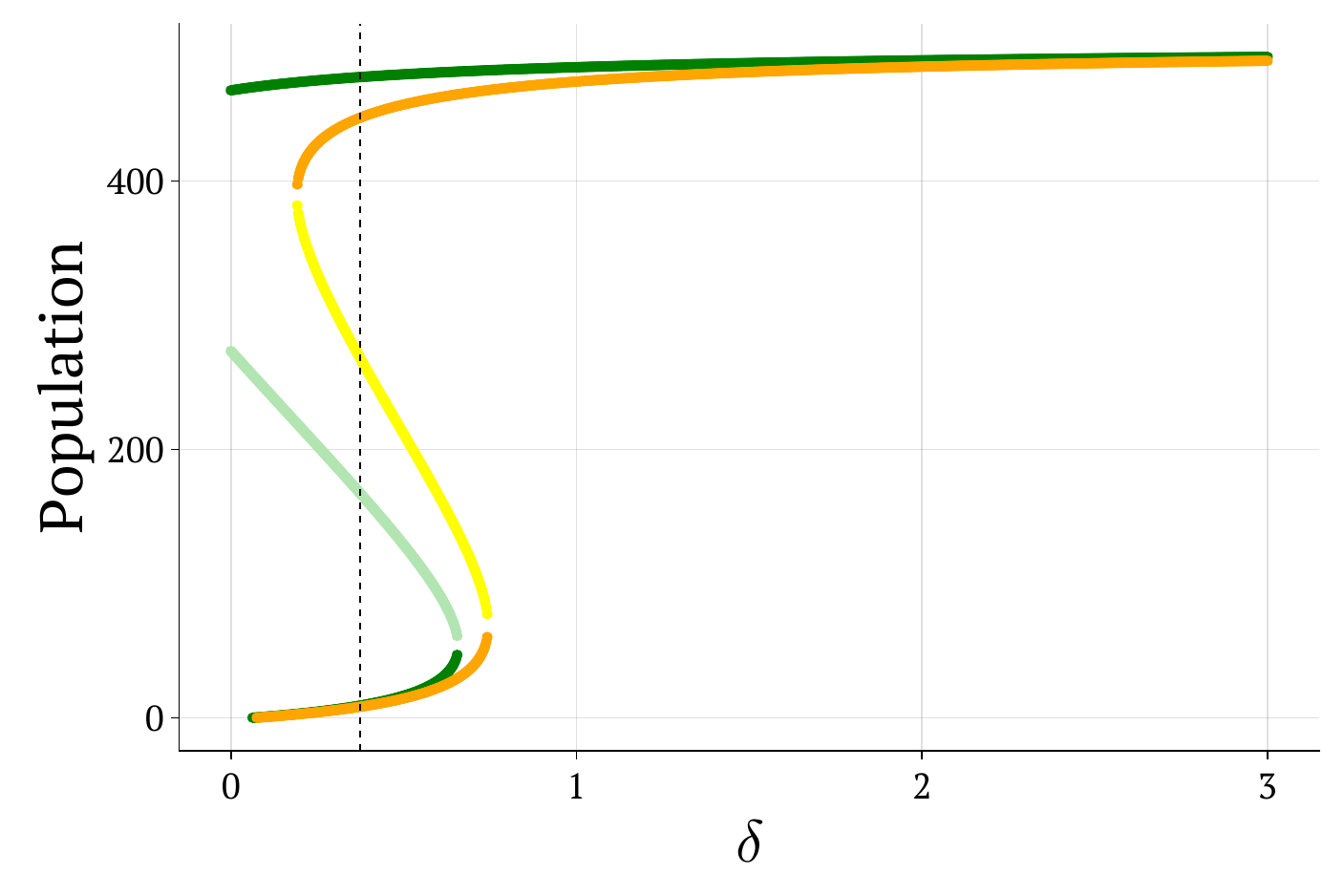}
    \label{fig:delta_1}
\end{subfigure} 
\begin{subfigure}{0.45\columnwidth}
 \caption{}
    \includegraphics[width=\textwidth]{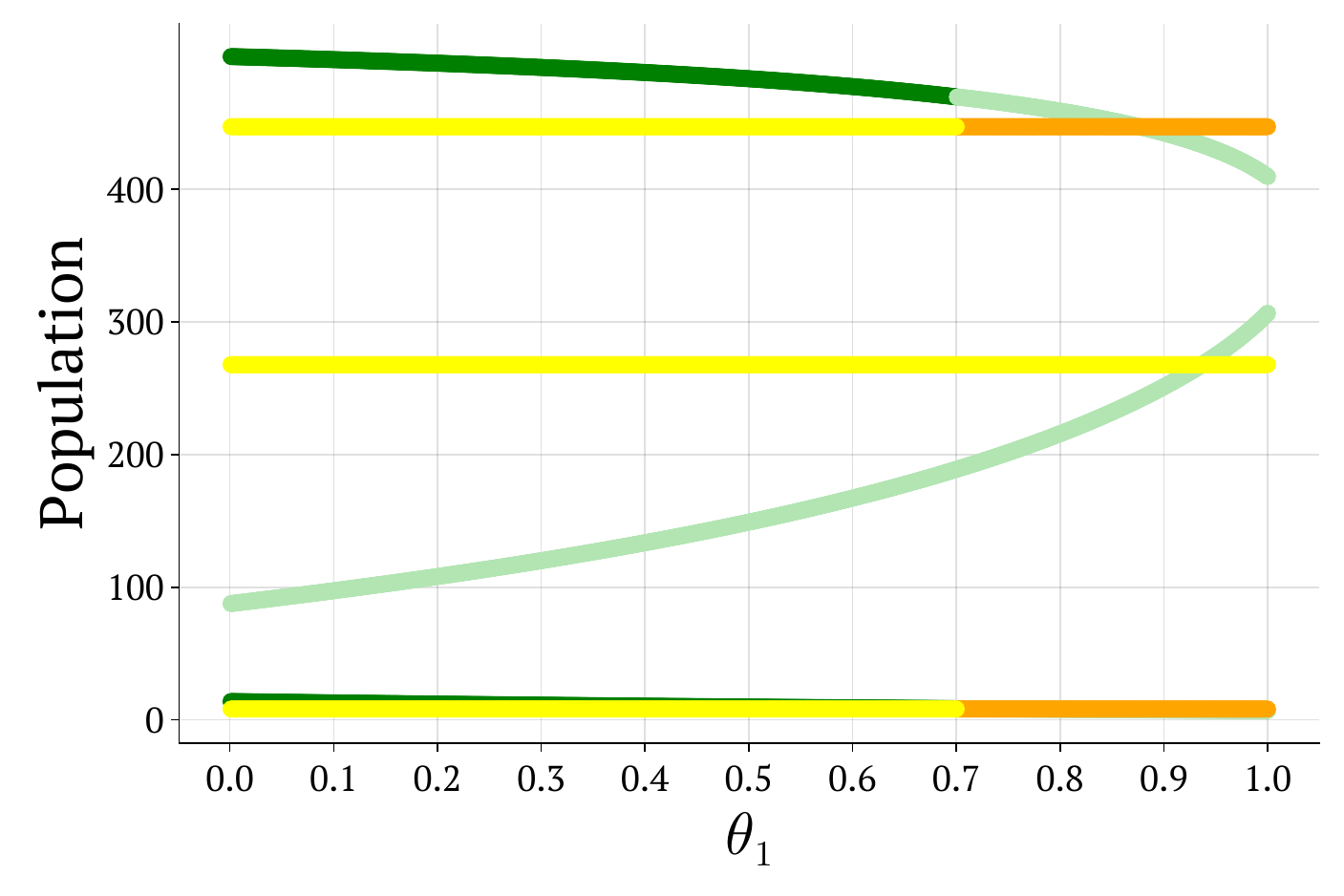}
    \label{fig:theta_1}
\end{subfigure}\\
\vspace{-0.6cm}
\begin{subfigure}{0.95\columnwidth}
  {\includegraphics[width=\linewidth]{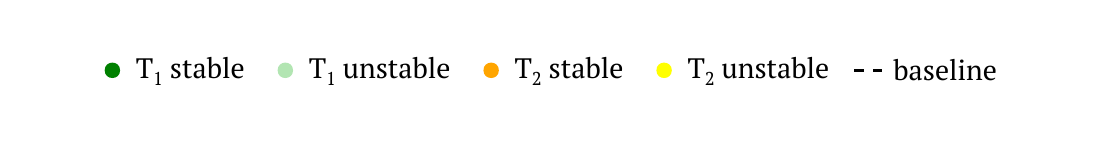}}
  \end{subfigure}
  \vspace{-1cm}
\caption{Bifurcation diagrams when $T_1$ employs immune checkpoint evasion. $\theta_1=0.6$ and $\theta_1=0.8$ for the $T_1$-only and $T_2$-only equilibria, respectively.}
\label{fig:bif_immune_checkpoint_evasion}
\end{figure}

\begin{figure}[!ht]
\centering
\captionsetup[subfigure]{justification=centering}
\begin{subfigure}{0.45\columnwidth}
 \caption{}
    \includegraphics[width=\textwidth]{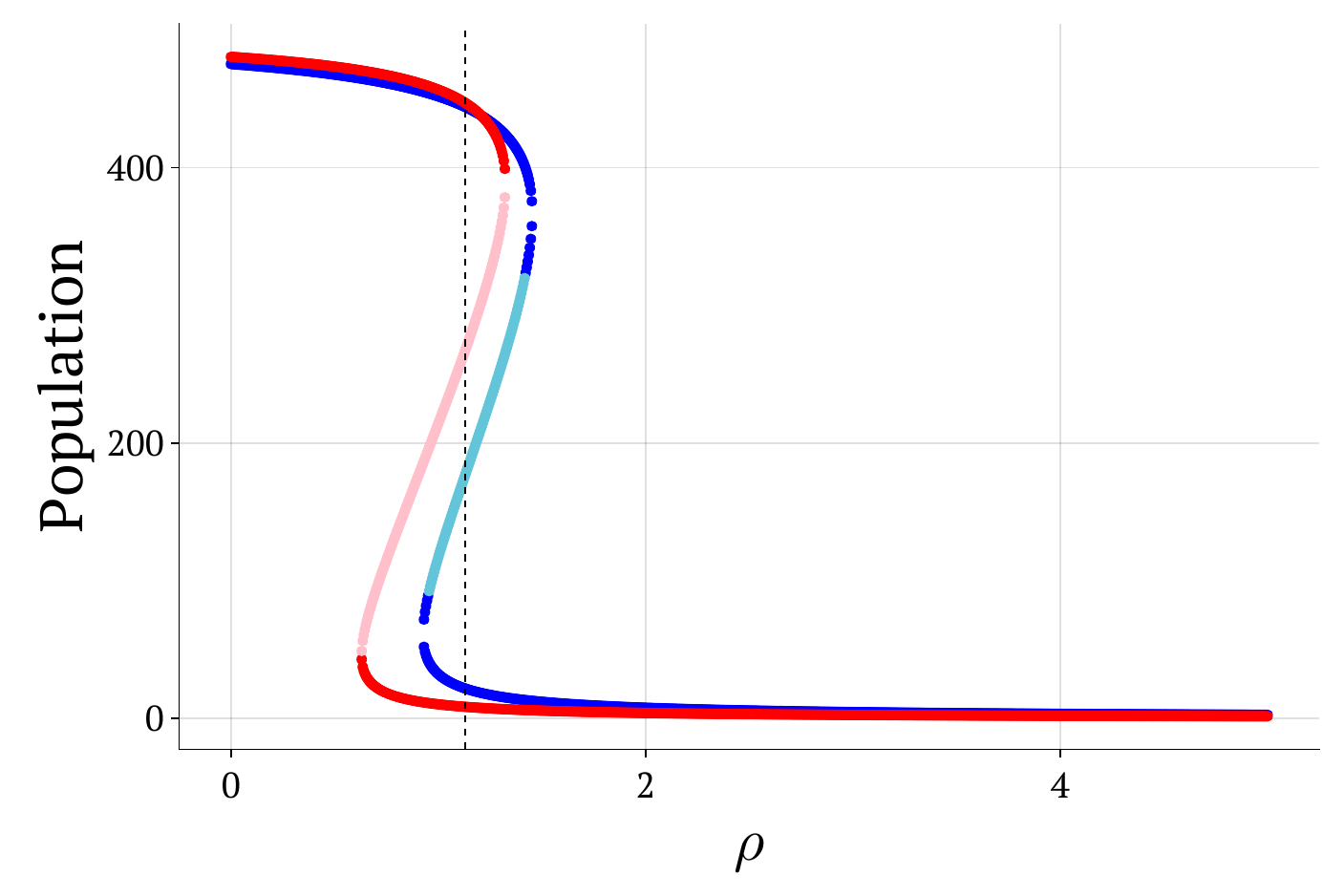}
    \label{fig:rho_2}
\end{subfigure}
\begin{subfigure}{0.45\columnwidth}
\caption{}
    \includegraphics[width=\textwidth]{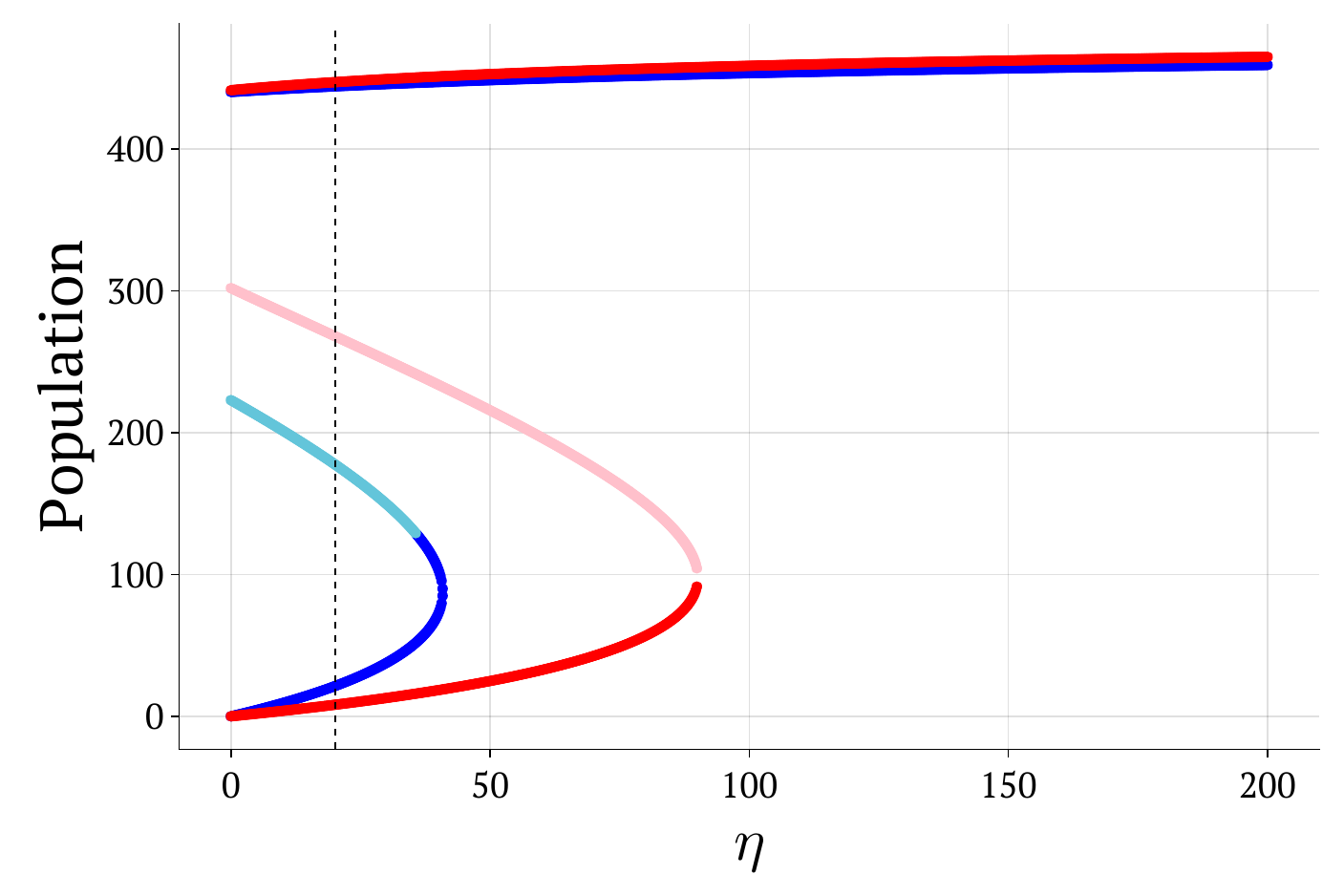}
    \label{fig:eta_2}
\end{subfigure}
\begin{subfigure}{0.45\columnwidth}
\caption{}
    \includegraphics[width=\textwidth]{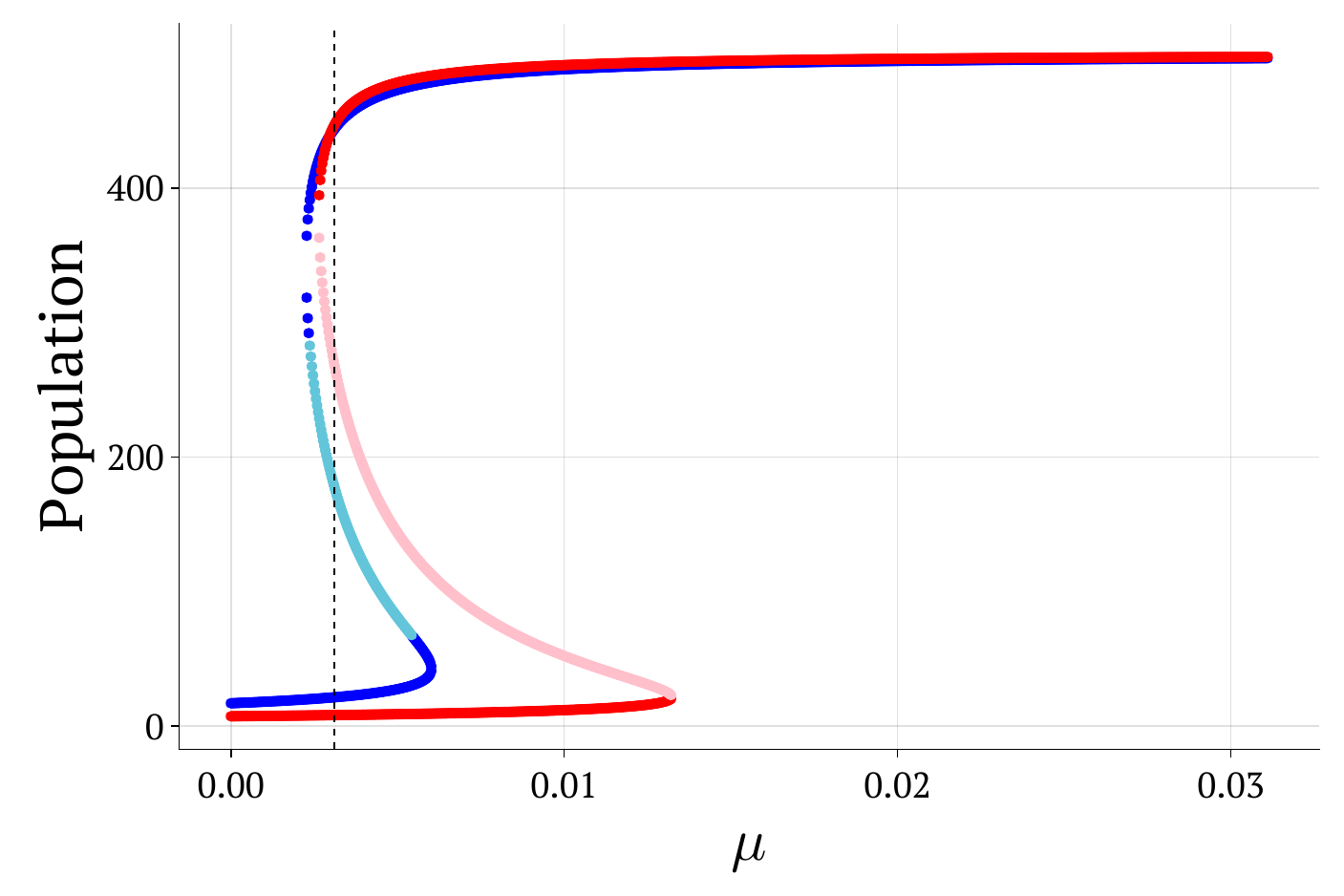}
    \label{fig:mu_2}
\end{subfigure}
\begin{subfigure}{0.45\columnwidth}
\caption{}
    \includegraphics[width=\textwidth]{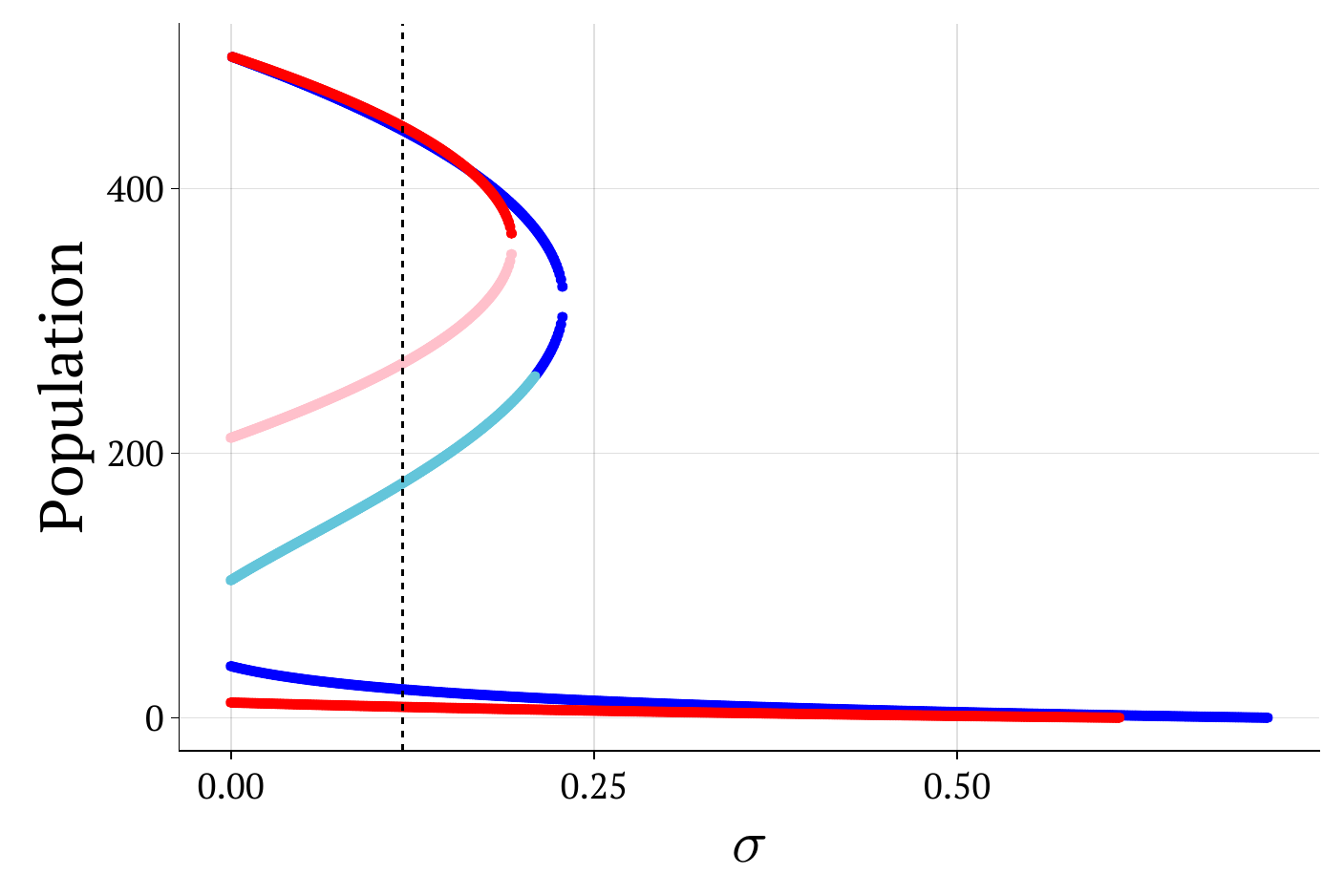}
    \label{fig:sigma_2}
\end{subfigure}
\begin{subfigure}{0.45\columnwidth}
\caption{}
    \includegraphics[width=\textwidth]{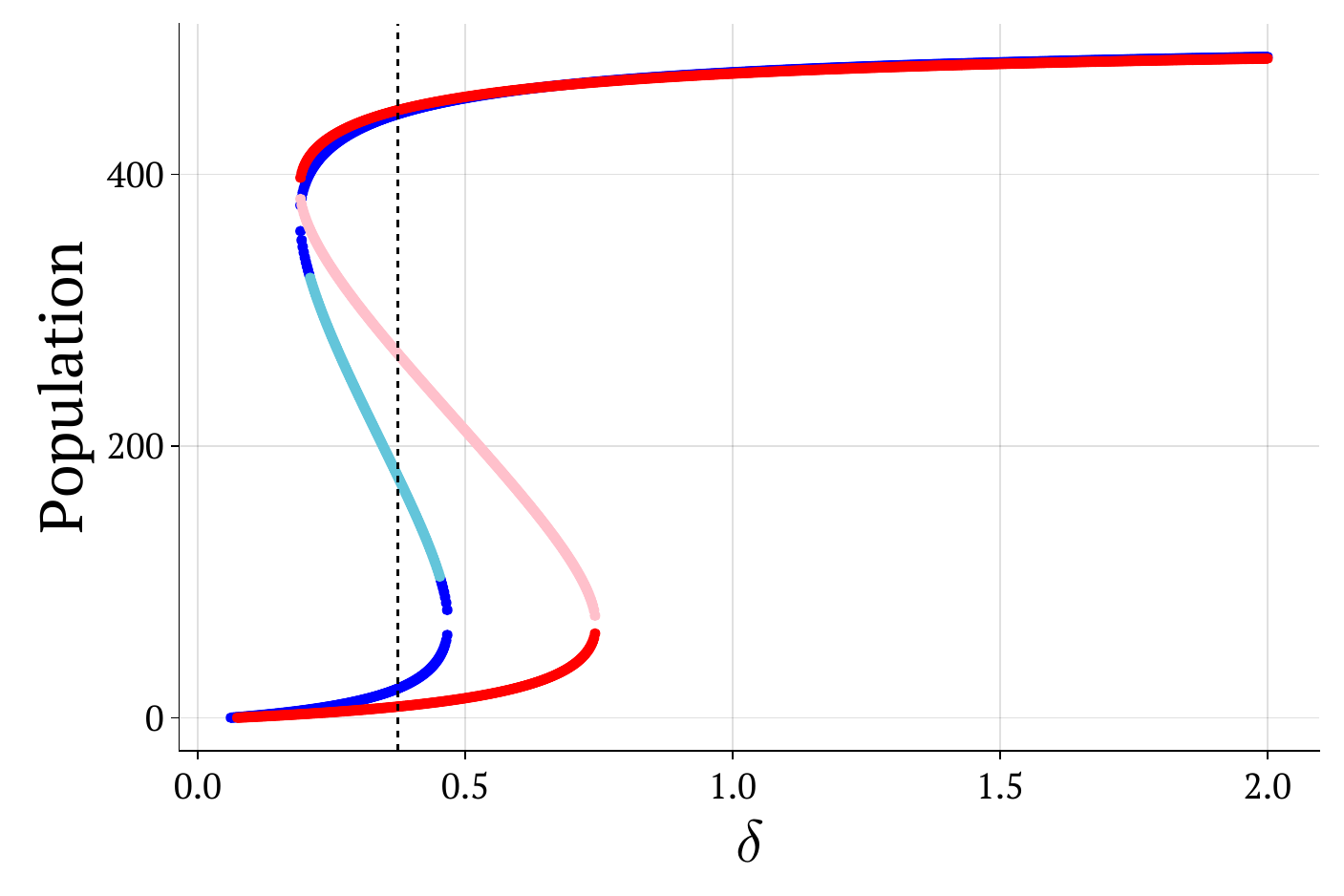}
    \label{fig:delta_2}
\end{subfigure}
\begin{subfigure}{0.45\columnwidth}
\caption{}
    \includegraphics[width=\textwidth]{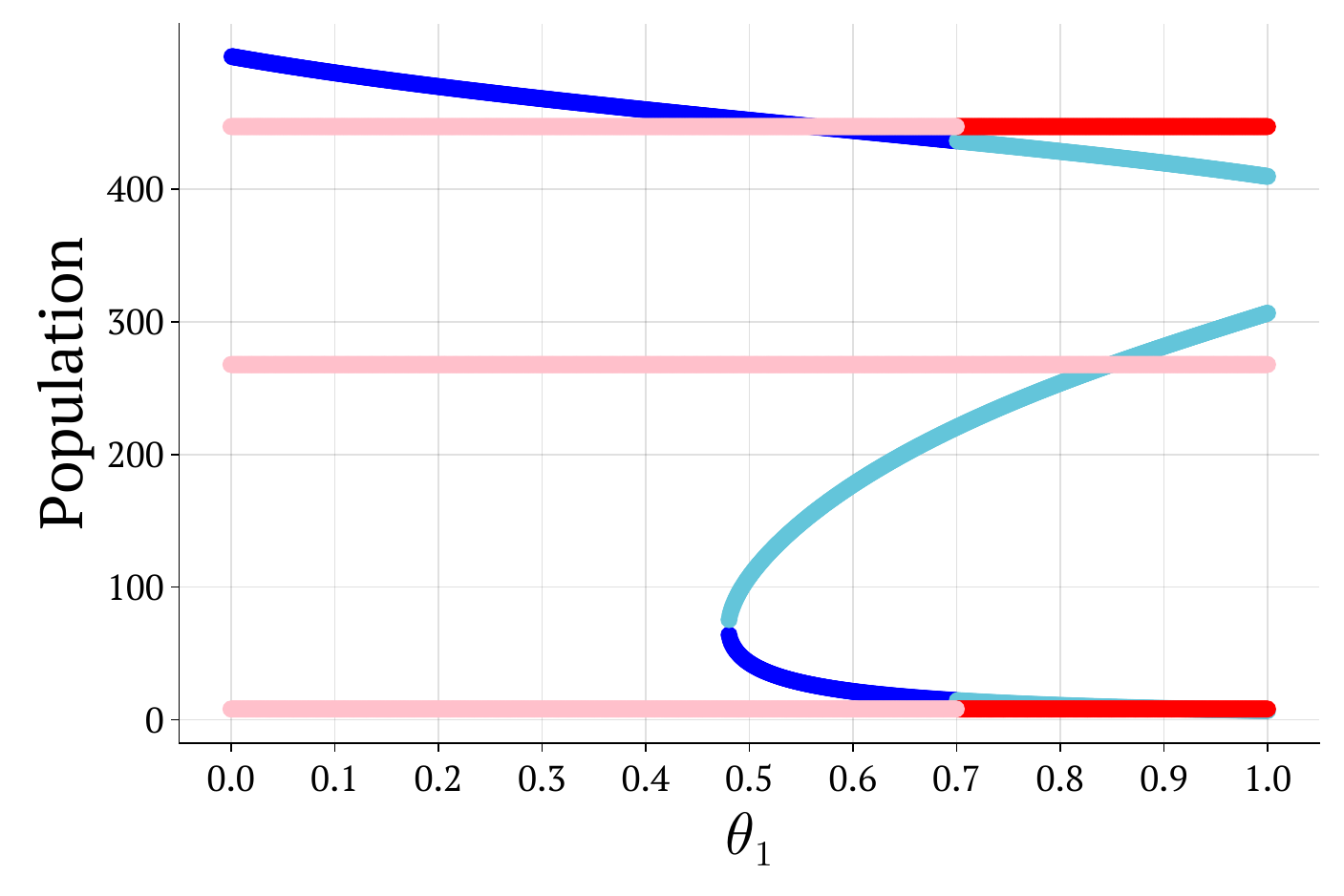}
    \label{fig:theta_2}
\end{subfigure}\\
\vspace{-0.6cm}
\begin{subfigure}{0.95\columnwidth}
  {\includegraphics[width=\linewidth]{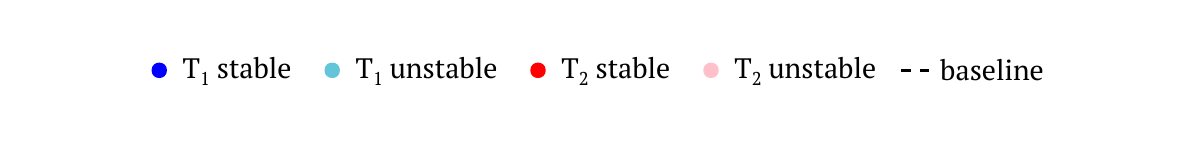}}
  \end{subfigure}
\vspace{-1cm}
\caption{Bifurcation diagrams when $T_1$ employs reduced antigen presentation. $\theta_1 = 0.6$ and $\theta_1 = 0.8$ for the $T_1$-only and $T_2$-only equilibria, respectively.}
\label{fig:bif_reduced_antigen_evasion}
\end{figure}

Here we consider the case where there is no chemotherapy ($\chi=0$) and thus $T_2$ is not chemo-resistant ($\theta_2=1$ and $\zeta_2=0$). We plot bifurcation diagrams for immune resistance tumors with immune checkpoint regulation and reduced antigen presentation in Figures \ref{fig:bif_immune_checkpoint_evasion} and \ref{fig:bif_reduced_antigen_evasion}, respectively. These figures provide insights into which biological processes most strongly determine treatment success in each scenario, and they identify effective therapeutic strategies that shift key parameter values. We set $\theta_1 = 0.6$ to obtain the $T_1$-only equilibrium and $\theta_1 = 0.8$ to obtain the $T_2$-only equilibrium. Under checkpoint evasion, two distinct scenarios arise depending on the value of $\phi$. When $\phi < 0.003$, the same therapeutic strategies are effective for both the $T_1$-only and $T_2$-only equilibria. However, when $\phi > 0.003$, no feasible modulation of immune-related parameters can control the $T_1$-only equilibrium, except by altering $\theta_1$. To study this scenarios, we set $\phi = \mu = 0.00311$, which lies beyond the identified threshold and highlights the altered dynamical behaviors. All remaining parameters were selected according to the baseline values listed in Table~\ref{tbl:param}. 

For each bifurcation diagram, the corresponding parameter was varied from zero up to ten times its baseline value while the remaining parameters are fixed to their baseline values. Although extending the parameter range beyond this interval could produce additional qualitative results, such extreme values are unlikely to be biologically realistic. And, in clinical settings, excessively large treatment intensities are neither feasible nor safe, as they may lead to severe toxicity and adverse effects for patients. Therefore, the parameter ranges considered here reflect therapeutically meaningful and physiologically plausible intervention levels.

Consider first the case where the tumor adopts an immune checkpoint regulation strategy (Figure \ref{fig:bif_immune_checkpoint_evasion}). For the $T_1$-only equilibria, a large tumor can only be prevented if $\mu$ is sufficiently low. Immune checkpoint regulation confers on $T_1$ a sufficient competitive advantage over both $T_2$ and the effector cell population $E$, allowing $T_1$ to dominate and stabilize at a large population size. The most effective therapeutic strategy in this case is the use of immune checkpoint inhibitor drugs. By blocking inhibitory receptors on tumor cells through antigen-mediated mechanisms, these therapies reduce the effector cell inactivation rate induced by $T_1$. As a result, the cytotoxic activity of the immune response is restored, limiting the growth and persistence of the $T_1$ population. On the other hand, low $T_2$-only equilibria can be achieved with either a sufficiently large $\rho$ or $\sigma$ or sufficiently small $\eta$ or $\delta$. These results indicate that when immune suppression is primarily functional, increasing immune cell abundance alone is insufficient. The limiting factor is the quality of the immune response rather than its quantity. Thus, therapies that restore cytotoxic function are predicted to be most effective.

Next, consider the reduced antigen presentation strategy, where $T_1$ decreases immune recognition without increasing exhaustion (Figure \ref{fig:bif_reduced_antigen_evasion}). Small $T_1$-only equilibria exist for high $\rho$ and $\sigma$ and low $\delta$. In this case, unlike the previous scenario in which $T_1$ employs an immune checkpoint regulation strategy, controlling a $T_1$-only population requires strengthening the effector immune response or reprogramming it to better recognize tumor cells. Mathematically, this corresponds to decreasing $\mu$ (the tumor-induced suppression rate) or increasing $\rho$ (the activation/proliferation rate of effector cells). From a therapeutic perspective, such tumors may be controlled using immune checkpoint inhibitors, CAR-T cell therapy, or cancer vaccines, all of which enhance immune recognition and activity against tumor cells.

Control of $T_2$-only equilibria when $T_1$ employs the reduced antigen presentation strategy is similar to the scenario where $T_1$ applies immune checkpoint regulation. In this case, the primary strategy is to increase the overall effector cell population. This can be achieved by increasing the natural birth rate $\sigma$ of effector cells or decreasing their natural death rate $\delta$, thereby strengthening the immune system’s baseline capacity to suppress tumor growth. Note that increasing $\sigma$ can also result in a small controlled tumor in the $T_1$-only equilibrium case. However, the required value of $\sigma$ to accomplish such a therapy is significantly larger than for the $T_2$-only case.

Note that the bifurcation diagrams with respect to $\eta$ (Figures \ref{fig:eta_1} and \ref{fig:eta_2}) provide limited insight for therapeutic design. However, they may be biologically relevant in the context of immuno-compromised individuals, where $\eta$ is significantly elevated and immune activation is substantially weakened. In contrast, the bifurcation diagrams for $\sigma$ and $\delta$ (Figures \ref{fig:sigma_2} and \ref{fig:delta_2}) clearly demonstrate that once the threshold condition in Proposition \ref{prop_1} is satisfied, the system transitions to the tumor-free equilibrium. This reflects the critical role of the ratio $\sigma/\delta$ in determining whether the immune system successfully eliminates the tumor. Under immune checkpoint evasion, the bifurcation diagram with respect to $\mu$ (Figure \ref{fig:mu_2}) indicates that changes in $\mu$ affect the stability of the high population equilibrium. In particular, as $\mu$ increases the equilibria change from a low population to a high population. Considering all plots suggests that, within the $T_1$-only regime, varying $\mu$ alone does not provide a clear pathway toward tumor elimination. In contrast, in the $T_2$-only regime, adjusting parameters such as $\sigma$ and $\delta$ can drive the system toward the tumor-free equilibrium, indicating a more feasible route to elimination.

\begin{table}[ht]
\centering
\begin{subtable}{\textwidth}
\centering
\caption{Checkpoint Evasion}
\begin{tabular}{c c c}
\toprule
Tumor Phenotype & Best Target & Therapeutic Focus \\
\midrule
$T_1$-only(immune evasive dominant)  & $\downarrow \mu$ 
& Checkpoint inhibitors \cite{Topalian12}\\
$T_2$-only(immune sensitive dominant) & $\uparrow \sigma$, $\downarrow \delta$ 
& Adoptive T-cell transfer\cite{Dudley02},\\ & & Cytotoxic Therapy \cite{Rosenberg14} \\
\bottomrule
\end{tabular}
\end{subtable}
\vspace{0.5cm}
\begin{subtable}{\textwidth}
\centering
\caption{Reduced Antigen Presentation}
\begin{tabular}{c c c}
\toprule
Tumor Phenotype & Best Target & Therapeutic Focus \\
\midrule
$T_1$-only (immune evasive dominant)  & $\uparrow \rho$, $\downarrow \mu$ 
& CAR-T therapy\cite{Maude14}, vaccines \\
$T_2$-only(immune sensitive dominant) & $\uparrow \sigma$, $\downarrow \delta$ 
& Adoptive T-cell transfer\cite{Dudley02},\\ & &  Cytotoxic Therapy \cite{Rosenberg14} \\
\bottomrule
\end{tabular}
\end{subtable}
\caption{Phenotype-guided immunotherapy strategies under two resistance mechanisms: (a) immune checkpoint evasion and (b) reduced antigen presentation. $T_1$-only and $T_2$-only denote dominance of the immune-evasive phenotype and the wild-type phenotype, respectively. For each dominant phenotype, the table identifies the most effective parameter targets predicted by the model and the corresponding therapeutic strategies. Arrows indicate the direction of modulation, where $\uparrow$ denotes increasing a parameter and $\downarrow$ denotes decreasing a parameter.}
\label{tbl:immunotherapy_guide}
\end{table}

Overall, the combined bifurcation analyses demonstrate that immune checkpoint regulation and reduced antigen presentation represent dynamically distinct mechanisms of tumor persistence as summarized in Table \ref{tbl:immunotherapy_guide}. Effective therapy must therefore target the dominant mechanism in each patient. The model provides a theoretical framework for rational, phenotype-guided immunotherapy design and explains why different tumors respond preferentially to different immunotherapy strategies. Finally, the bifurcation analyses highlight the importance of patient-specific parameter identification. Since stability thresholds depend on the ratio $\sigma/\delta$ and on $\max\{\alpha, \alpha_1/\theta_1\}$, small variations in immune competence or tumor immunogenicity may significantly alter therapeutic response. This supports a precision-medicine approach in which immunotherapy is selected based on measurable biomarkers such as PD-L1 expression, T-cell infiltration, exhaustion markers, or antigen presentation profiles.

\subsection{Bifurcation Diagrams for Chemo-resistance and Immuno-resistance} \label{results:bifurcation_chemo_resistant}

Although immune control may occasionally suppress tumor growth, immune dynamics alone are generally insufficient to fully eliminate the tumor. Thus, we consider the case where both chemotherapy and immunotherapy are applied. We find that chemotherapy significantly enlarges the parameter region corresponding to tumor control or tumor-free equilibria. It reduces tumor burden and directly targets resistant subpopulations, while immunotherapy enhances immune-mediated tumor suppression. However, unlike the previous scenario in which there is a knife-edge situation between $T_1$-only and $T_2$-only equilibria, coexistence of both types is supported over a range of parameter values.

\begin{figure}[!ht]
\centering
\captionsetup[subfigure]{justification=centering}
\begin{subfigure}{0.45\columnwidth}
\caption{}
    \includegraphics[width=\textwidth]{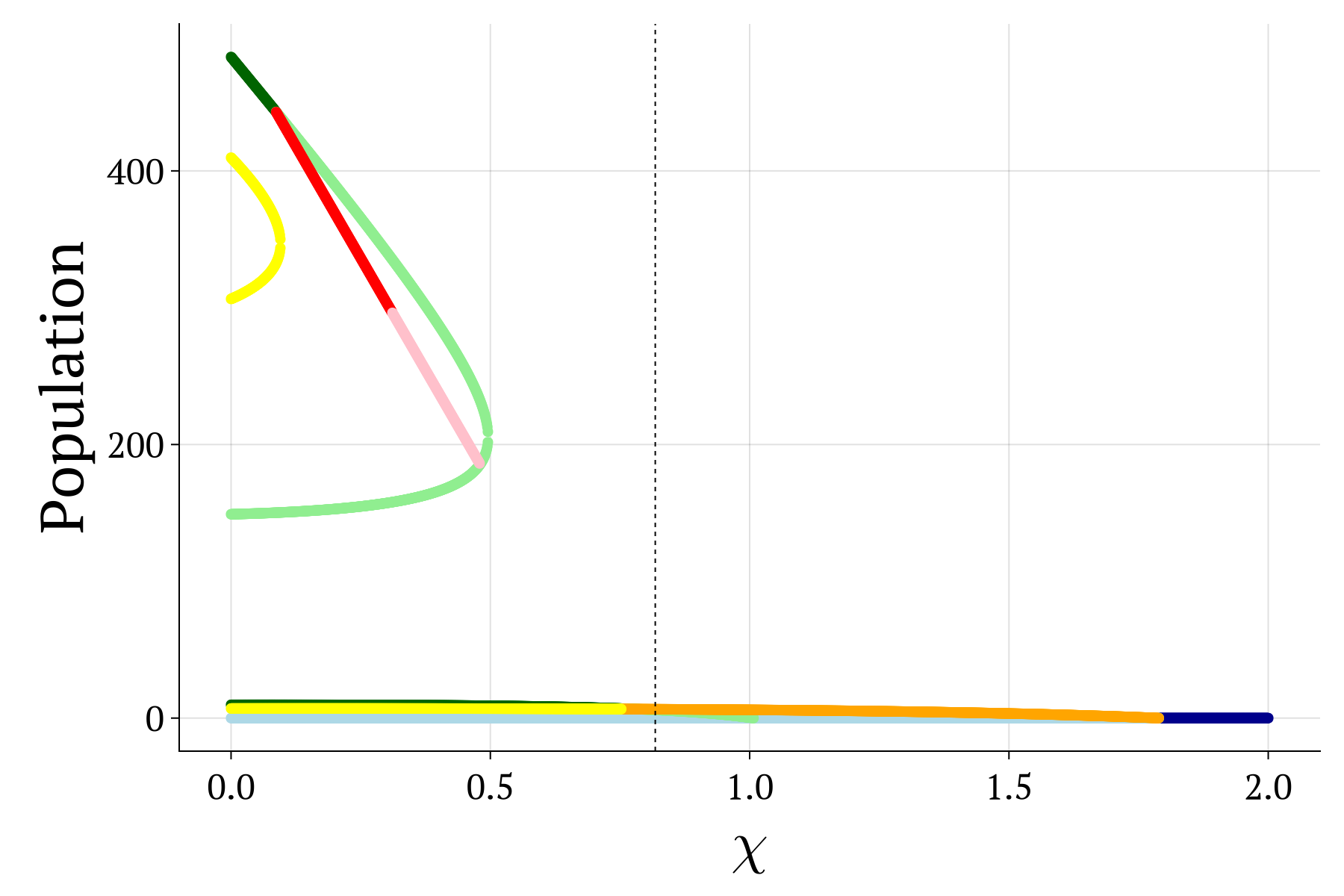}
\end{subfigure}
\begin{subfigure}{0.45\columnwidth}
\caption{}
    \includegraphics[width=\textwidth]{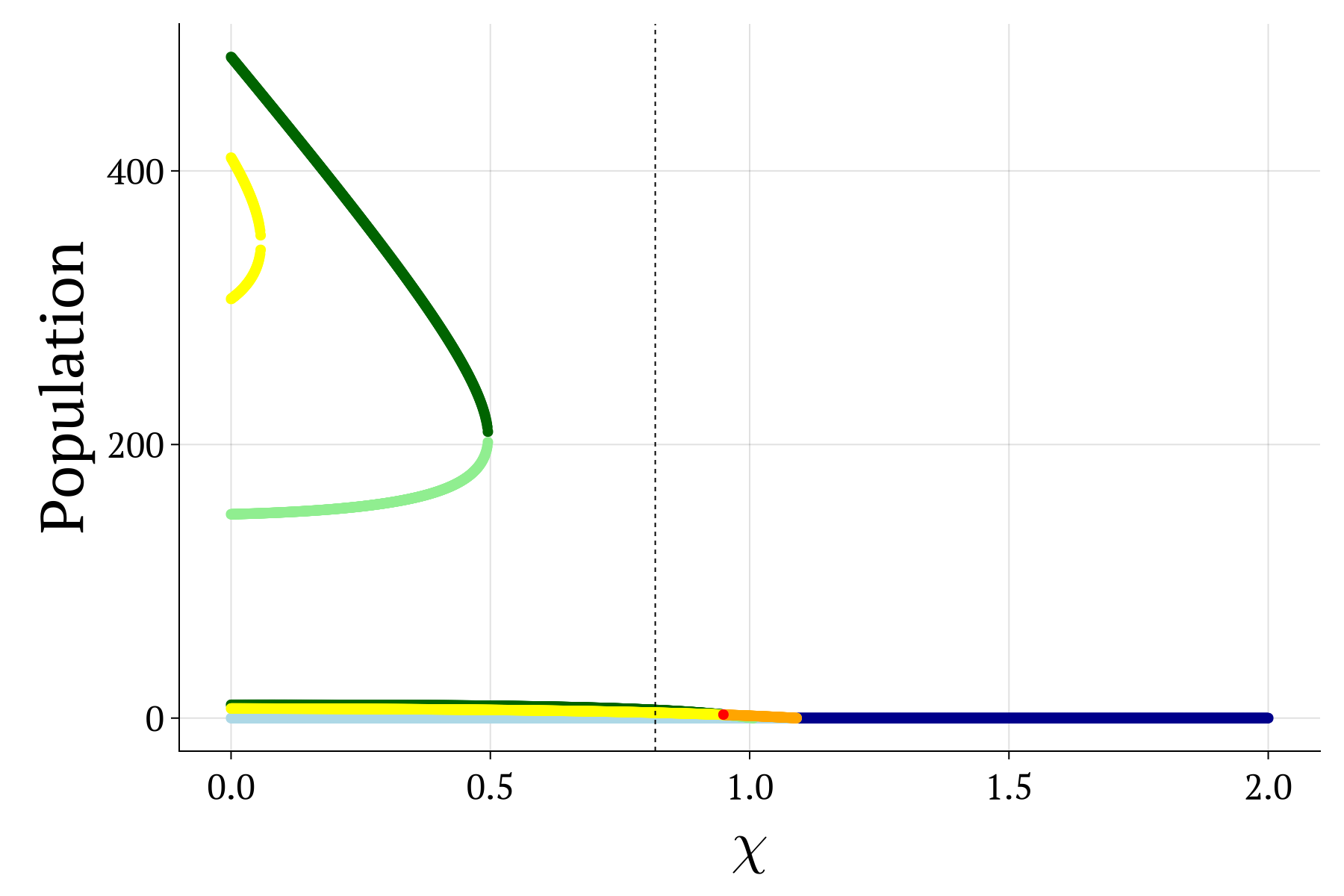}
\end{subfigure}
\begin{subfigure}{0.45\columnwidth}
\caption{}
    \includegraphics[width=\textwidth]{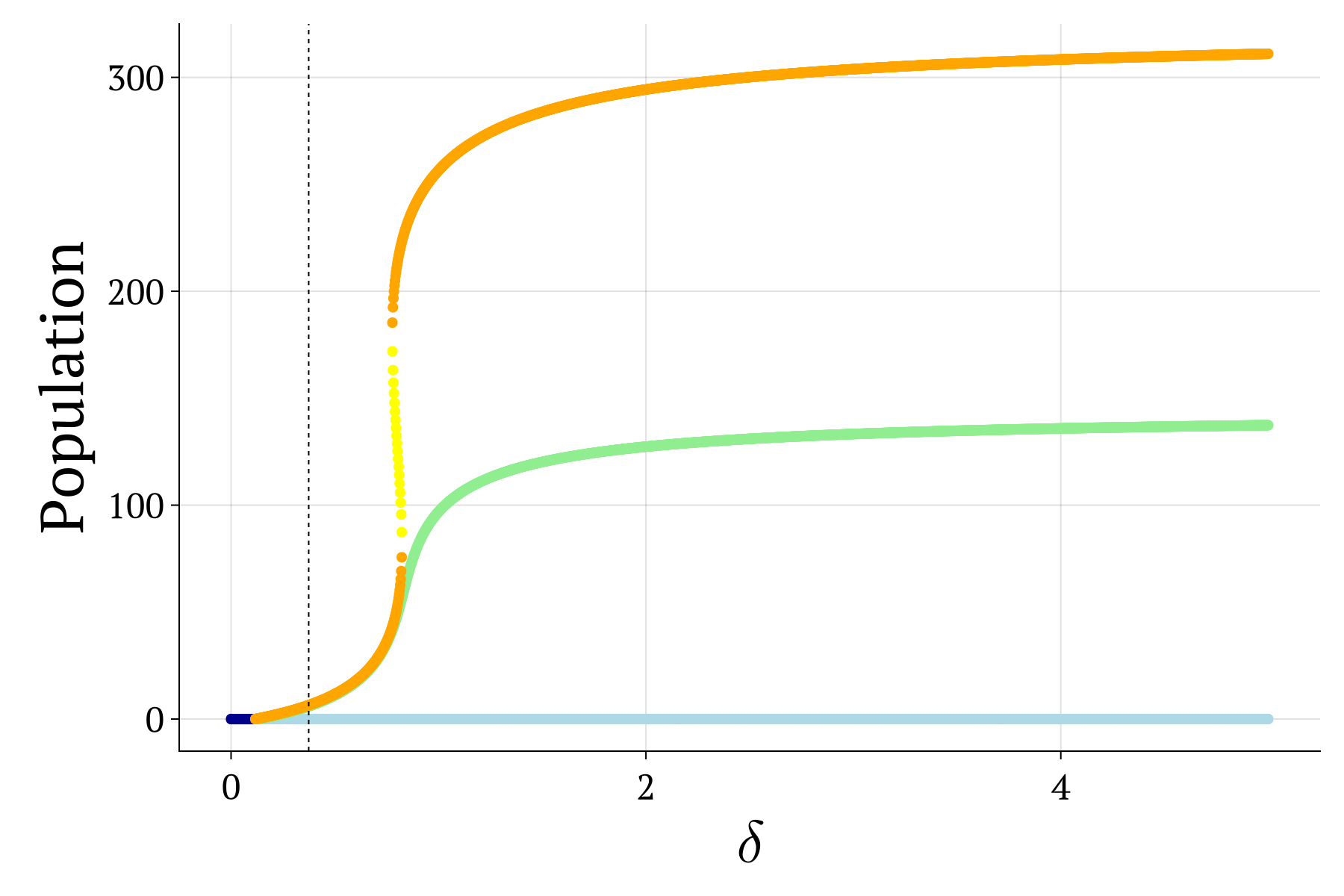}
\end{subfigure}
\begin{subfigure}{0.45\columnwidth}
\caption{}
    \includegraphics[width=\textwidth]{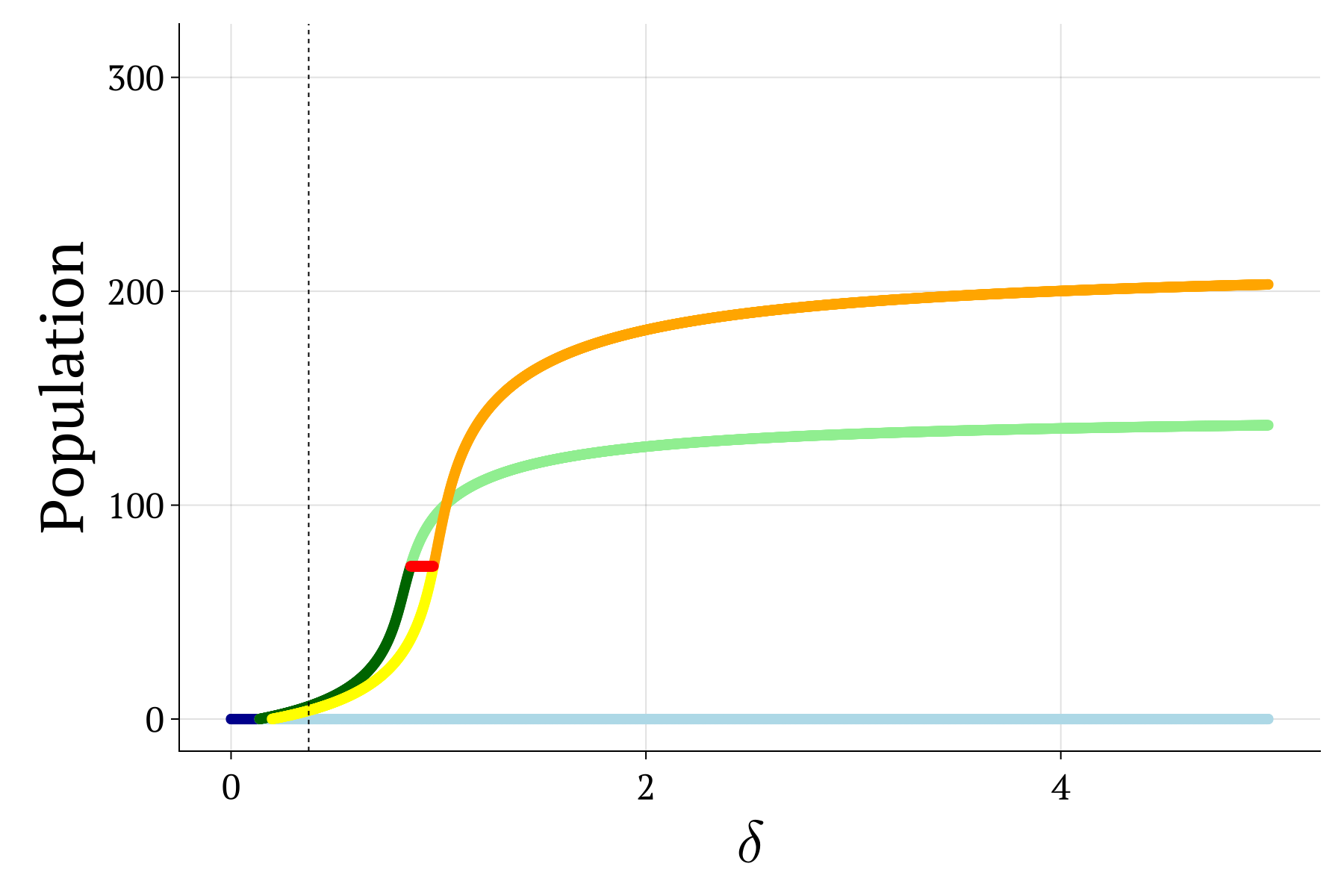}
\end{subfigure}
\begin{subfigure}{0.45\columnwidth}
\caption{}
    \includegraphics[width=\textwidth]{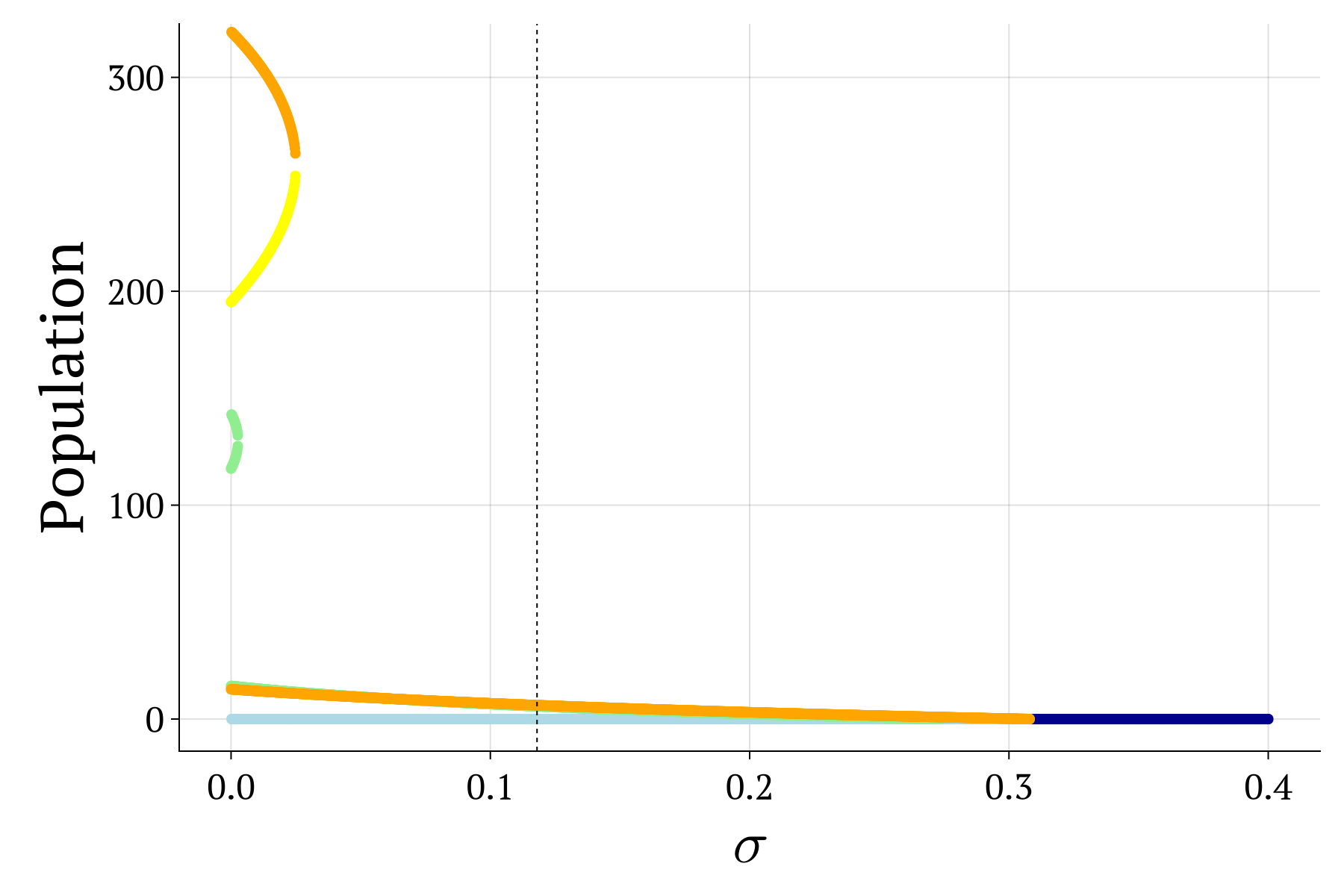}
\end{subfigure}
\begin{subfigure}{0.45\columnwidth}
\caption{}
    \includegraphics[width=\textwidth]{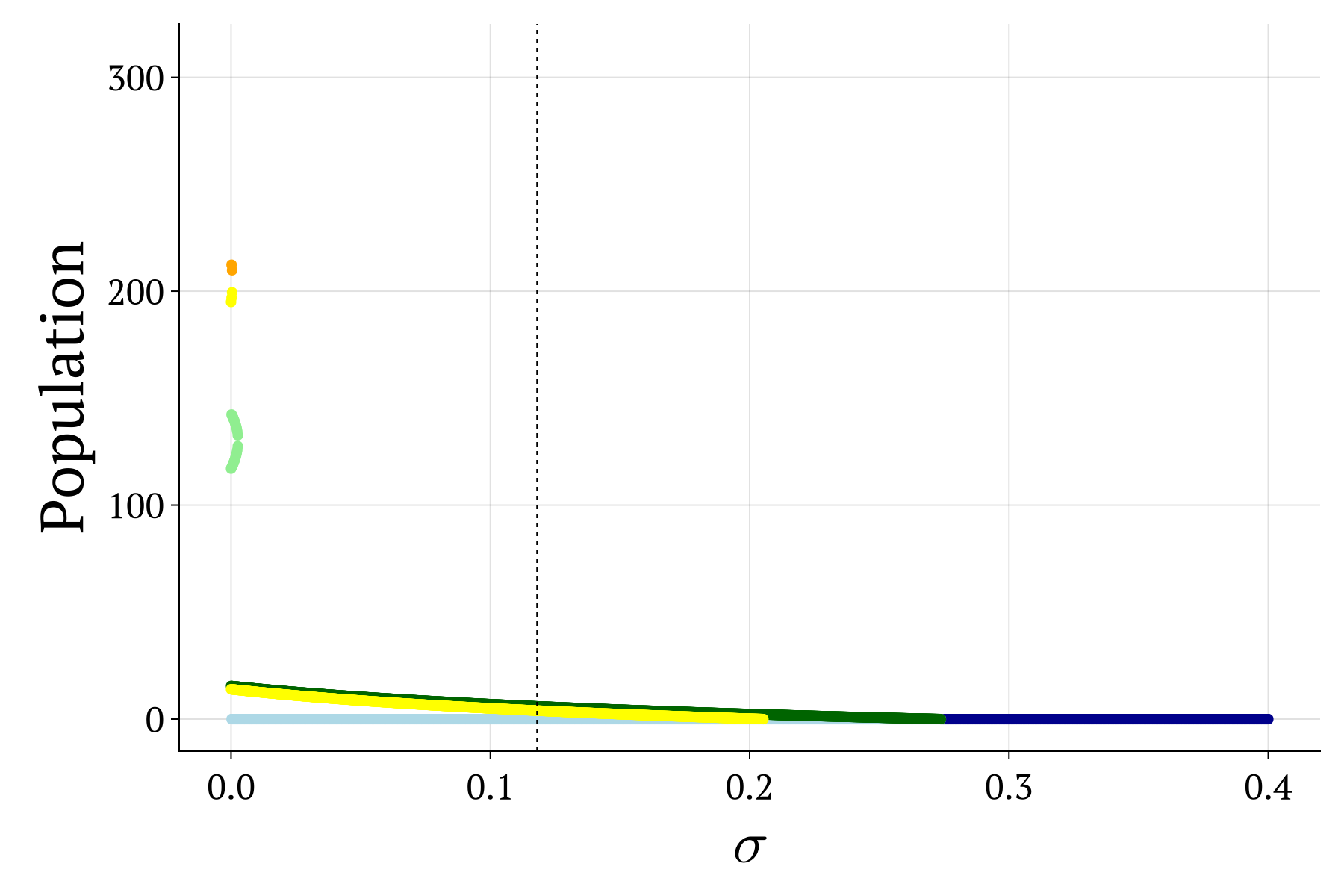}
\end{subfigure}\\
\begin{subfigure}{\columnwidth}
\includegraphics[width=\linewidth]{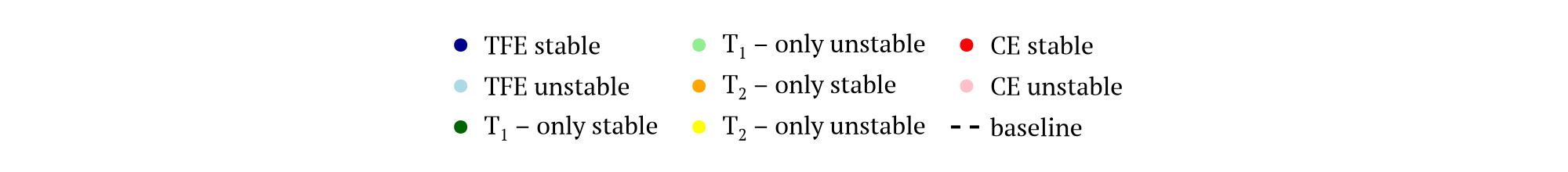}
\end{subfigure}
\vspace{-1cm}
\caption{Bifurcation diagrams for $\chi$, $\delta$, and $\sigma$ when $T_2$ is chemo-resistant and $T_1$ employs immune checkpoint evasion. $\theta_1=\theta_2=0.5$ and $\theta_1=0.5 < \theta_2=0.8$ in the left and right columns, respectively.}
\label{fig:bif_chemo_ICE}
\end{figure}

\begin{figure}[!ht]
\centering
\captionsetup[subfigure]{justification=centering}
\begin{subfigure}{0.45\columnwidth}
\caption{}
    \includegraphics[width=\textwidth]{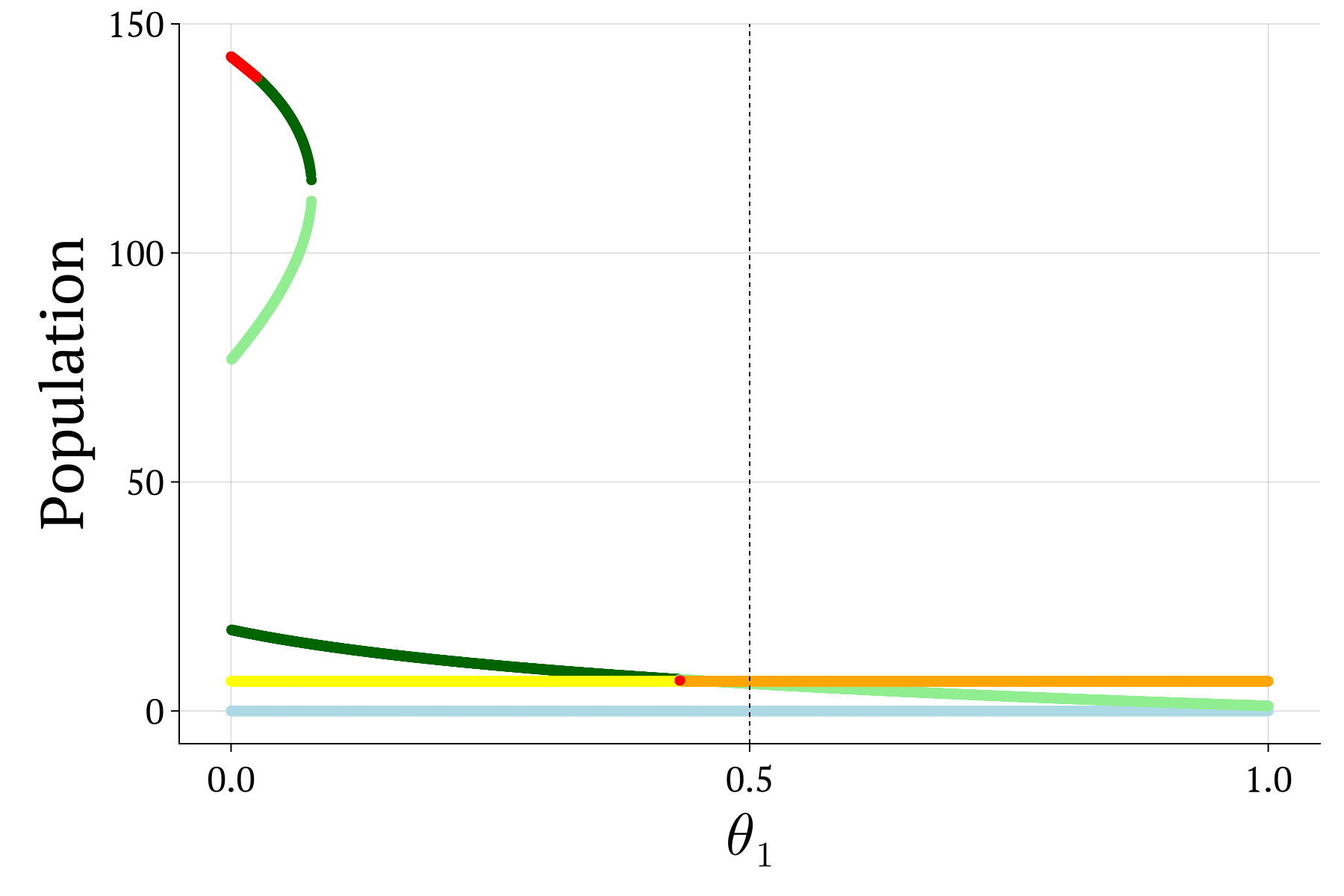}
\end{subfigure}
\begin{subfigure}{0.45\columnwidth}
\caption{}
    \includegraphics[width=\textwidth]{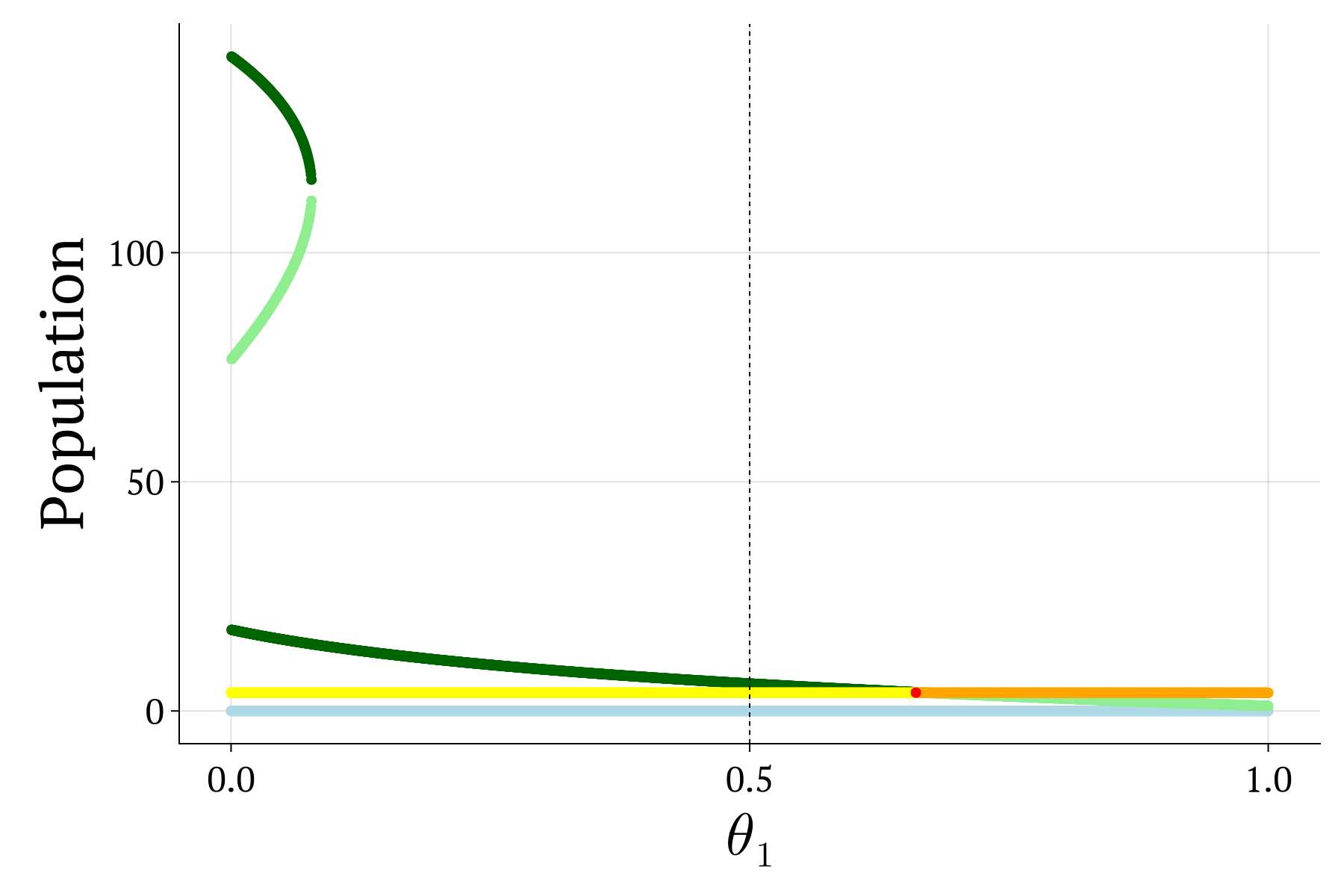}
\end{subfigure}
\begin{subfigure}{0.45\columnwidth}
\caption{}
    \includegraphics[width=\textwidth]{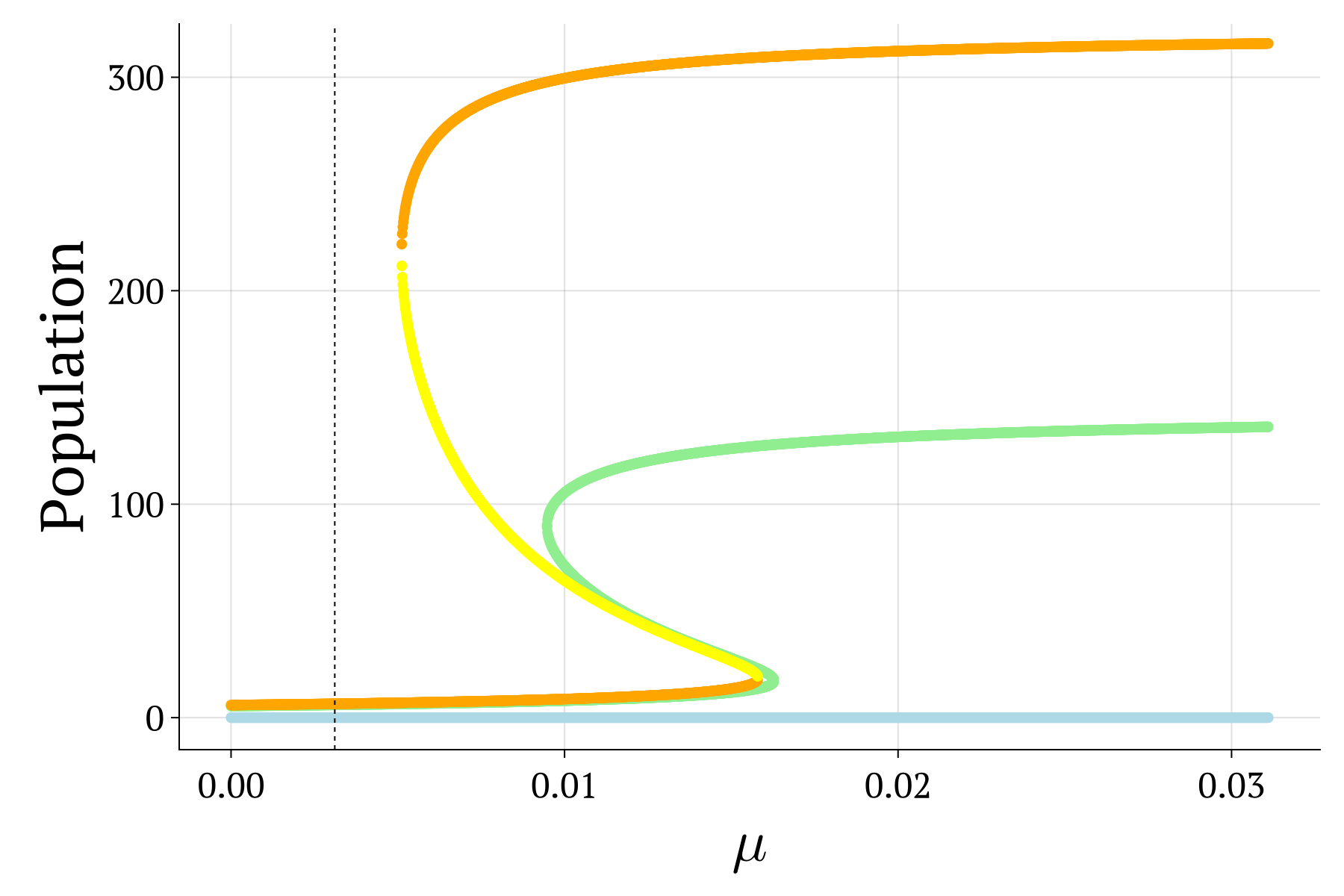}
\end{subfigure}
\begin{subfigure}{0.45\columnwidth}
    \includegraphics[width=\textwidth]{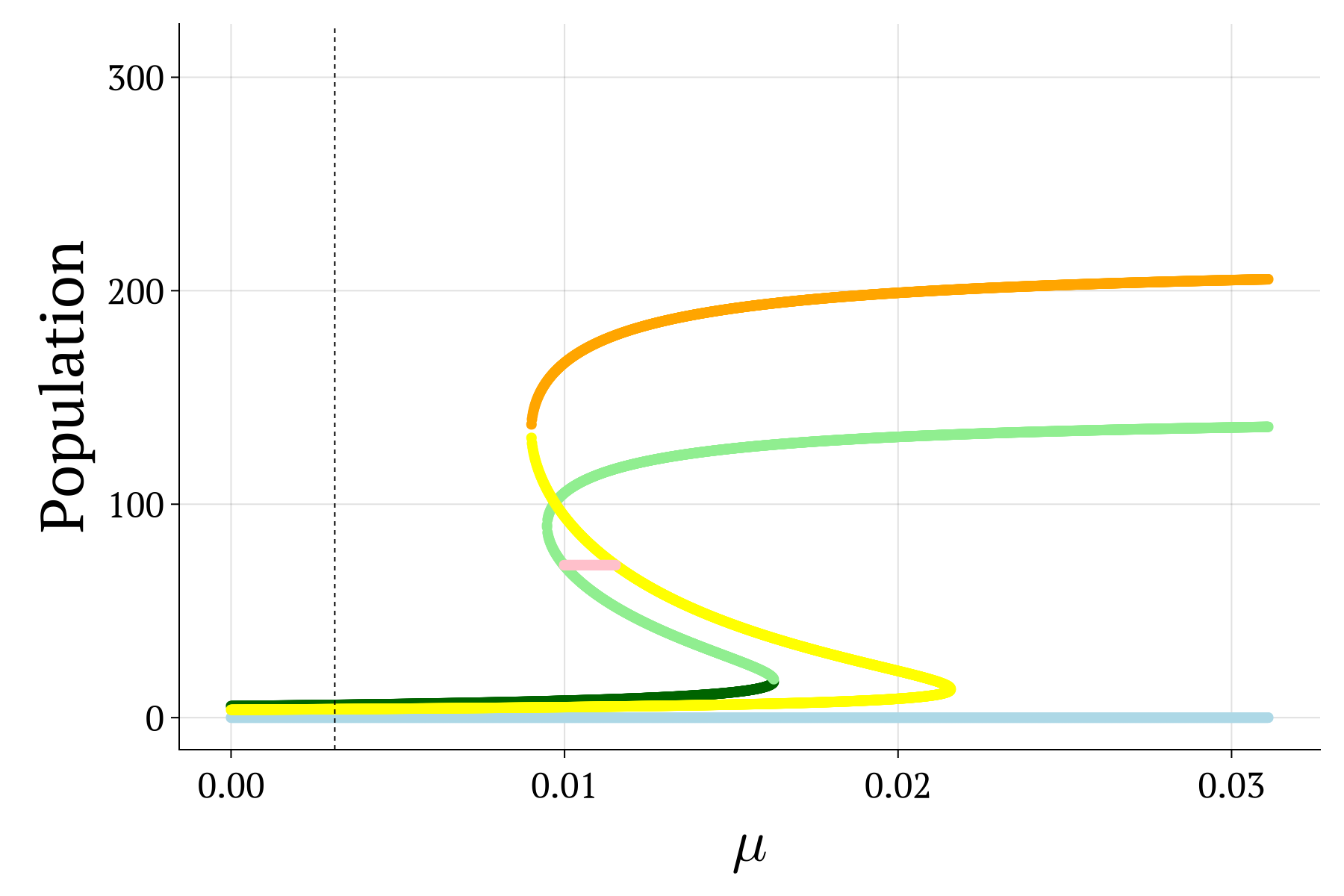}
\end{subfigure}
\begin{subfigure}{0.45\columnwidth}
\caption{}
    \includegraphics[width=\textwidth]{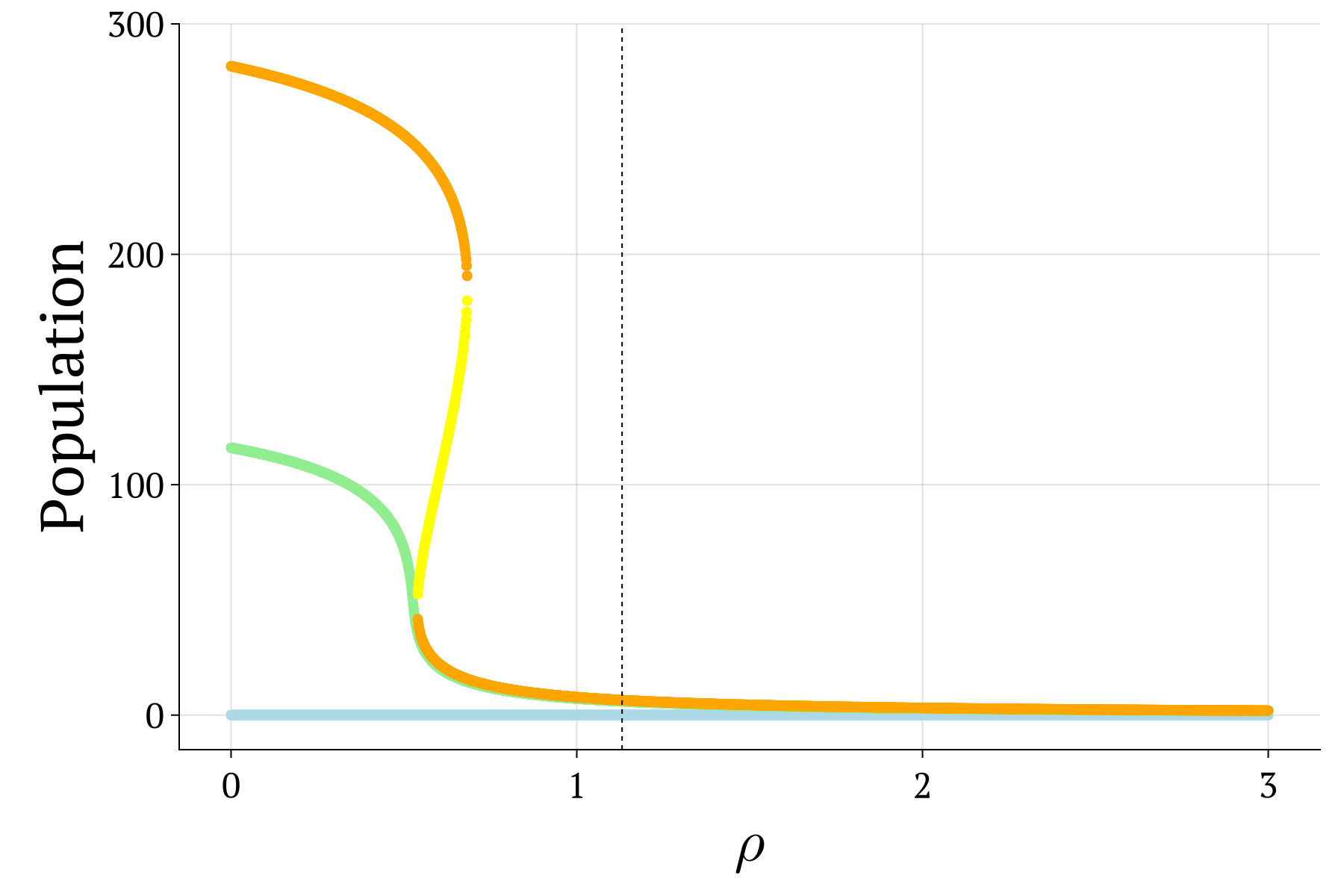}
\end{subfigure}
\begin{subfigure}{0.45\columnwidth}
\caption{}
    \includegraphics[width=\textwidth]{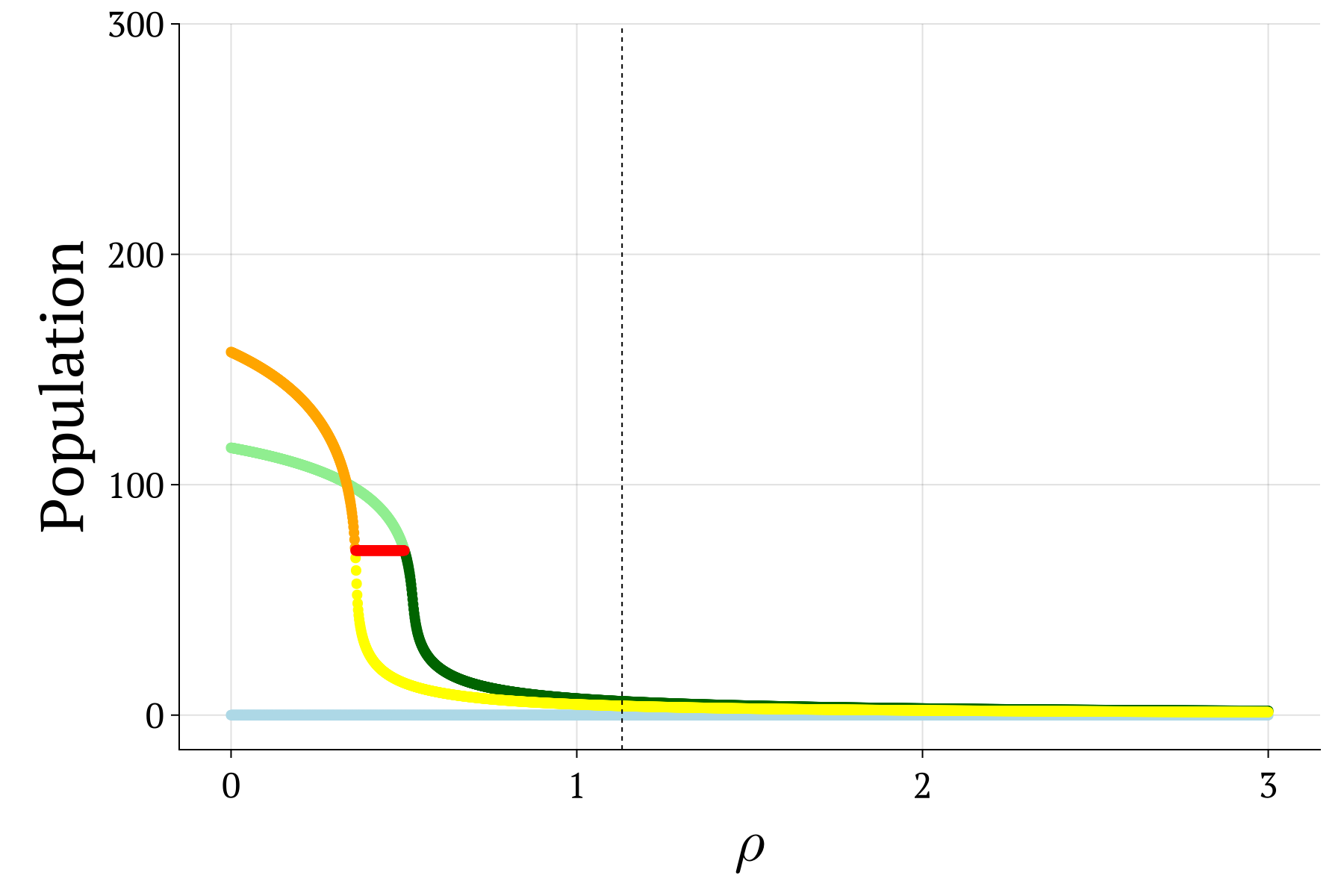}
\end{subfigure}\\
\begin{subfigure}{\columnwidth}
\includegraphics[width=\linewidth]{figures/Chemo_bifurcation_legend.png}
  \end{subfigure}
\vspace{-1cm}
\caption{Bifurcation diagrams for $\theta_1$, $\mu$, and $\rho$ when $T_2$ is chemo-resistant and $T_1$ employs immune checkpoint evasion. Except when varying $\theta_1$, $\theta_1=\theta_2=0.5$ and $\theta_1=0.5 < \theta_2=0.8$ in the left and right columns, respectively.}
\label{fig:bif_immuno_ICE}
\end{figure}

\begin{figure}[!ht]
\centering
\captionsetup[subfigure]{justification=centering}
\begin{subfigure}{0.45\columnwidth}
 \caption{}
    \includegraphics[width=\textwidth]{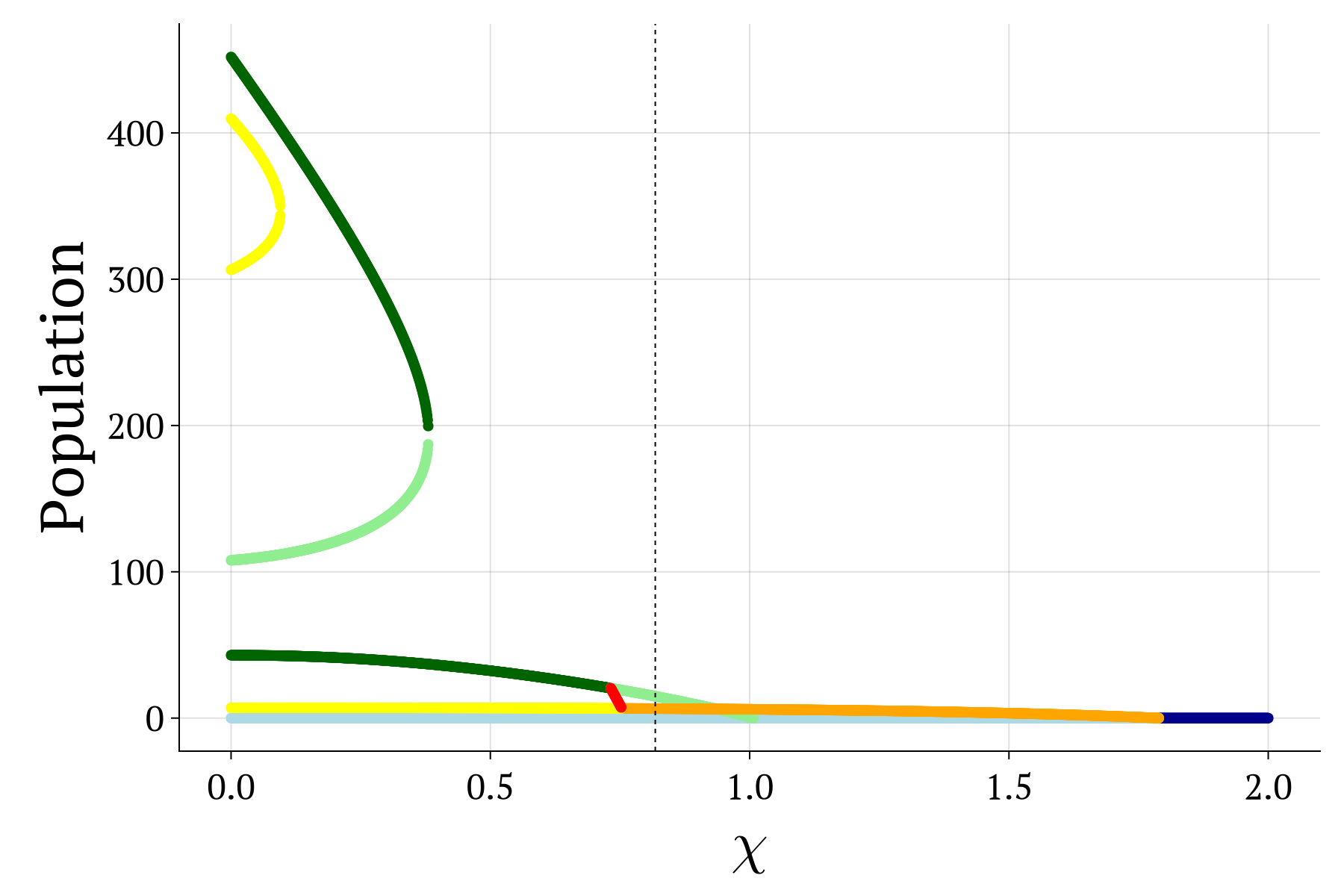}
\end{subfigure}
\begin{subfigure}{0.45\columnwidth}
 \caption{}
    \includegraphics[width=\textwidth]{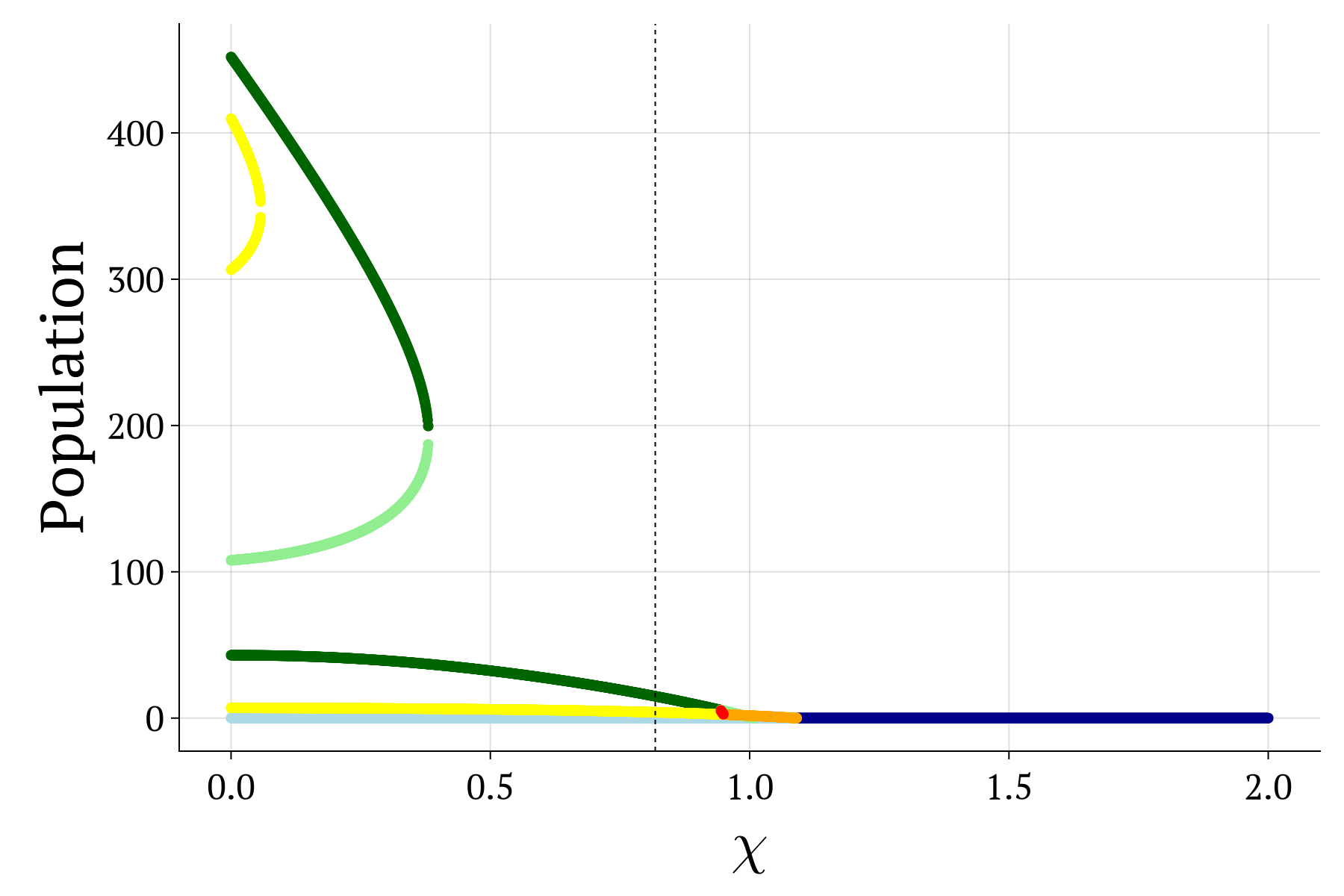}
\end{subfigure}
\begin{subfigure}{0.45\columnwidth}
 \caption{}
    \includegraphics[width=\textwidth]{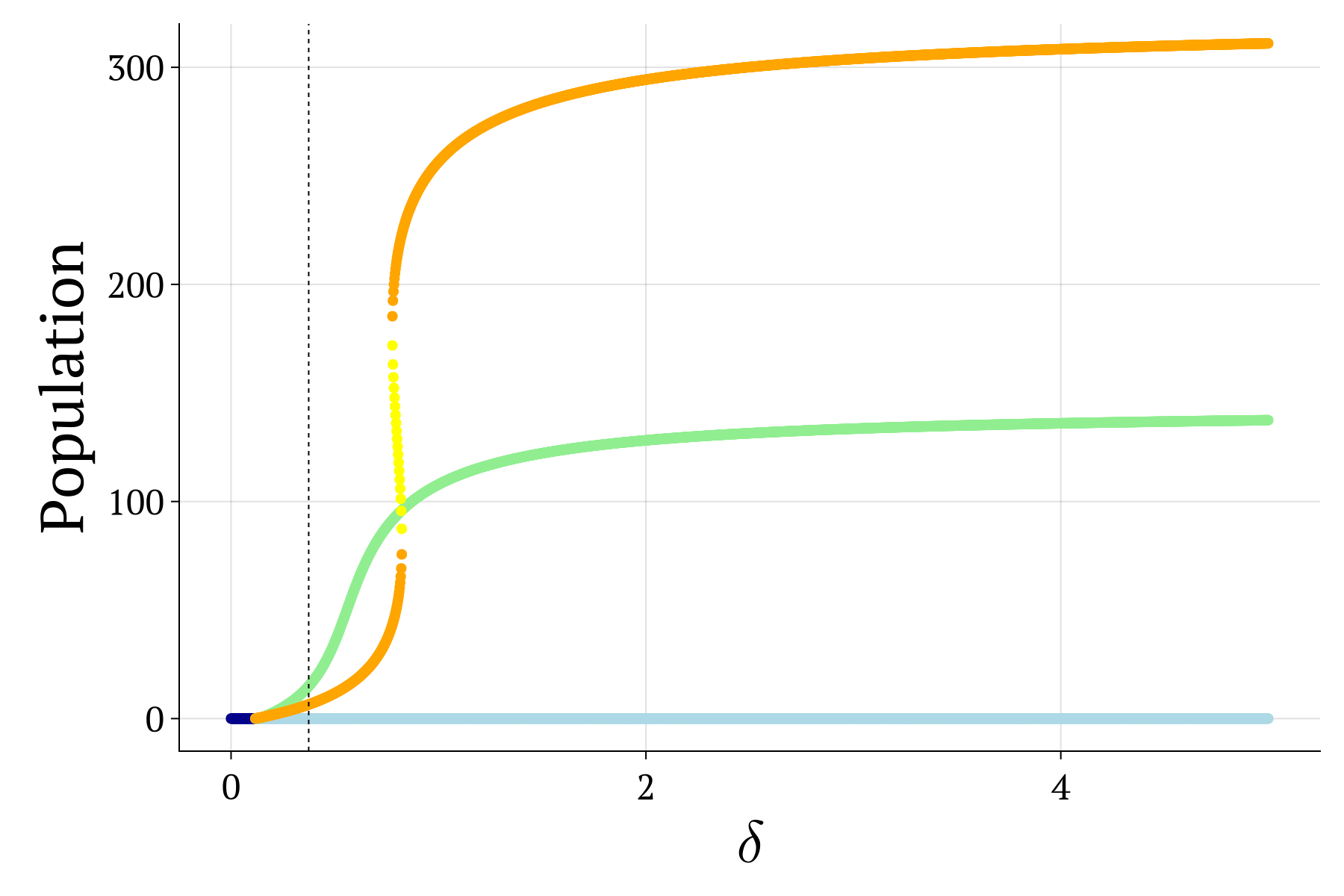}
\end{subfigure}
\begin{subfigure}{0.45\columnwidth}
 \caption{}
    \includegraphics[width=\textwidth]{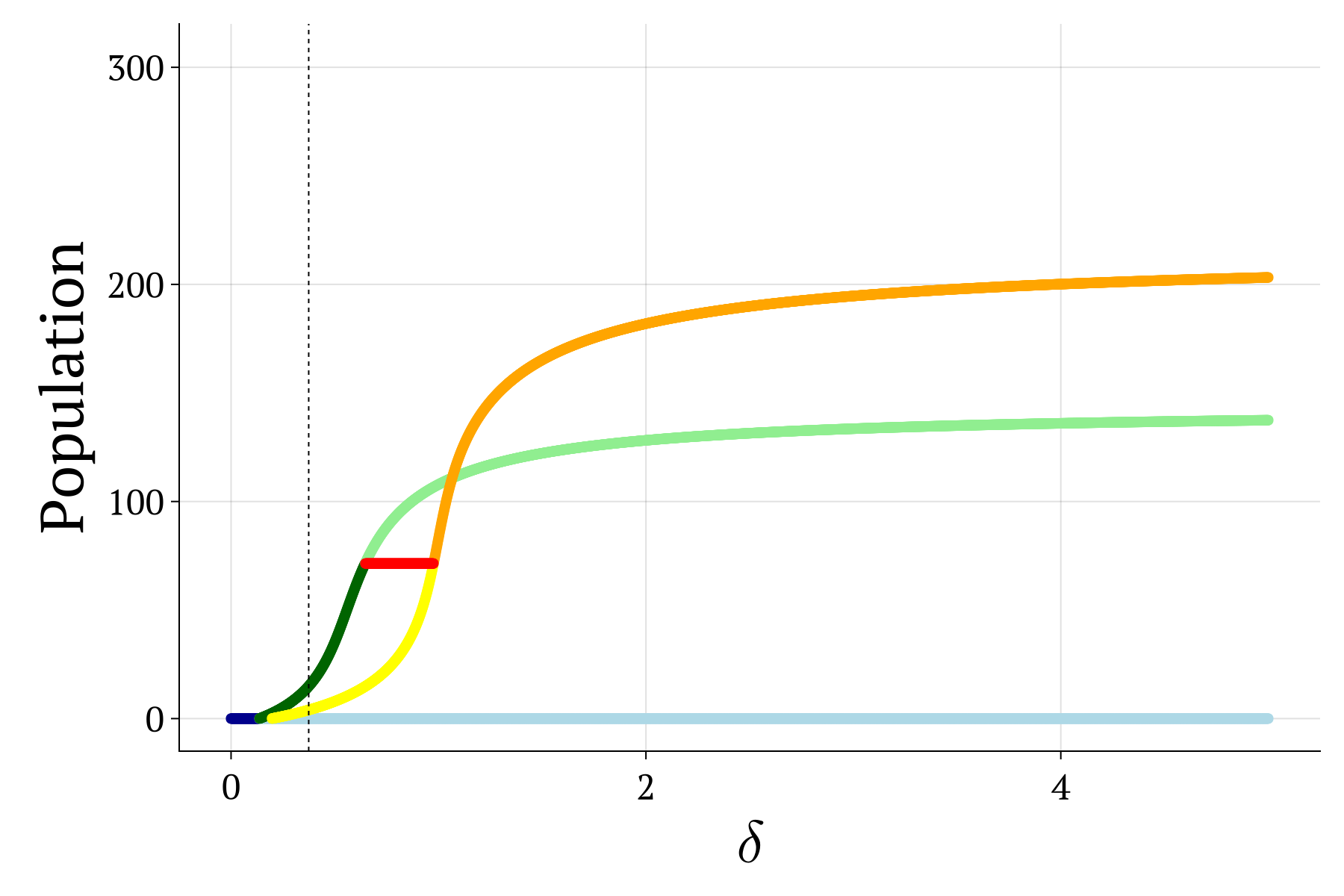}
\end{subfigure}
\begin{subfigure}{0.45\columnwidth}
 \caption{}
    \includegraphics[width=\textwidth]{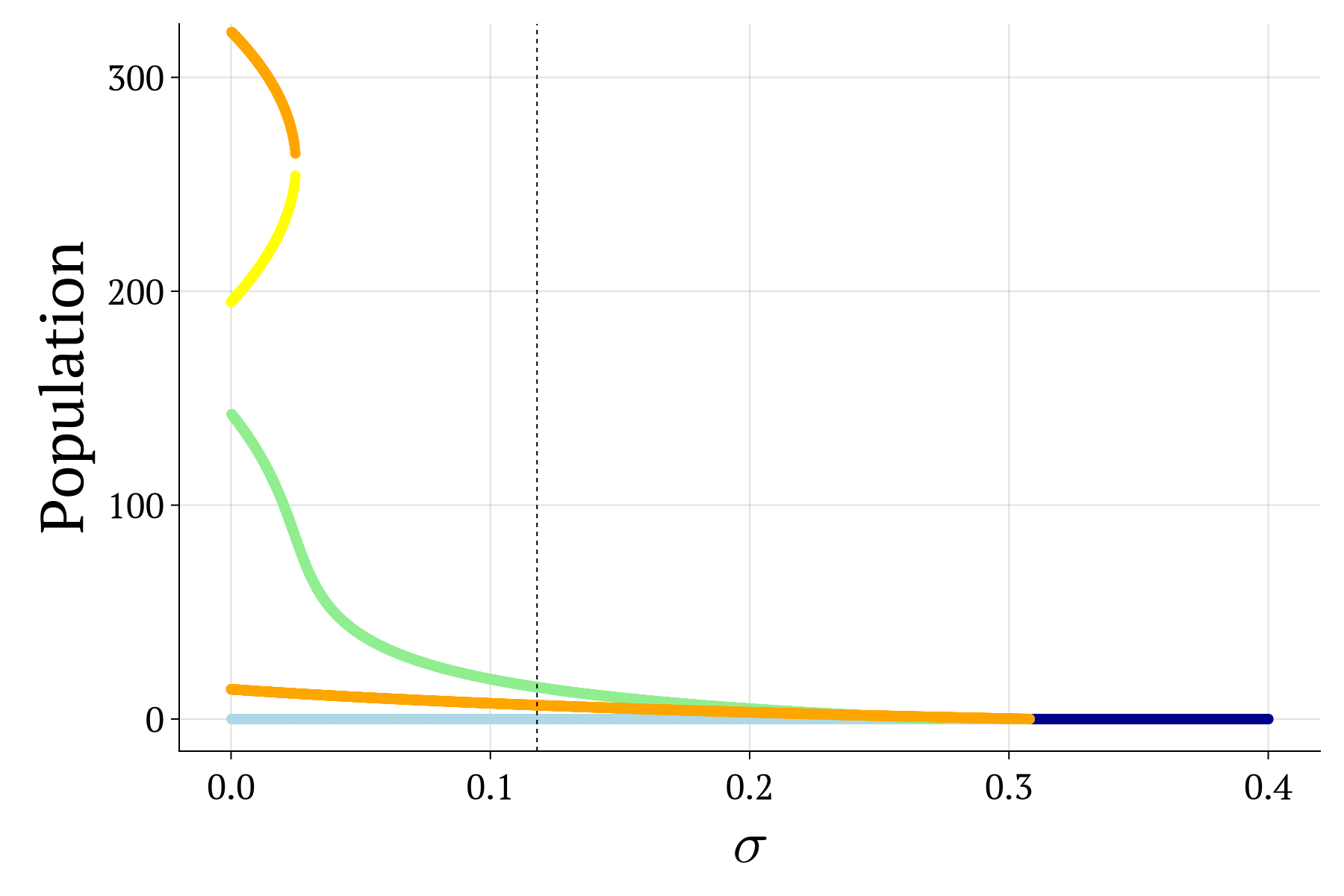}
\end{subfigure}
\begin{subfigure}{0.45\columnwidth}
 \caption{}
    \includegraphics[width=\textwidth]{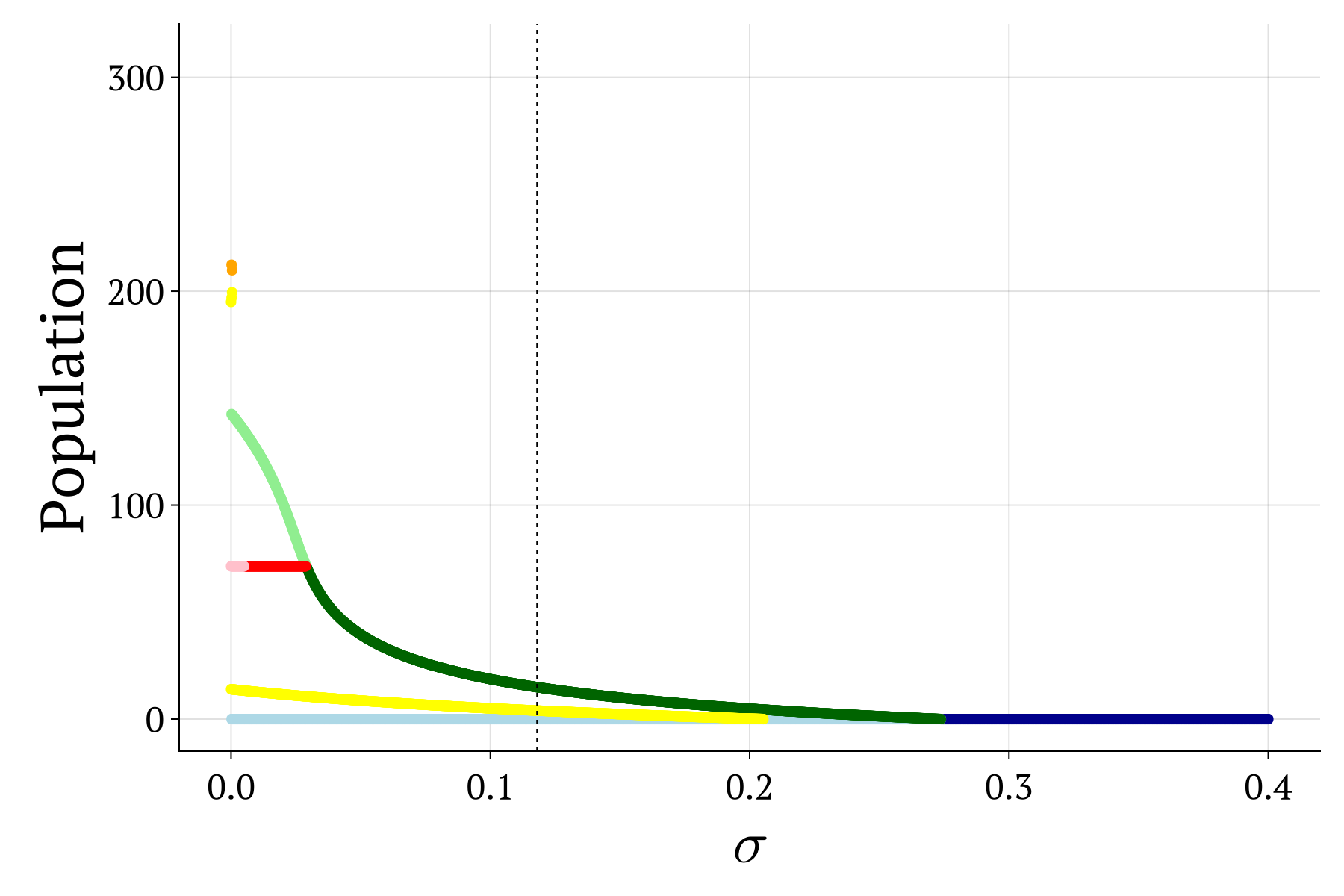}
\end{subfigure}\\
\begin{subfigure}{\columnwidth}
\includegraphics[width=\linewidth]{figures/Chemo_bifurcation_legend.png}
  \end{subfigure}
\vspace{-1cm}
\caption{Bifurcation diagrams for $\chi$, $\delta$, and $\sigma$ when $T_2$ is chemo-resistant and $T_1$ employs reduced antigen presentation. $\theta_1=\theta_2=0.5$ and $\theta_1=0.5 < \theta_2=0.8$ in the left and right columns, respectively.}
\label{fig:bif_chemo_RAP}
\end{figure}

\begin{figure}[!ht]
\centering
\captionsetup[subfigure]{justification=centering}
\begin{subfigure}{0.45\columnwidth}
 \caption{}
    \includegraphics[width=\textwidth]{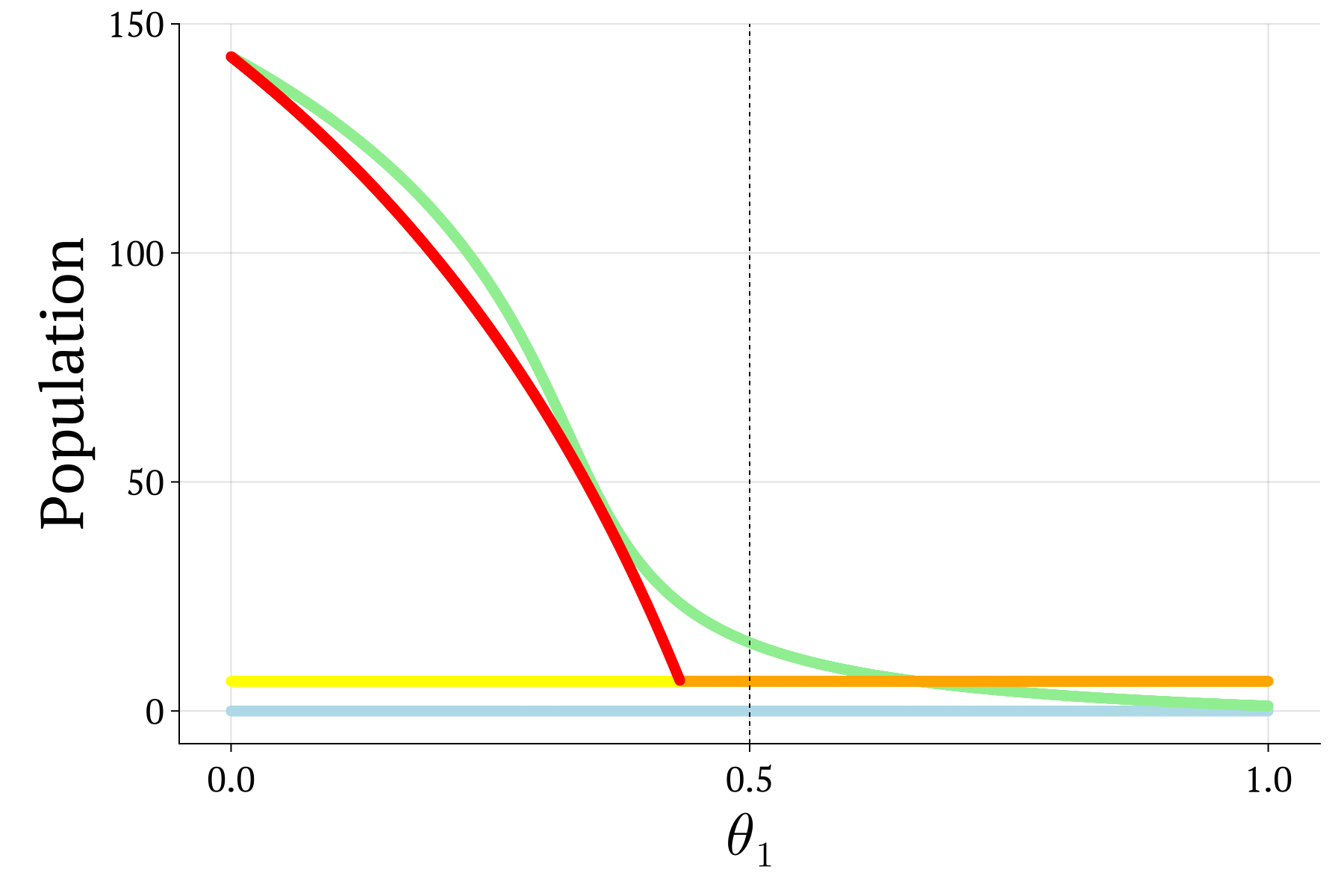}
\end{subfigure}
\begin{subfigure}{0.45\columnwidth}
 \caption{}
    \includegraphics[width=\textwidth]{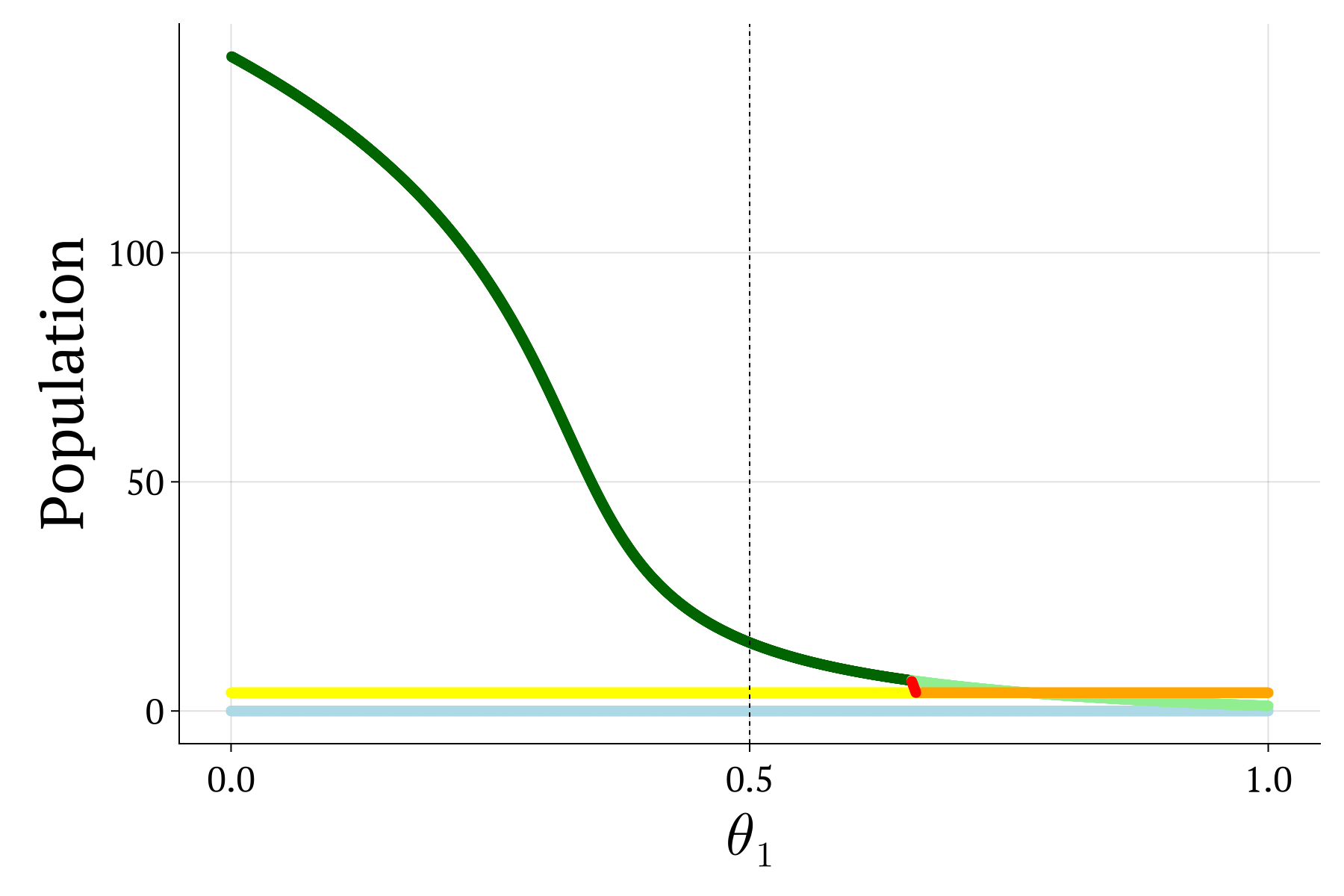}
\end{subfigure}
\begin{subfigure}{0.45\columnwidth}
 \caption{}
    \includegraphics[width=\textwidth]{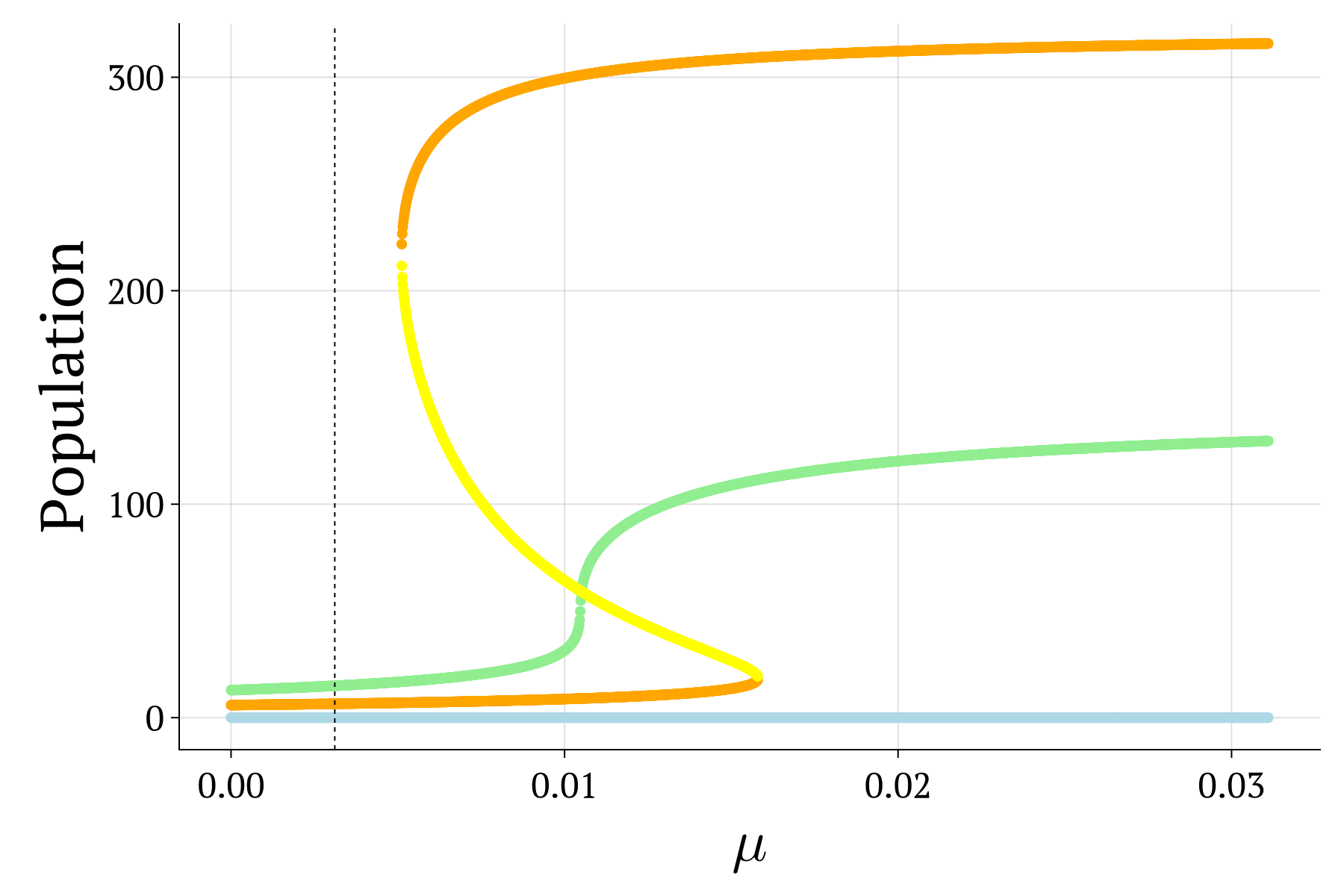}
\end{subfigure}
\begin{subfigure}{0.45\columnwidth}
 \caption{}
    \includegraphics[width=\textwidth]{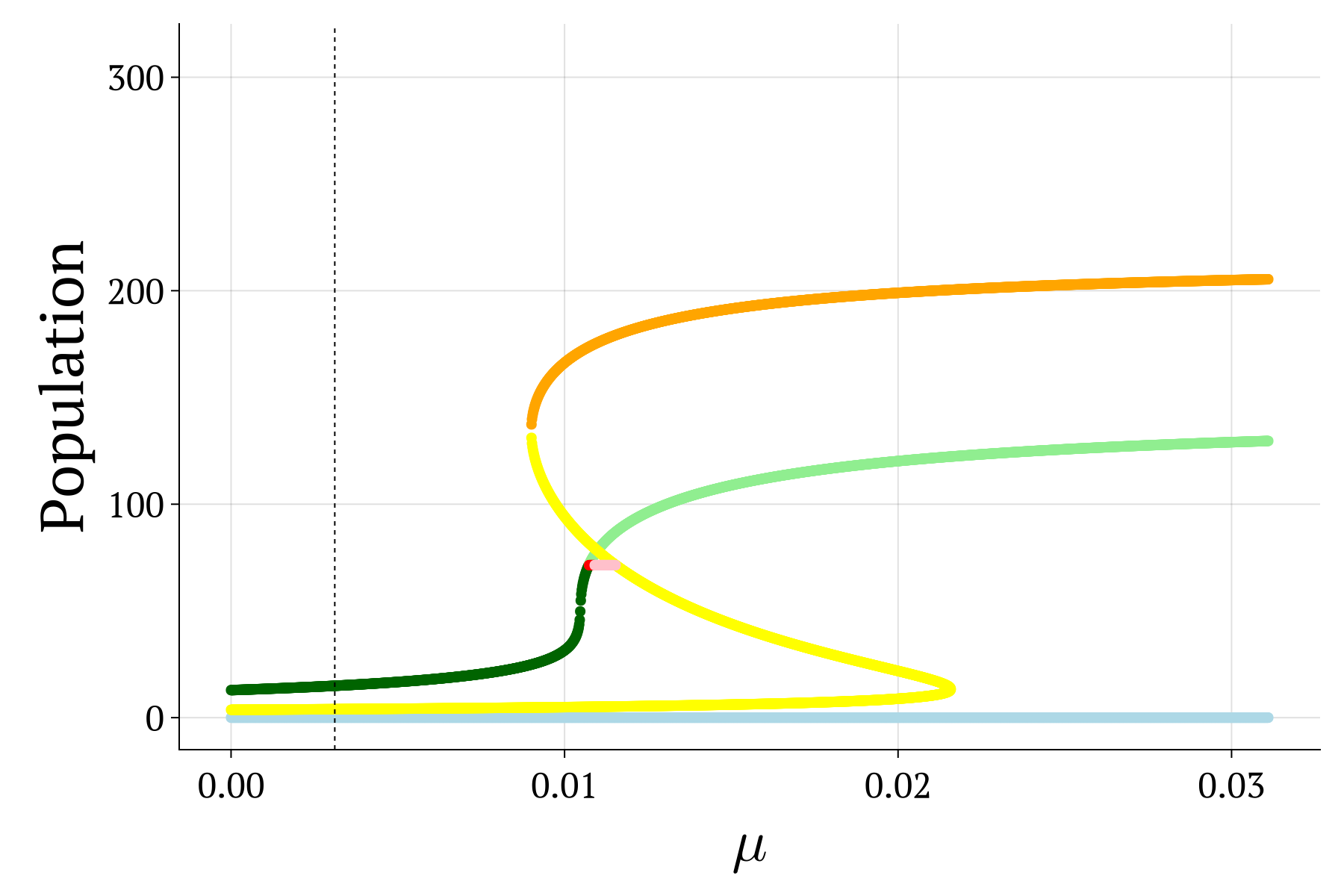}
\end{subfigure}
\begin{subfigure}{0.45\columnwidth}
 \caption{}
    \includegraphics[width=\textwidth]{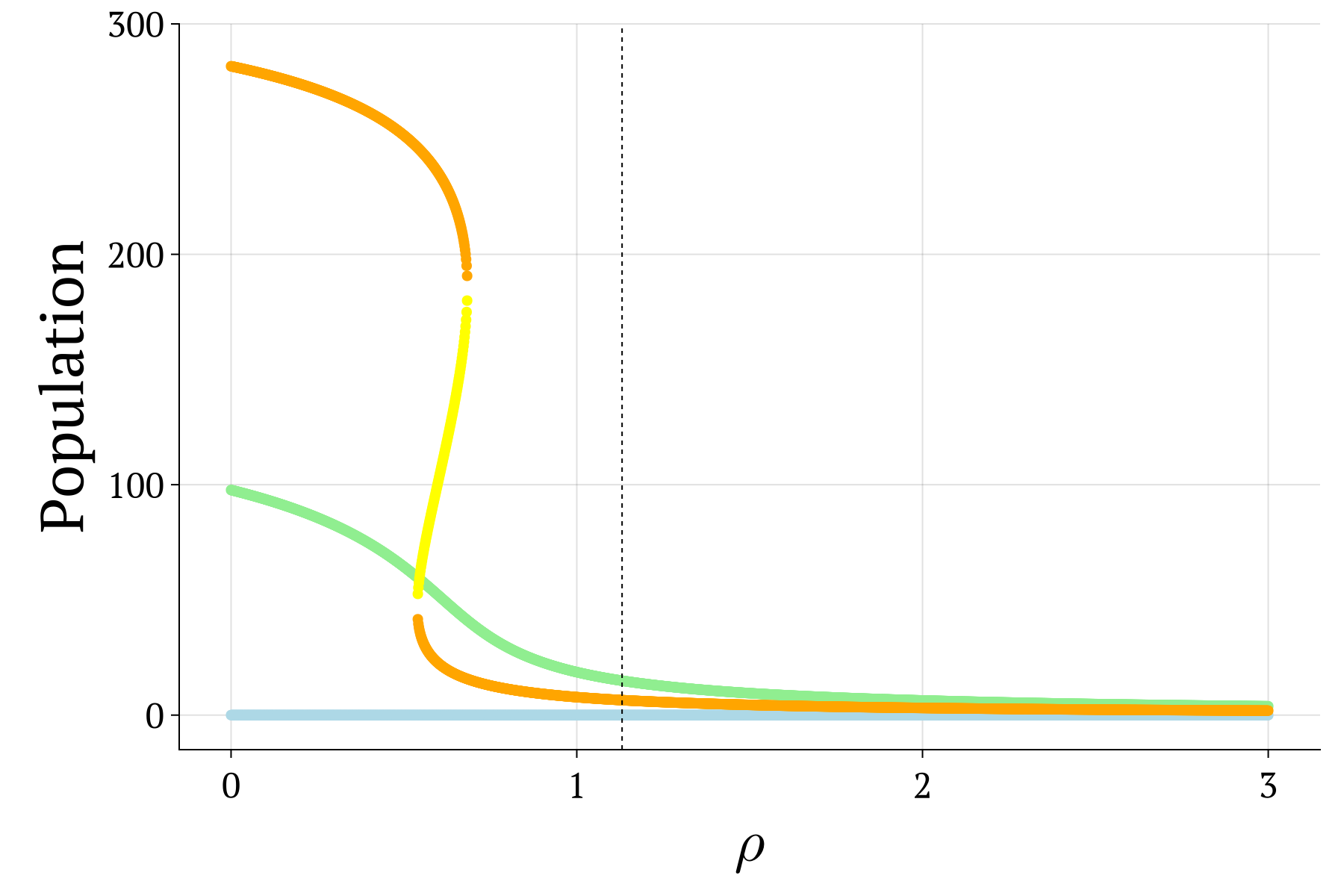}
\end{subfigure}
\begin{subfigure}{0.45\columnwidth}
 \caption{}
    \includegraphics[width=\textwidth]{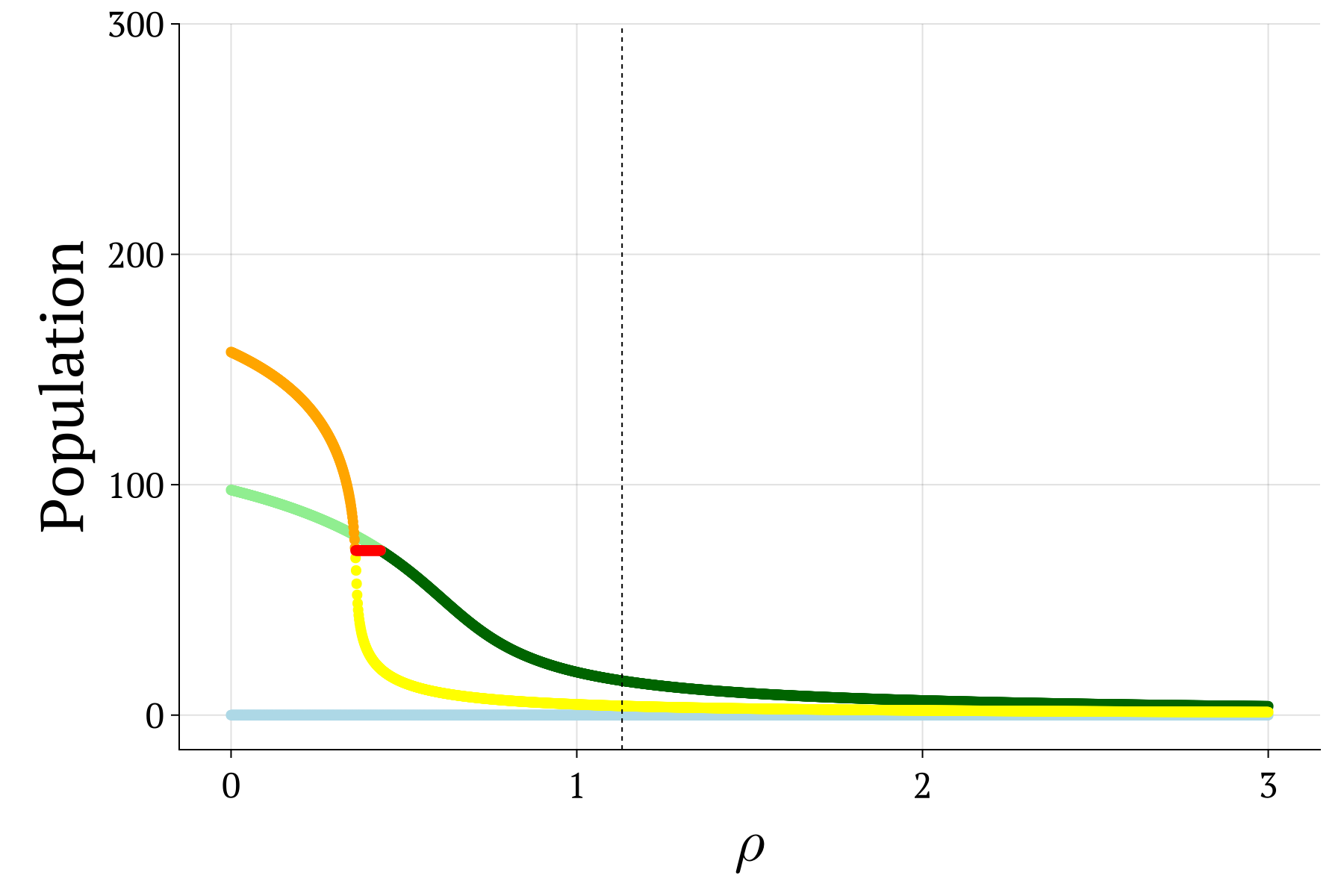}
\end{subfigure}\\
\begin{subfigure}{\columnwidth}
\includegraphics[width=\linewidth]{figures/Chemo_bifurcation_legend.png}
  \end{subfigure}
\vspace{-1cm}
\caption{Bifurcation diagrams for $\theta_1$, $\mu$, and $\rho$ when $T_2$ is chemo-resistant and $T_1$ employs reduced antigen presentation. Except when varying $\theta_1$, $\theta_1=\theta_2=0.5$ and $\theta_1=0.5 < \theta_2=0.8$ in the left and right columns, respectively.}
\label{fig:bif_immuno_RAP}
\end{figure}

Using the baseline parameter values listed in Table \ref{tbl:param}, we observe that the $T_2$-only equilibrium is stable. Recall that $\theta_2$ represents the benefit of becoming resistant for $T_2$. When $\theta_2$ is small, the benefit of resistance is high. In this case, fewer $T_2$ cells are killed by chemotherapy, allowing $T_2$ to dominate. Sufficiently high $\theta_2$ and low $\theta_1$, however, allow $T_1$ to dominate. Fixing $\theta_1=0.5$, we analyze the $T_2$-only case by setting $\theta_2=0.5$ and the $T_1$-only case by setting $\theta_2 = 0.8$. We then investigate how variations in chemotherapy-related and immunotherapy-related parameters influence system dynamics under these scenarios, including both immune checkpoint inhibition or reduced antigen presentation strategies for $T_1$. Figures \ref{fig:bif_chemo_ICE} and \ref{fig:bif_immuno_ICE} display the results when $T_1$ employs immune checkpoint inhibition, and Figures \ref{fig:bif_chemo_RAP} and \ref{fig:bif_immuno_RAP} display the results for when it employs reduced antigen presentation. Bifurcation diagrams for $\chi$, $\delta$, $\sigma$, $\theta_1$, $\mu$, and $\rho$ are depicted. Bifurcation diagrams for the remaining parameters are provided in Appendix \ref{app:bifurcation_diagrams}. Separating the figures in this manner improves clarity while preserving the complete analysis. Regarding the chemotherapy kill rate coefficient of effectors $\omega$, it is important to note that it is taken from human data (see \cite{de09}), and not from mouse models as in \cite{kuznetsov94}.

Consider first the case where $T_1$ employs immune checkpoint evasion. High values of $\chi$ control or eliminate the tumor (Figure \ref{fig:bif_chemo_ICE}). The tumor free equilibrium (TFE) is the sole stable state for lower values of $\chi$ when $\theta_2 = 0.8$ than when $\theta_2=0.5$. Sufficiently reducing $\chi$ below the baseline value results in the $T_1$-only low population equilibrium being stable. Further reduction results in a bifurcation. When $\theta_2=0.8$, both large and small $T_1$-only equilibria are stable. When $\theta_2=0.5$, a large coexistence equilibria (CE) may also exist. This scenario is particularly concerning, since such a heterogeneous population may be able to effectively adapt to therapies, immunological or chemical. The effects of varying $\delta$ and $\sigma$ are similar to the results in Section \ref{results:bifurcation_immuno_resistant}: lower $\delta$ and higher $\sigma$ result in control if not elimination of the tumor. Therefore, one potential combination therapy is combing chemotherapy with immuno-therapies that increase the presentation of effector cells by changing $\delta$ and $\sigma$.

Next we consider the parameters most tied to immunotherapy, $\theta_1$, $\mu$, and $\rho$, and immune checkpoint evasion (Figure \ref{fig:bif_immuno_ICE}). If $\theta_1$ is very low, then large population $T_1$ equilibria can be stable. Increasing $\mu$ can create a bistable system with stable high and low population equilibria. This case is nearest the baseline parameter values when $\theta_2=0.5$. In contrast, decreasing $\rho$ creates bistability for $\theta_2=0.5$. And, further decreasing it creates a single stable high population equilibrium. Bistability is no observed for $\theta_2=0.8$. However, coexistence of $T_1$ and $T_2$ is occurs over a narrow range and tumors can still be sizable. These results show how changing immune-resistance related parameters are beneficial for controlling or eliminating tumors. When $T_1$ dominates under the immune checkpoint inhibition strategy, tumor control is primarily achieved by decreasing immune exhaustion ($\mu$) or increasing immune activation ($\rho$ and $\sigma$). In this case, checkpoint inhibitors combined with moderate chemotherapy provide the most effective control strategy.

Next, we consider $T_1$ employing reduced antigen presentation. Figure \ref{fig:bif_chemo_RAP} depicts bifurcation diagrams for $\chi$, $\delta$, and $\sigma$. When $\chi$ is high or $\theta_2=0.8$, the results are similar to the immune checkpoint evasion case. However, for small $\chi$ and $\theta_2=0.5$, there is bistability between large and small $T_1$-only tumors. There is no coexistence between tumor types and large tumors become stable for a higher $\chi$. These results suggest that reduced antigen presentation is more robust against relatively weak chemotherapy. The results for $\delta$ and $\sigma$ are similar to those in Figure \ref{fig:bif_chemo_ICE}. However, the region of $\delta$ where there is coexistence is larger, and, more importantly, stable coexistence and $T_1$-only equilibria exist for low $\sigma$. Again, $T_1$ with reduced antigen presentation is more resistant to chemotherapy.

Under reduced antigen presentation, coexistence occurs when both $\theta_1$ and $\theta_2$ are low (Figure \ref{fig:bif_immuno_RAP}). When $\theta_2=0.8$, the $T_1$-only equilibrium is generally larger than under immune checkpoint evasion. However, for very low $\theta_1$, the latter exhibits bistability with a stable high population $T_1$-only equilibrium. The remaining bifurcation diagrams, $\mu$ and $\rho$, are similar to the diagrams for immune checkpoint evasion.

Increasing immune recruitment ($\sigma$), decreasing immune death ($\delta$), or enhancing activation ($\rho$) improves treatment outcomes. These results suggest that therapies such as CAR-T cells, cancer vaccines, or adoptive T-cell transfer may be particularly effective when combined with chemotherapy. In contrast, Section \ref{results:bifurcation_immuno_resistant} shows that increasing presentation of effector cell is not significantly helpful in controlling $T_1$ in absence of chemotherapy. Overall, these results demonstrate that combination immuno-chemotherapy provides greater flexibility and improved tumor control compared to single-therapy approaches. Moreover, the effectiveness of treatment depends strongly on the dominant tumor phenotype and immune-evasion mechanism, supporting phenotype-guided combination therapy strategies.

\section{Discussion}


In this work, we investigated the dynamics of tumors interacting with an immune response and therapeutic interventions. Building upon the framework of \cite{kuznetsov94}, our model incorporates tumor heterogeneity by distinguishing between effector-resistant and effector-sensitive tumor cell subpopulations. We characterized the conditions under which tumor subpopulation persist, are eliminated, or are controlled. We also examined combination immuno-chemo therapy and demonstrated that combined treatment strategies improve tumor control compared to single therapy. Chemotherapy reduces tumor burden and targets sensitive populations, while immunotherapy enhances immune-mediated tumor suppression. When applied together, these therapies expand the parameter region exhibiting tumor-free or low population equilibria. Our results also demonstrate that modest parameter changes can induced qualitative transitions between immune control and tumor escape, highlighting the nonlinear nature of tumor-immune dynamics.

Our bifurcation analyses show that therapeutic efficacy depends on both the immune evasion mechanism and dominant tumor phenotype. We considered two resistance strategies: immune checkpoint regulation and reduced antigen presentation. When the immune-resistant subpopulation employing immune checkpoint regulation dominates, tumor control is highly sensitive to the immune exhaustion rate, indicating that checkpoint inhibitors are the most effective therapy. In contrast, when reduced antigen presentation drives resistance, stability depends more strongly on effector activation and exhaustion, supporting strategies such as checkpoint inhibitors, CAR-T therapy, or cancer vaccines. On the other hand, when the immune-sensitive phenotype dominants, boosting effector recruitment and persistence becomes critical. Adoptive T-cell transfer or cytotoxic therapy are then the preferred therapies. These results highlight why non-personalized, uniform treatment often fails. By explicitly accounting for tumor heterogeneity, the results support tailoring therapy based on dominant tumor phenotype and provides a theoretical framework for such phenotype-specific immunotherapy. The bifurcation results further highlight how treatment-related parameters influence tumor stability and transitions between equilibria. We identified parameter thresholds separating tumor elimination, coexistence, and dominance regimes. These findings demonstrate that relatively small changes in immune or treatment parameters may lead to significant shifts in tumor behavior. Such sensitivity supports a medicine approach in which therapy selection is guided by measurable biological markers and tumor composition.

Based on the bifurcation analyses, we propose sequential treatment strategies guided by tumor composition. These strategies aim to minimize tumor burden while preventing resistance from becoming dominant. When treatment begins in a $T_1$-dominant state, chemotherapy may first reduce tumor burden, followed by immunotherapy to maintain control. Alternatively, immunotherapy may initially reduce tumor size, after which chemotherapy can eliminate residual resistant cells. For $T_2$-dominant tumors, chemotherapy may reduce tumor burden but can also lead to resistant escape; therefore, immunotherapy is typically applied afterward to stabilize the system. Finally, if parameters predict coexistence of both types of resistance, the best strategy is first to apply a chemotherapy thereby eliminating $T_1$ cells. Remaining is the low population $T_2$-only equilibrium. By applying immunotherapy, we can potentially eliminate the tumor. However, if the tumor is only controlled, the emergence of $T_1$ mutants could result in resurgence of the tumor.

Several extensions remain for future work. For one, using optimal control methods may help determine treatment schedules that minimize tumor burden while reducing toxicity. Additionally, studying alternative immune evasion strategies may further improve understanding of resistance dynamics and therapeutic design. More specific models scenarios could also be explored. For example, microenvironment feedback loops, particularly in cancers such as breast or prostate carcinoma, could be considered. Furthermore, including oncologic viral therapy \cite{wodarz12, jenner18, storey20, pooladvand21, baabdulla24} could provide insight into how viral replication, immune activation, and tumor resistance coevolve. Finally, emerging machine learning and deep learning approaches \cite{gallagher24} may enable personalized adaptive therapy strategies informed by patient-specific parameter estimation \cite{rodriguezmessan21}. Coupling data-driven prediction with mechanistic modeling represents a promising direction toward clinically implementable, individualized treatment protocols.

Overall, this work provides a mathematical framework for adaptive combination therapy in heterogeneous tumors. By accounting for tumor heterogeneity, immune response, and treatment interactions, the model offers insights into designing treatment strategies that improve long-term tumor control while delaying resistance emergence.

\subsection*{Code and Data Availability}
Code to run the simulations is available at github.com/NazMokari/Adaptive-Cancer-Therapy.

\bibliography{adaptive_therapy}
\bibliographystyle{abbrv}

\appendix

\section{Further Bifurcation Diagrams} \label{app:bifurcation_diagrams}

Here we presents additional bifurcation diagrams supporting the results discussed in Section \ref{results:bifurcation_chemo_resistant}. These figures illustrate how variations in model parameters influence tumor dynamics under combination therapy. As discussed in Section \ref{results:bifurcation_chemo_resistant}, immune control alone is often insufficient to eliminate tumors. Therefore, we consider the combined effects of chemotherapy and immunotherapy. In contrast to the immunotherapy-only case, the presence of chemotherapy stabilizes coexistence equilibria, requiring separate analysis of the $T_1$-only and $T_2$-only regimes.

\begin{figure}[!ht]
\centering
\captionsetup[subfigure]{justification=centering}
\begin{subfigure}{0.45\columnwidth}
 \caption{}
    \includegraphics[width=\textwidth]{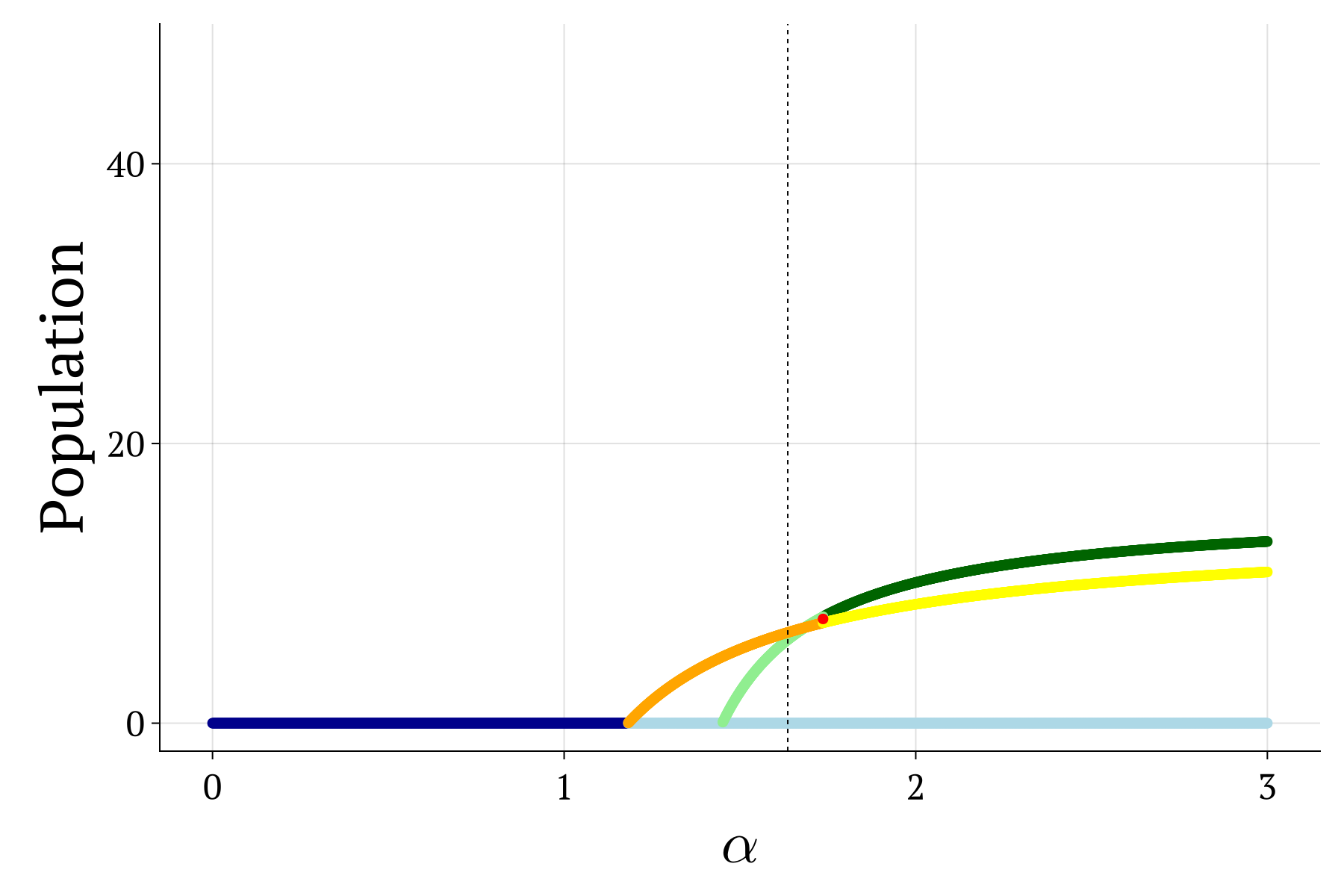}
\end{subfigure}
\begin{subfigure}{0.45\columnwidth}
 \caption{}
    \includegraphics[width=\textwidth]{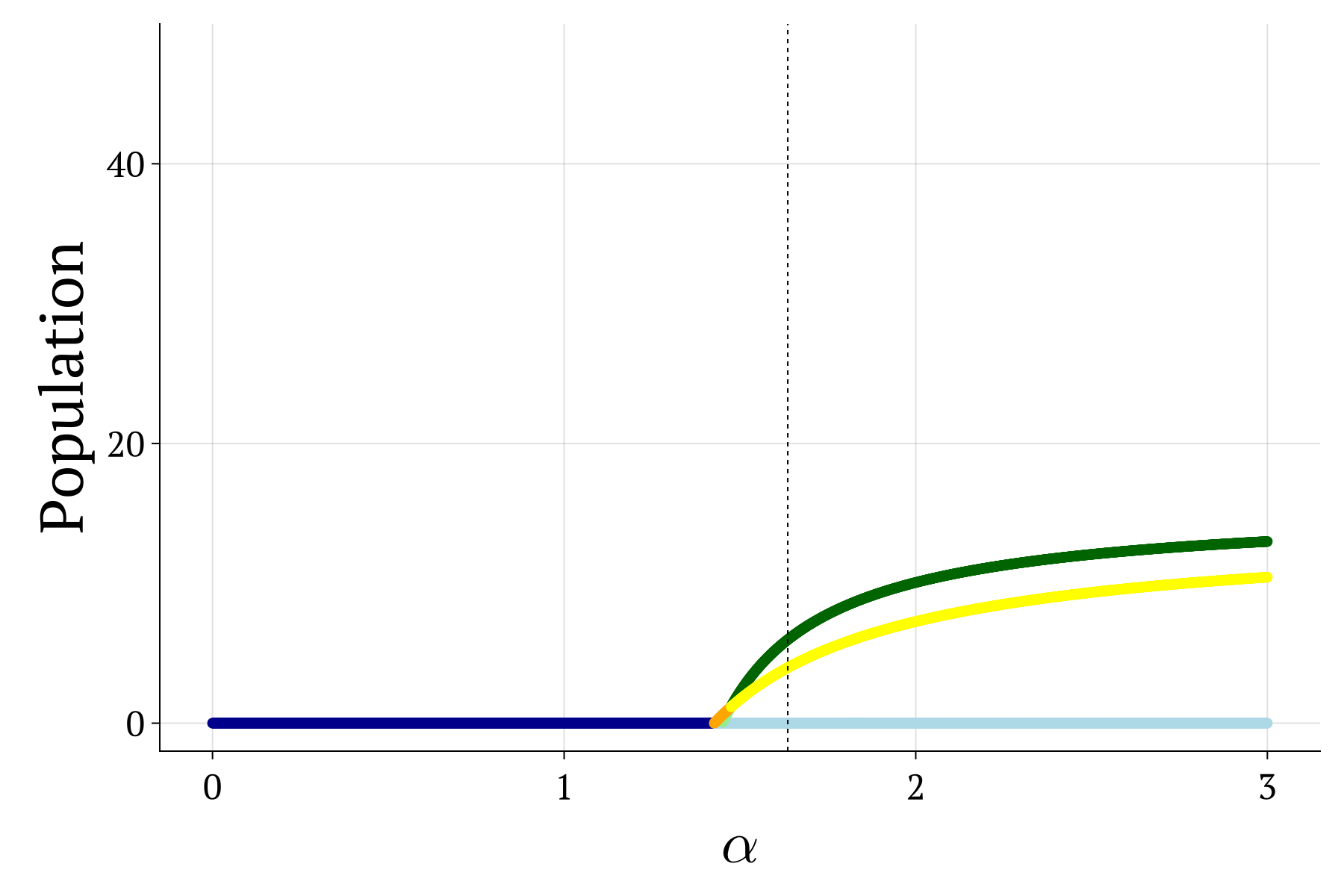}
\end{subfigure}
\begin{subfigure}{0.45\columnwidth}
 \caption{}
    \includegraphics[width=\textwidth]{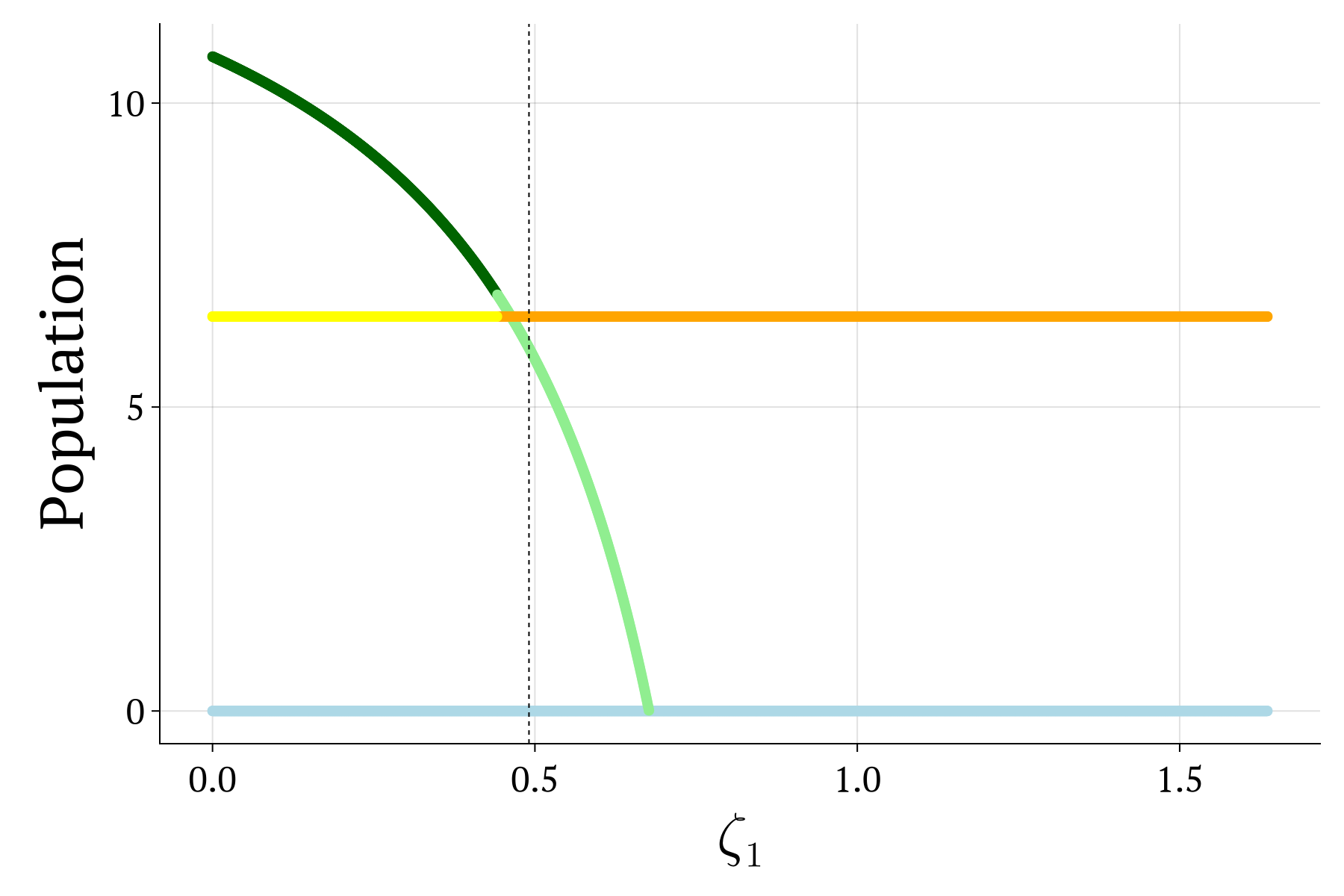}
\end{subfigure}
\begin{subfigure}{0.45\columnwidth}
 \caption{}
    \includegraphics[width=\textwidth]{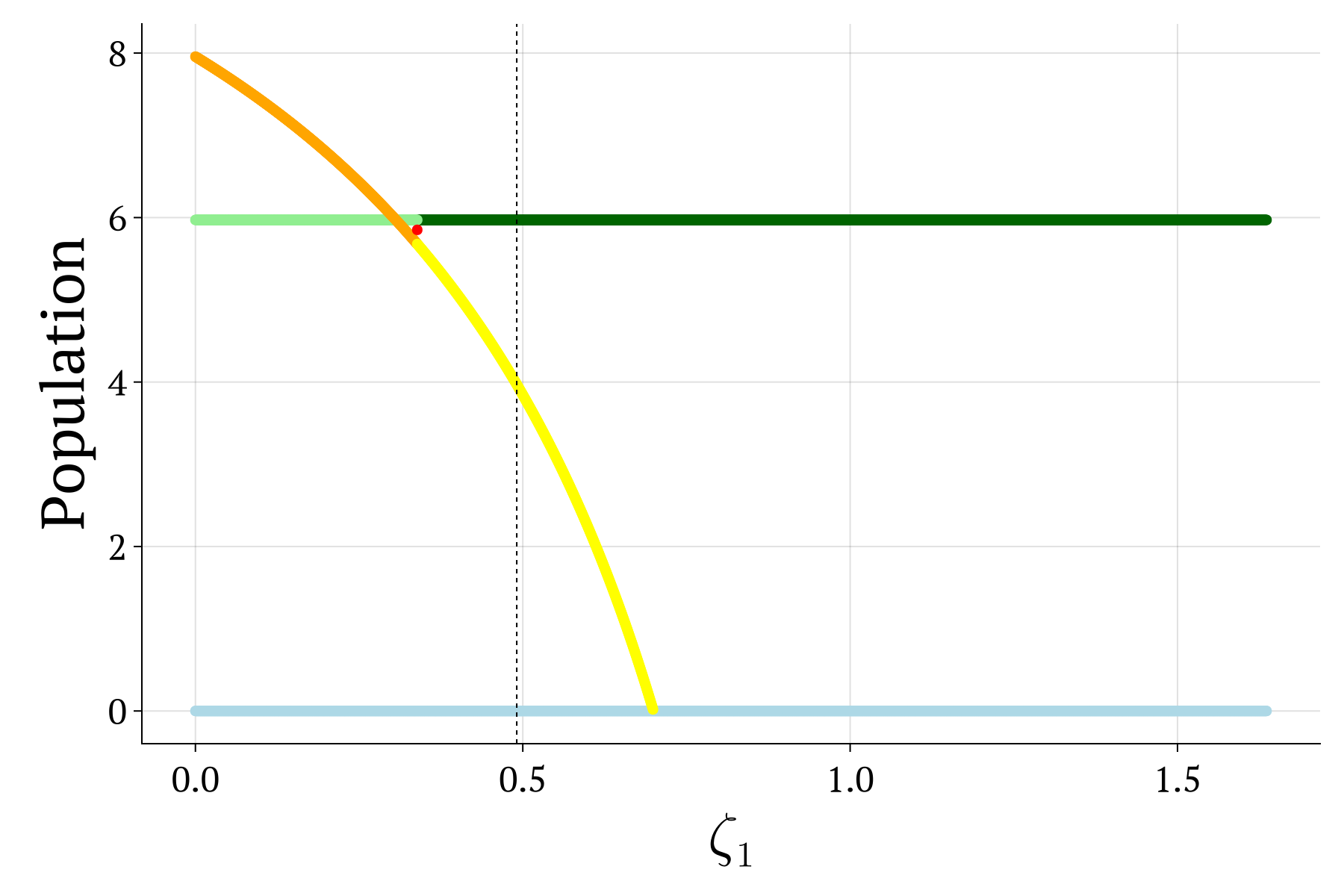}
\end{subfigure}
\begin{subfigure}{0.45\columnwidth}
 \caption{}
    \includegraphics[width=\textwidth]{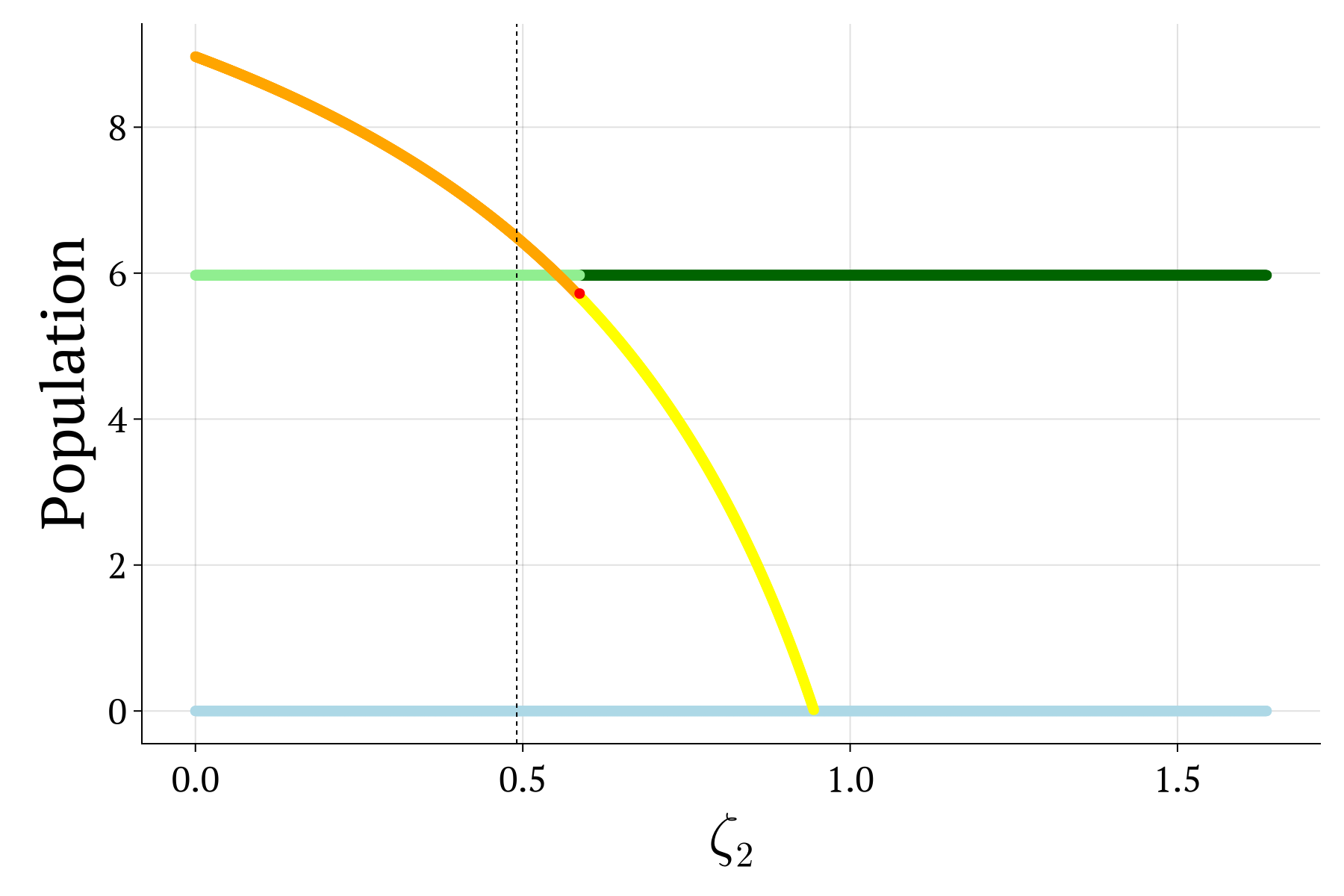}
\end{subfigure}
\begin{subfigure}{0.45\columnwidth}
 \caption{}
    \includegraphics[width=\textwidth]{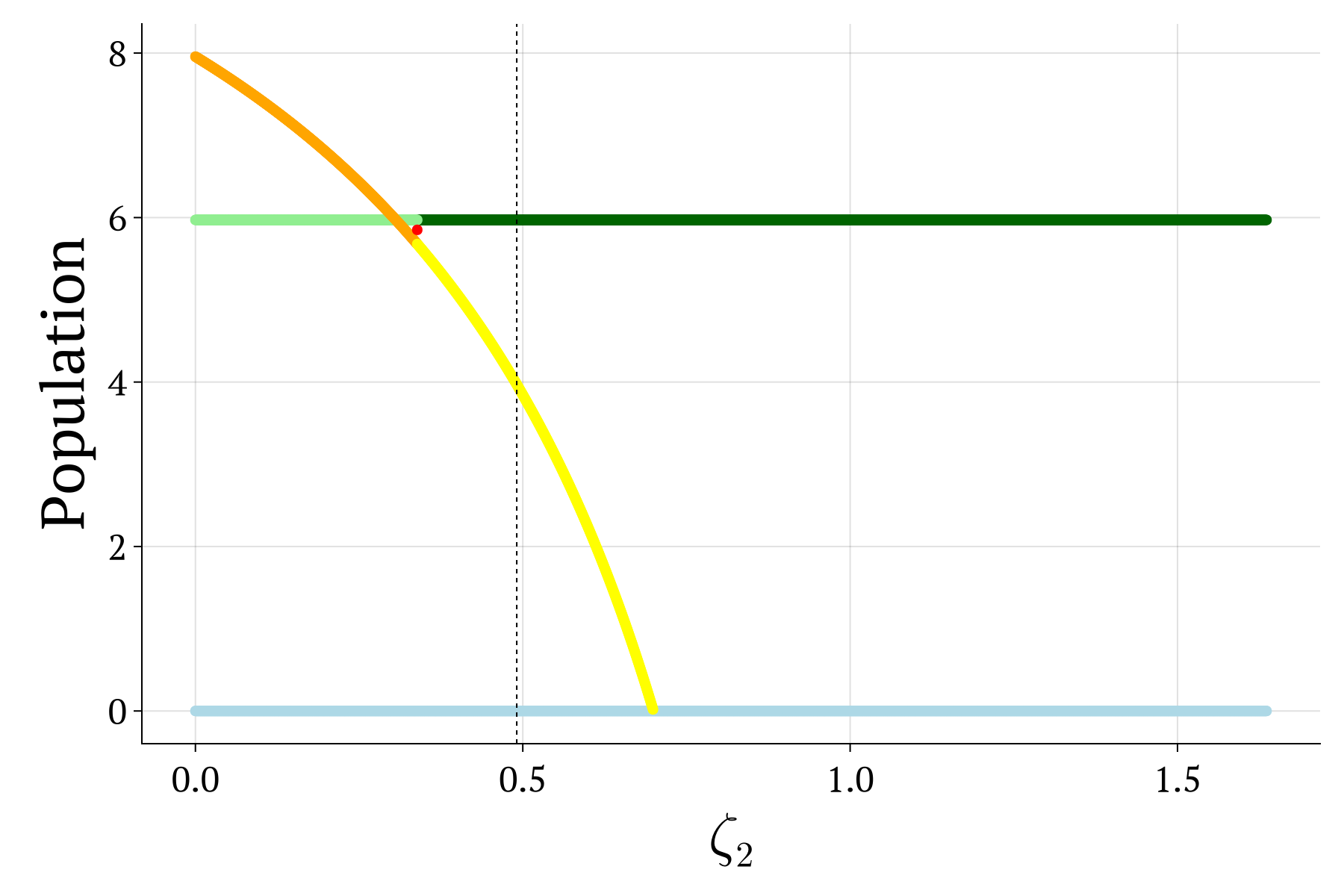}
\end{subfigure}\\
\begin{subfigure}{\columnwidth}
\includegraphics[width=\linewidth]{figures/Chemo_bifurcation_legend.png}
  \end{subfigure}
\caption{Bifurcation diagrams for $\alpha$ and $\zeta_i$ when $T_2$ is chemo-resistant and $T_1$ employs immune checkpoint evasion. $\theta_1=\theta_2=0.5$ and $\theta_1=0.5 < \theta_2=0.8$ in the left and right columns, respectively.}
\label{fig:bif_chemo_ICE_cost}
\end{figure}

\begin{figure}[!ht]
\centering
\captionsetup[subfigure]{justification=centering}
\begin{subfigure}{0.45\columnwidth}
 \caption{}
    \includegraphics[width=\textwidth]{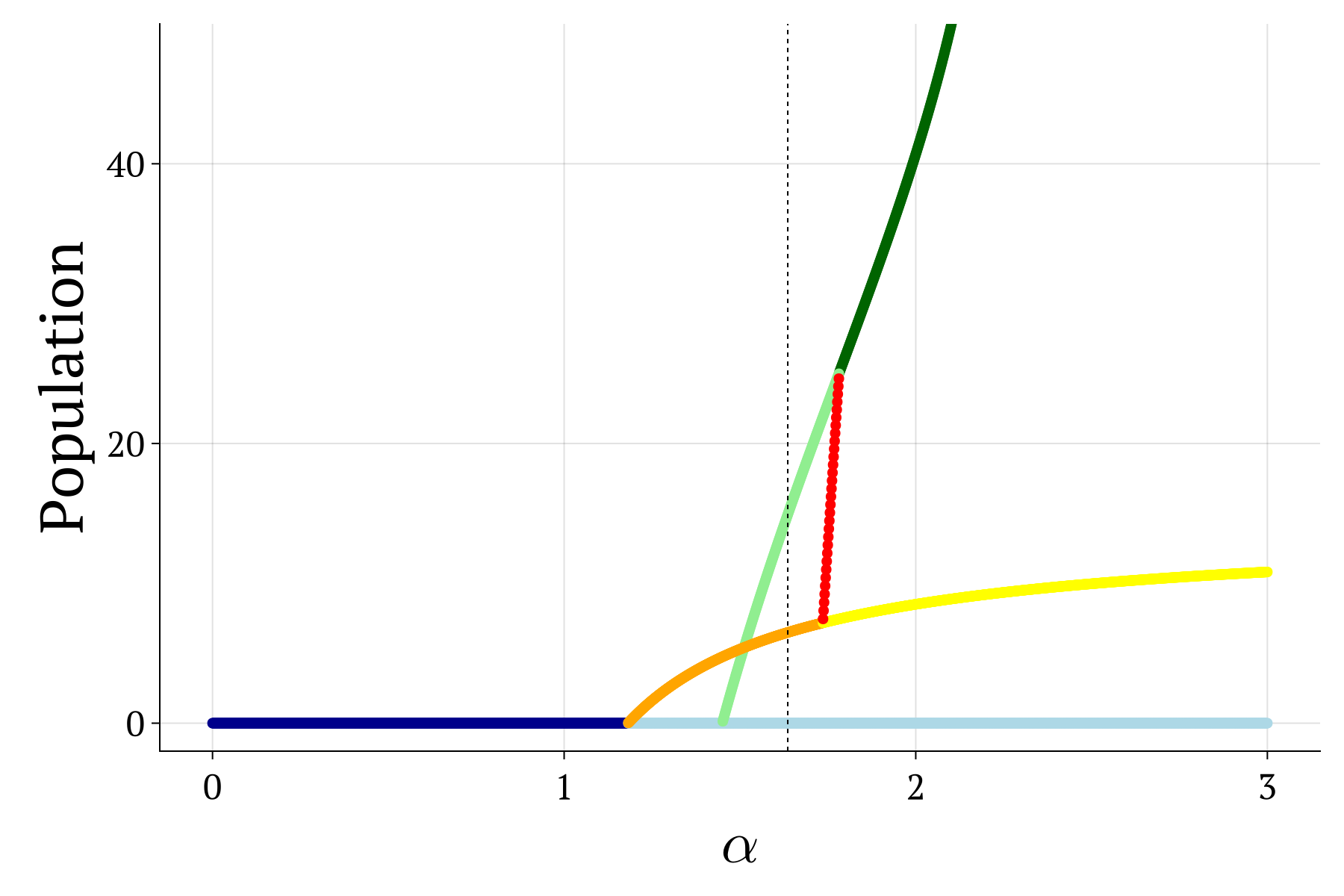}
\end{subfigure}
\begin{subfigure}{0.45\columnwidth}
 \caption{}
    \includegraphics[width=\textwidth]{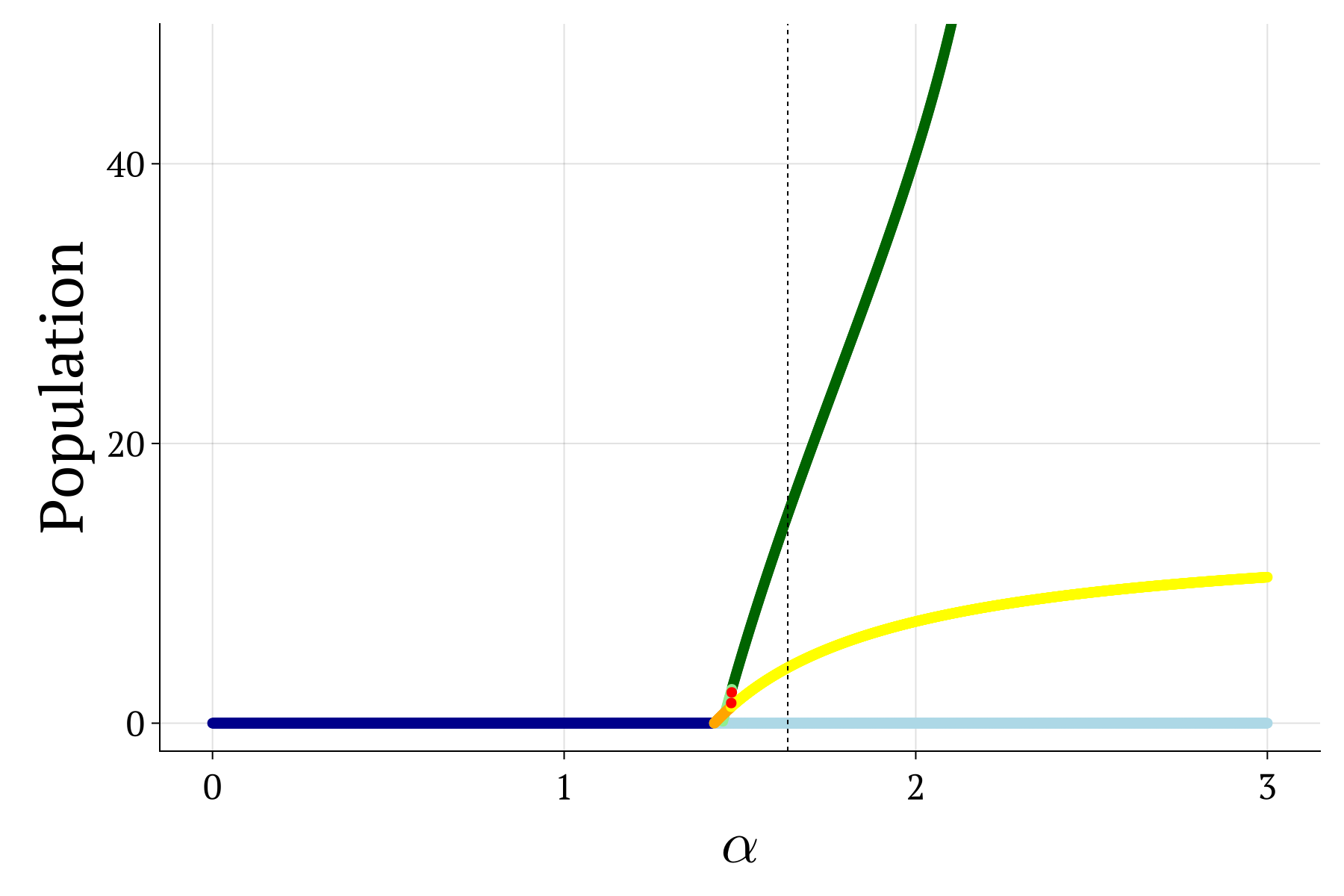}
\end{subfigure}
\begin{subfigure}{0.45\columnwidth}
 \caption{}
    \includegraphics[width=\textwidth]{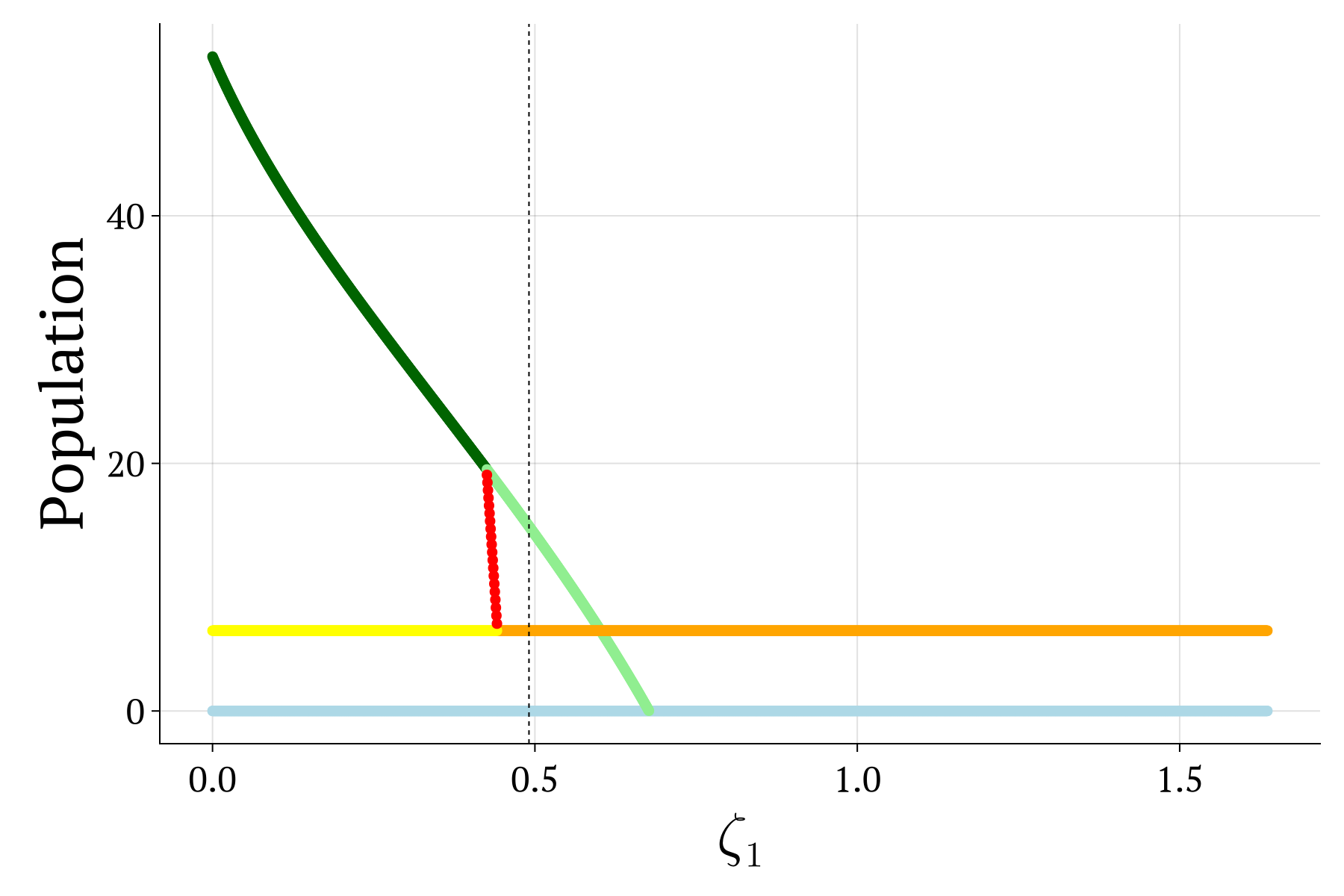}
\end{subfigure}
\begin{subfigure}{0.45\columnwidth}
 \caption{}
    \includegraphics[width=\textwidth]{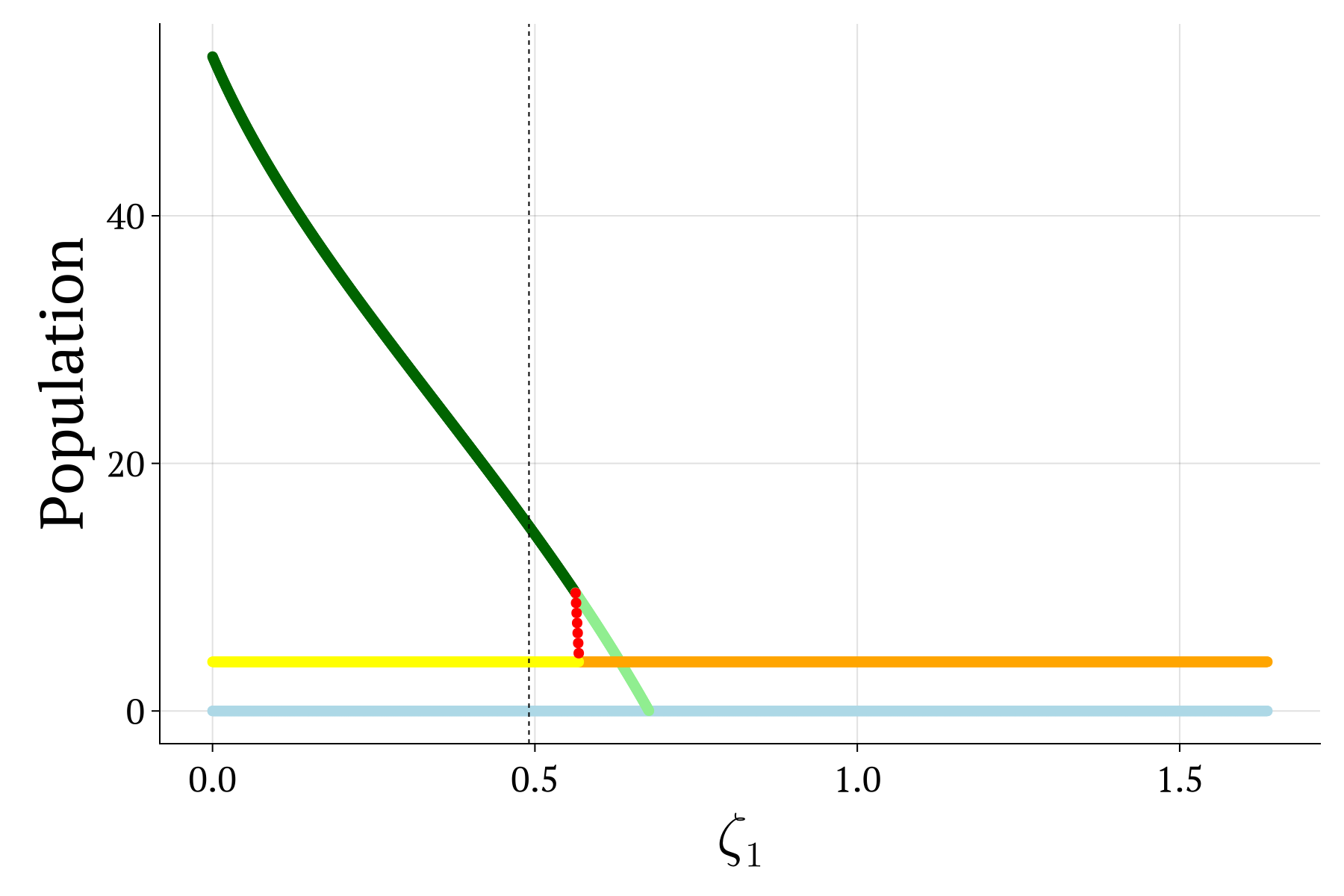}
\end{subfigure}
\begin{subfigure}{0.45\columnwidth}
 \caption{}
    \includegraphics[width=\textwidth]{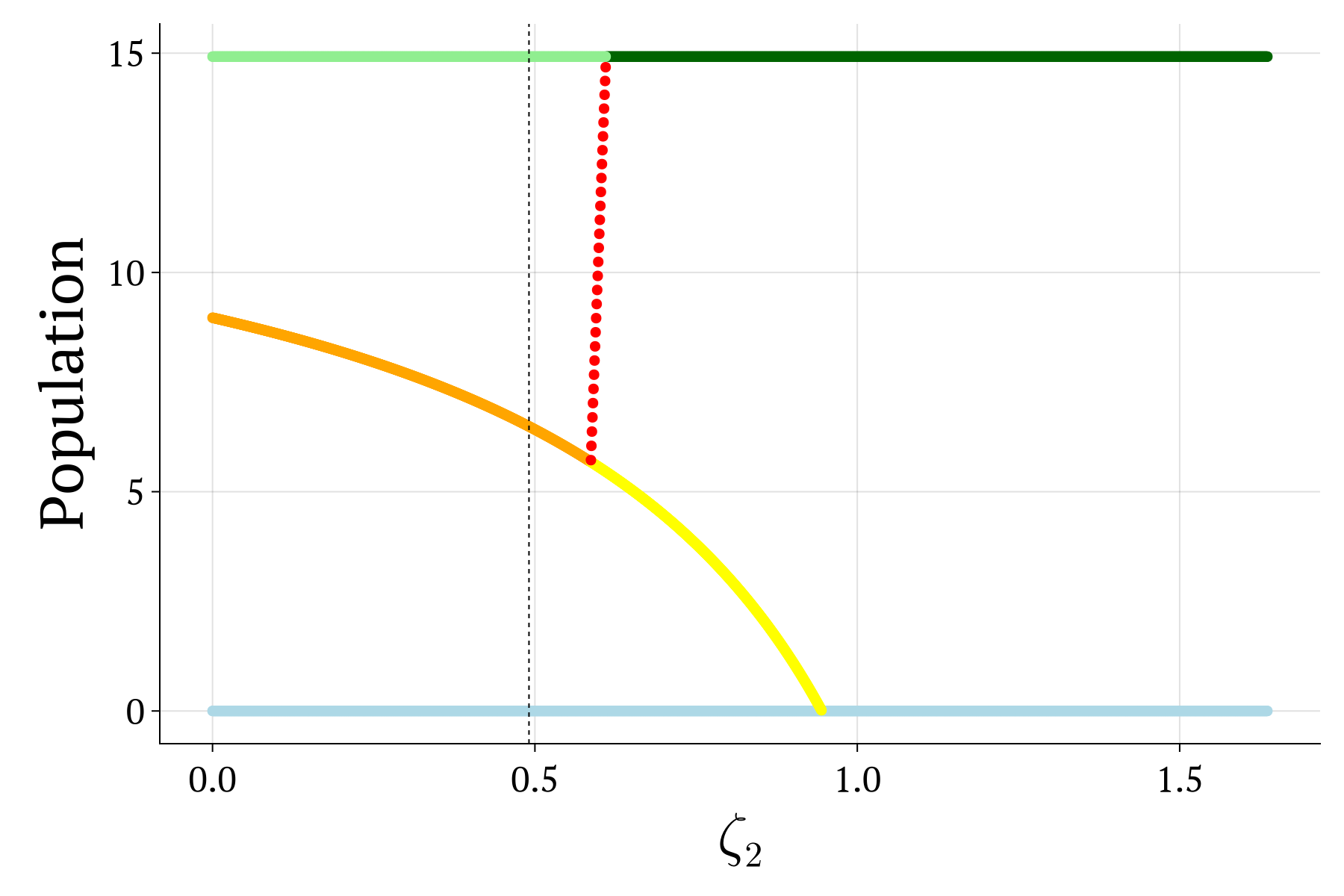}
\end{subfigure}
\begin{subfigure}{0.45\columnwidth}
 \caption{}
    \includegraphics[width=\textwidth]{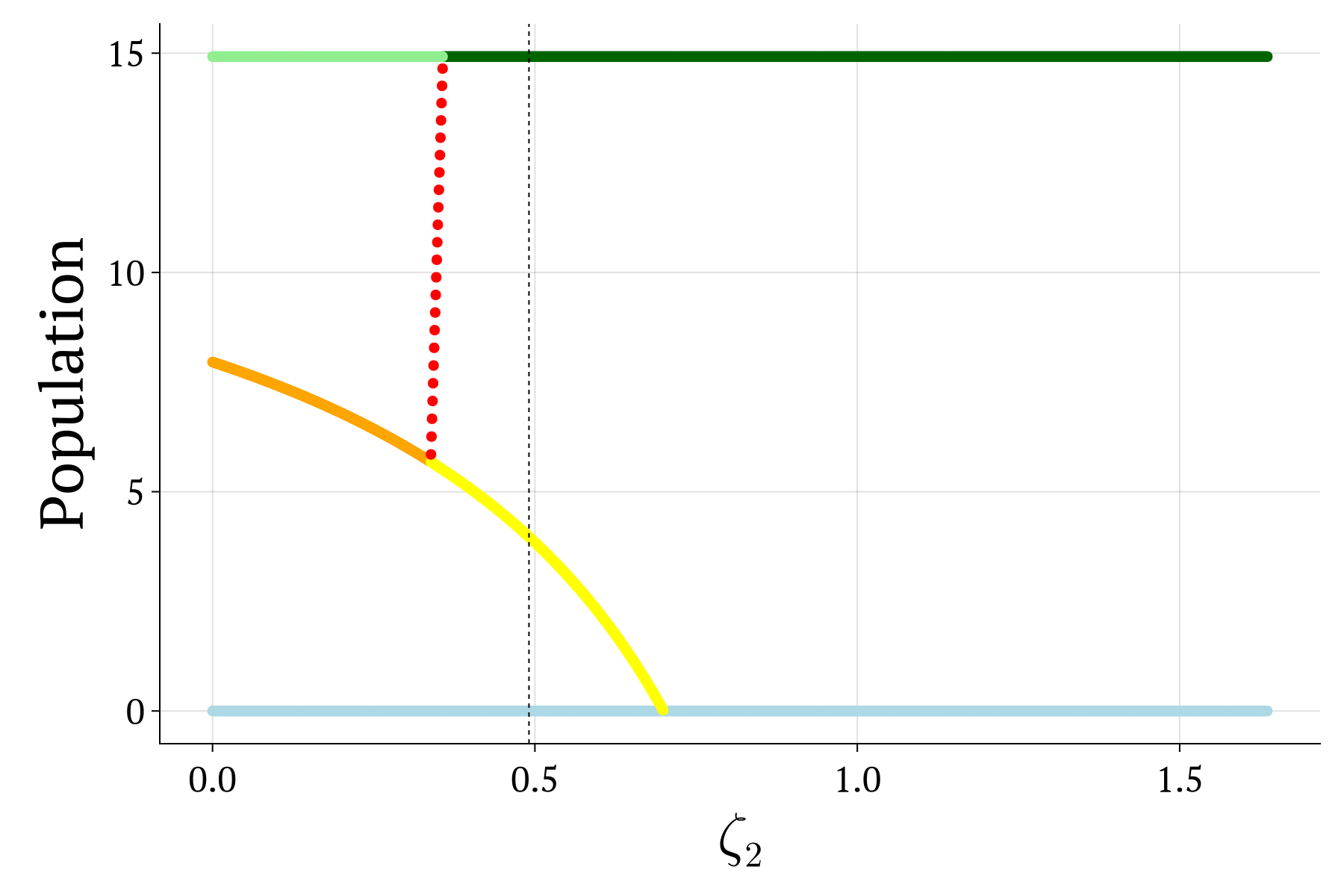}
\end{subfigure}\\
\begin{subfigure}{\columnwidth}
\includegraphics[width=\linewidth]{figures/Chemo_bifurcation_legend.png}
  \end{subfigure}
    \vspace{-1cm}
\caption{Bifurcation diagrams for $\omega$, $\phi$, and $\eta$ when $T_2$ is chemo-resistant and $T_1$ employs reduced antigen presentation. $\theta_1=\theta_2=0.5$ and $\theta_1=0.5 < \theta_2=0.8$ in the left and right columns, respectively.}
\label{fig:bif_chemo_RAP_cost}
\end{figure}

Using the baseline parameter values from Table\ref{tbl:param}, we first analyze Figures \ref{fig:bif_chemo_ICE_cost} and \ref{fig:bif_chemo_RAP_cost}, which show how the cost and benefit of resistance affect system outcomes. We begin with the growth rate $\alpha$. For large values of $\alpha$, the $T_1$ population becomes stable. However, for smaller values of $\alpha$, both tumor types incur costs that outweigh the benefits of resistance, causing them to lose their competitive advantage and driving the system toward the tumor-free equilibrium (TFE). As $\alpha$ decreases further, the equilibrium transitions from $T_1$-only to $T_2$-only and eventually to a TFE. The parameters $\zeta_i$ represent the cost of resistance. When $\zeta_i$ is small, the corresponding tumor population $T_i$ has a greater opportunity to dominate.

\begin{figure}[!ht]
\centering
\captionsetup[subfigure]{justification=centering}
\begin{subfigure}{0.45\columnwidth}
 \caption{}
    \includegraphics[width=\textwidth]{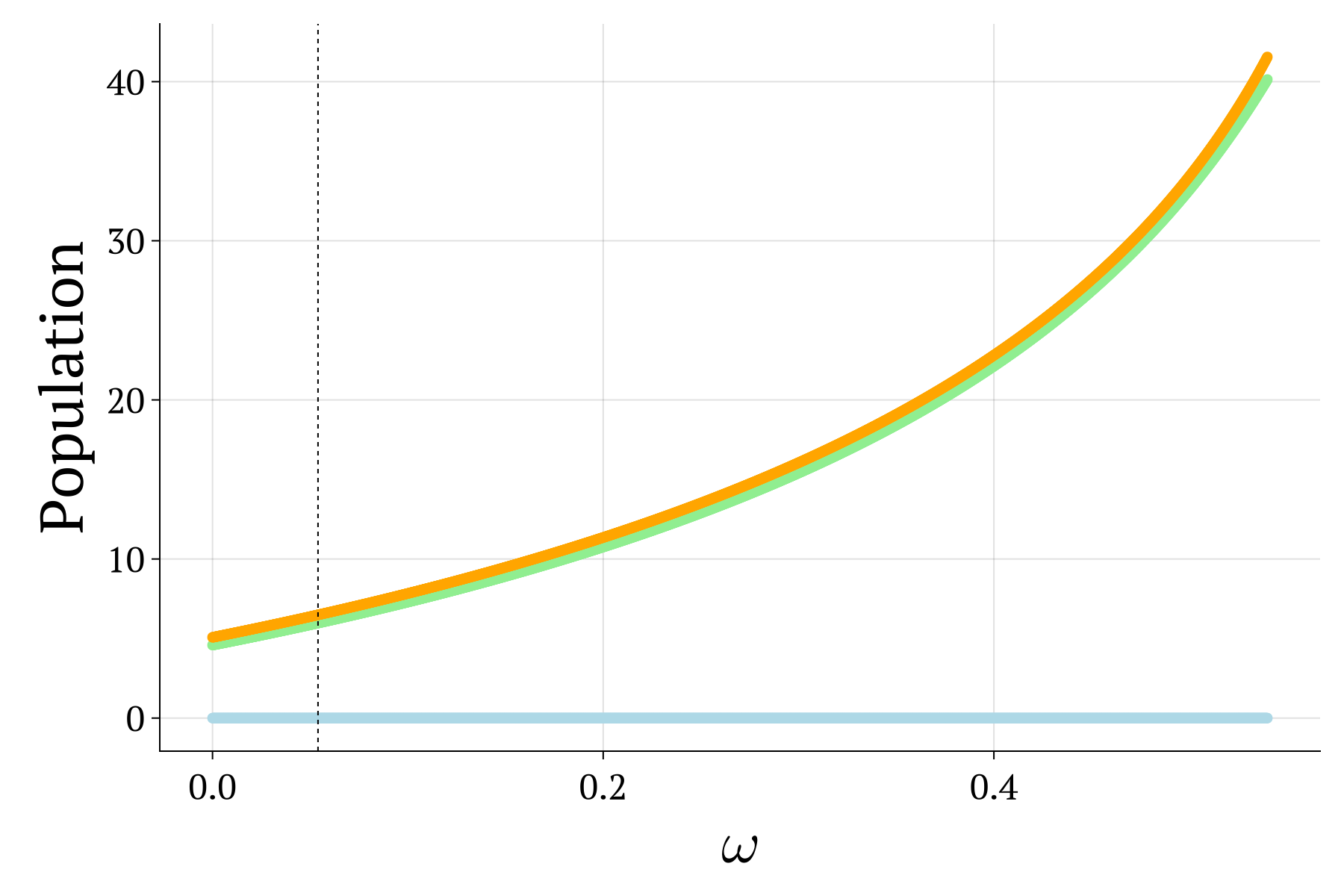}
\end{subfigure}
\begin{subfigure}{0.45\columnwidth}
 \caption{}
    \includegraphics[width=\textwidth]{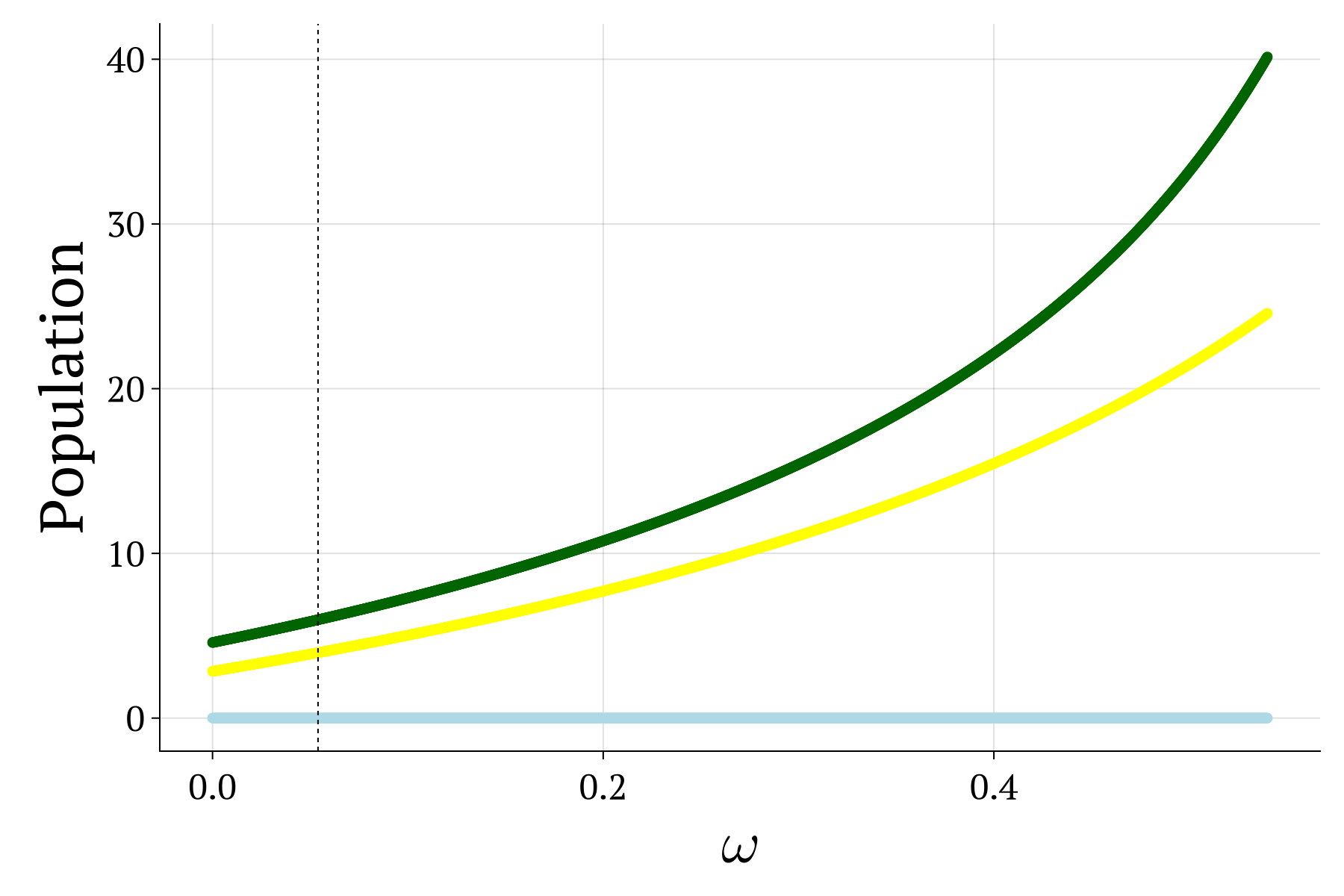}
\end{subfigure}
\begin{subfigure}{0.45\columnwidth}
 \caption{}
    \includegraphics[width=\textwidth]{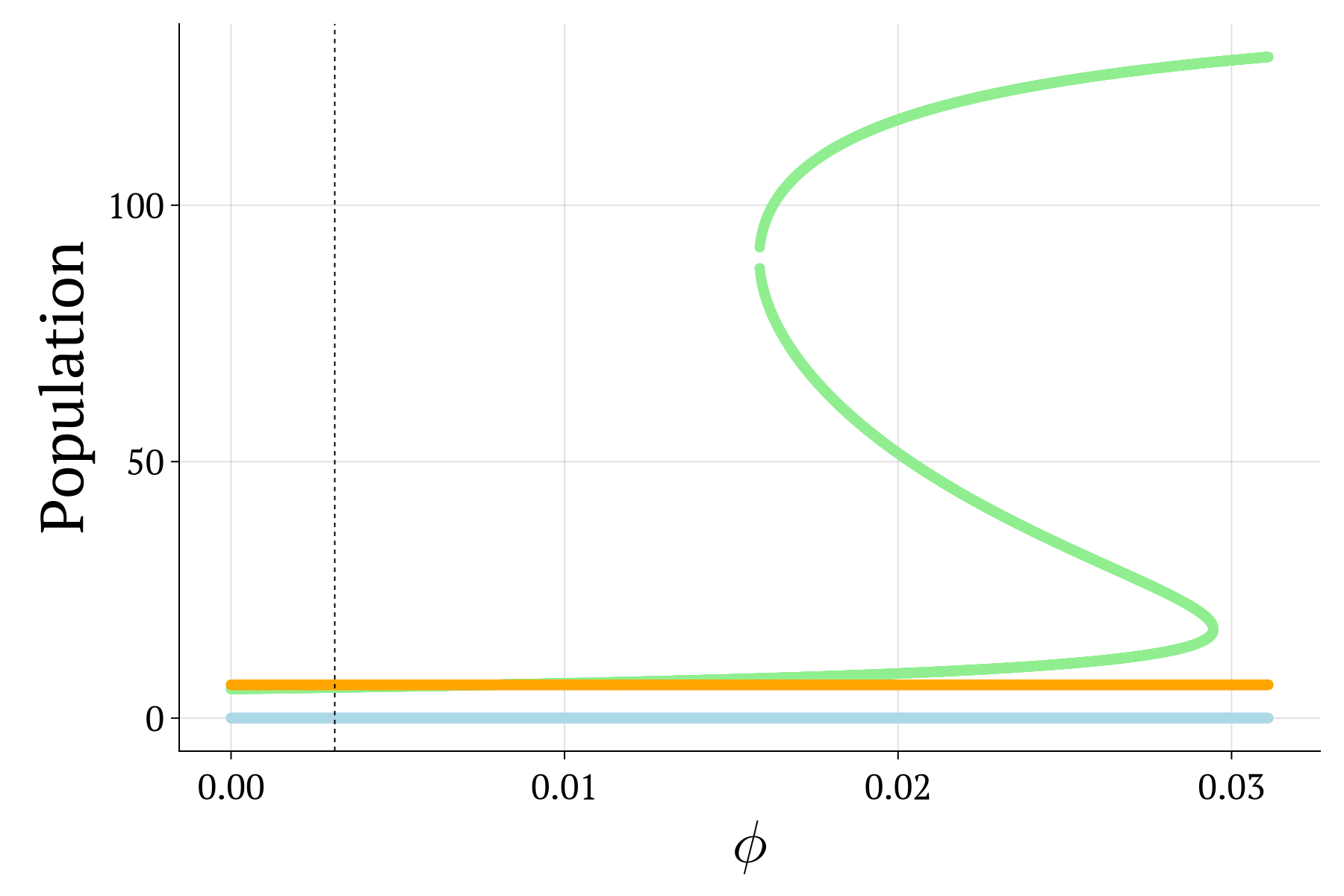}
\end{subfigure}
\begin{subfigure}{0.45\columnwidth}
 \caption{}
    \includegraphics[width=\textwidth]{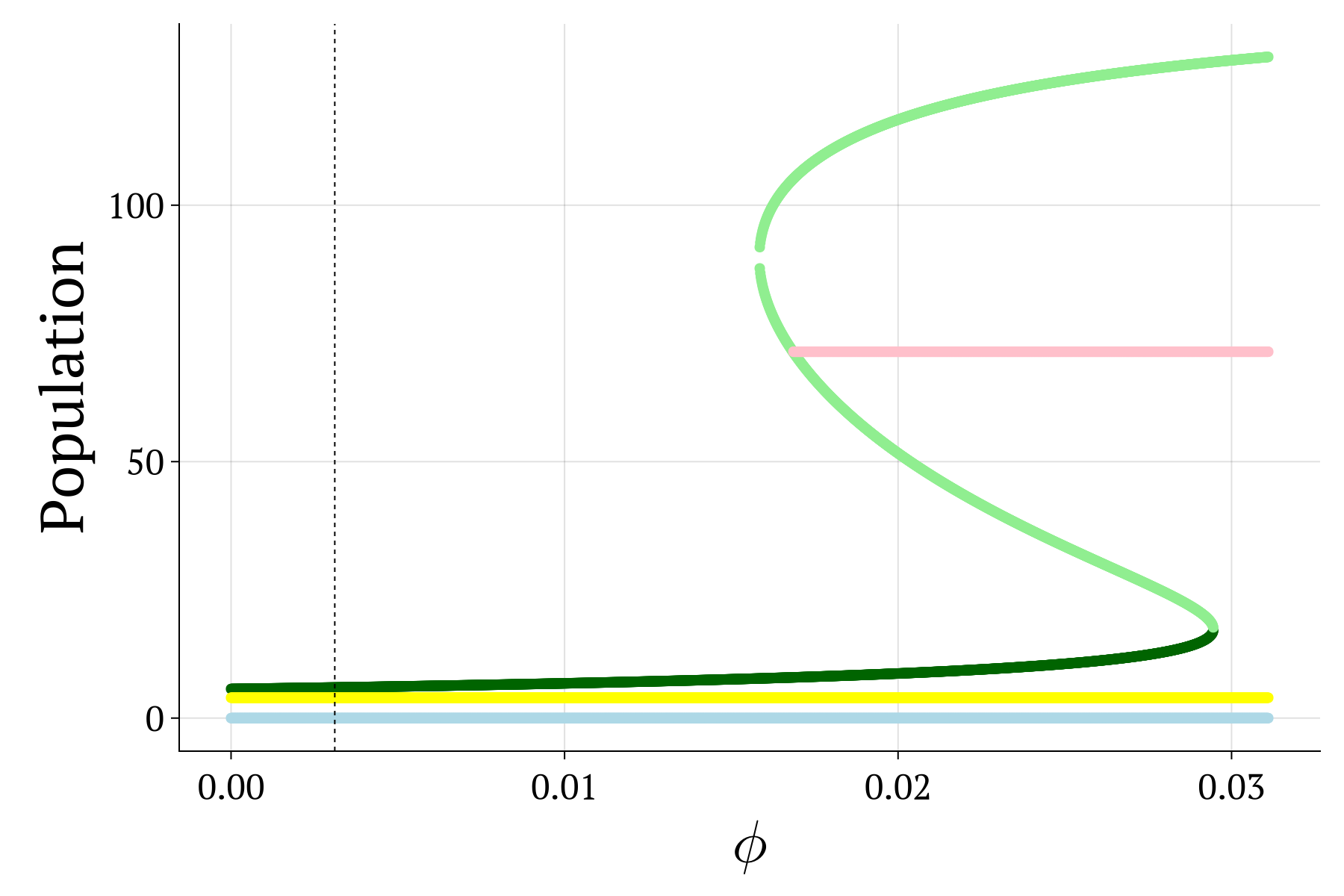}
\end{subfigure}
\begin{subfigure}{0.45\columnwidth}
 \caption{}
    \includegraphics[width=\textwidth]{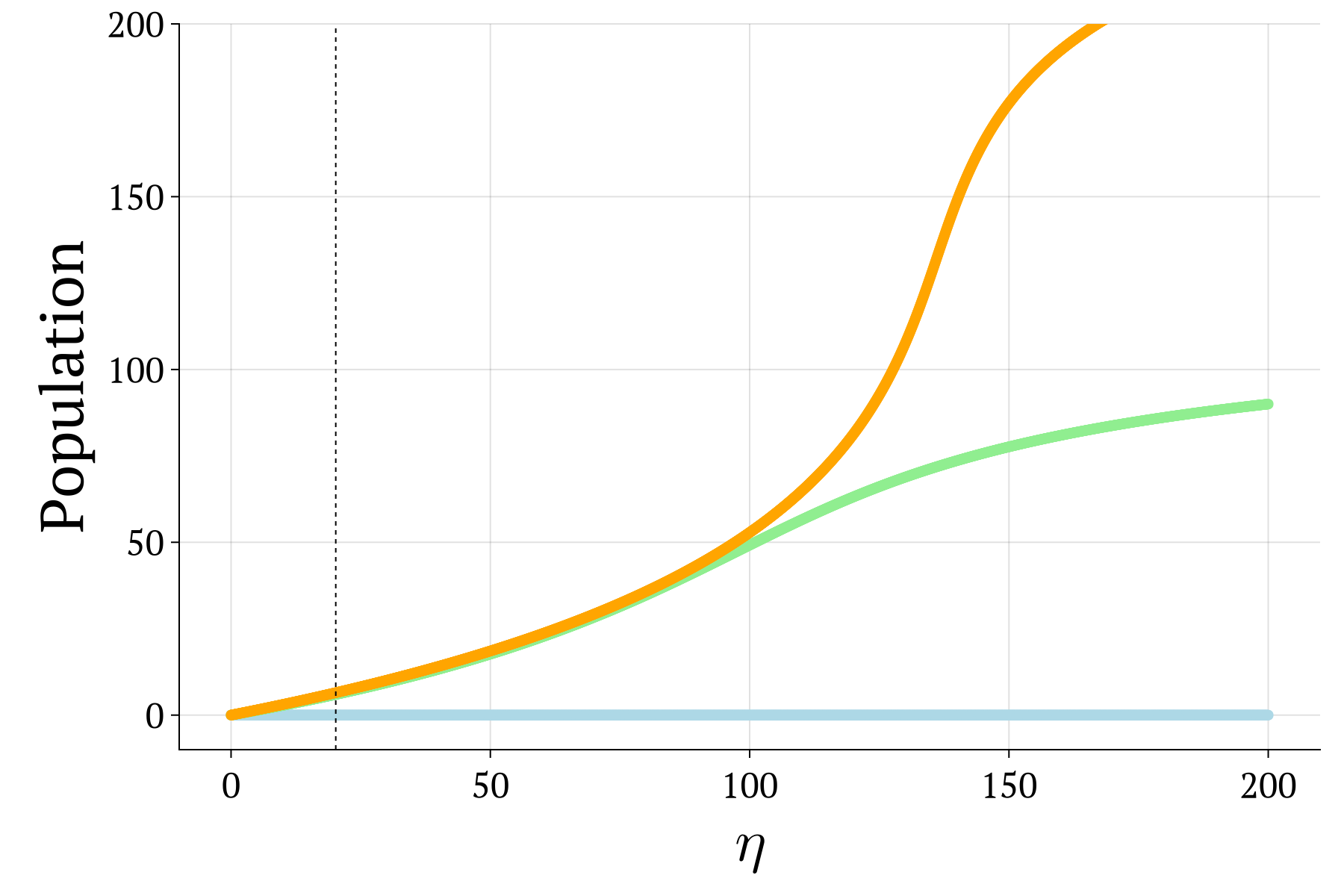}
\end{subfigure}
\begin{subfigure}{0.45\columnwidth}
 \caption{}
    \includegraphics[width=\textwidth]{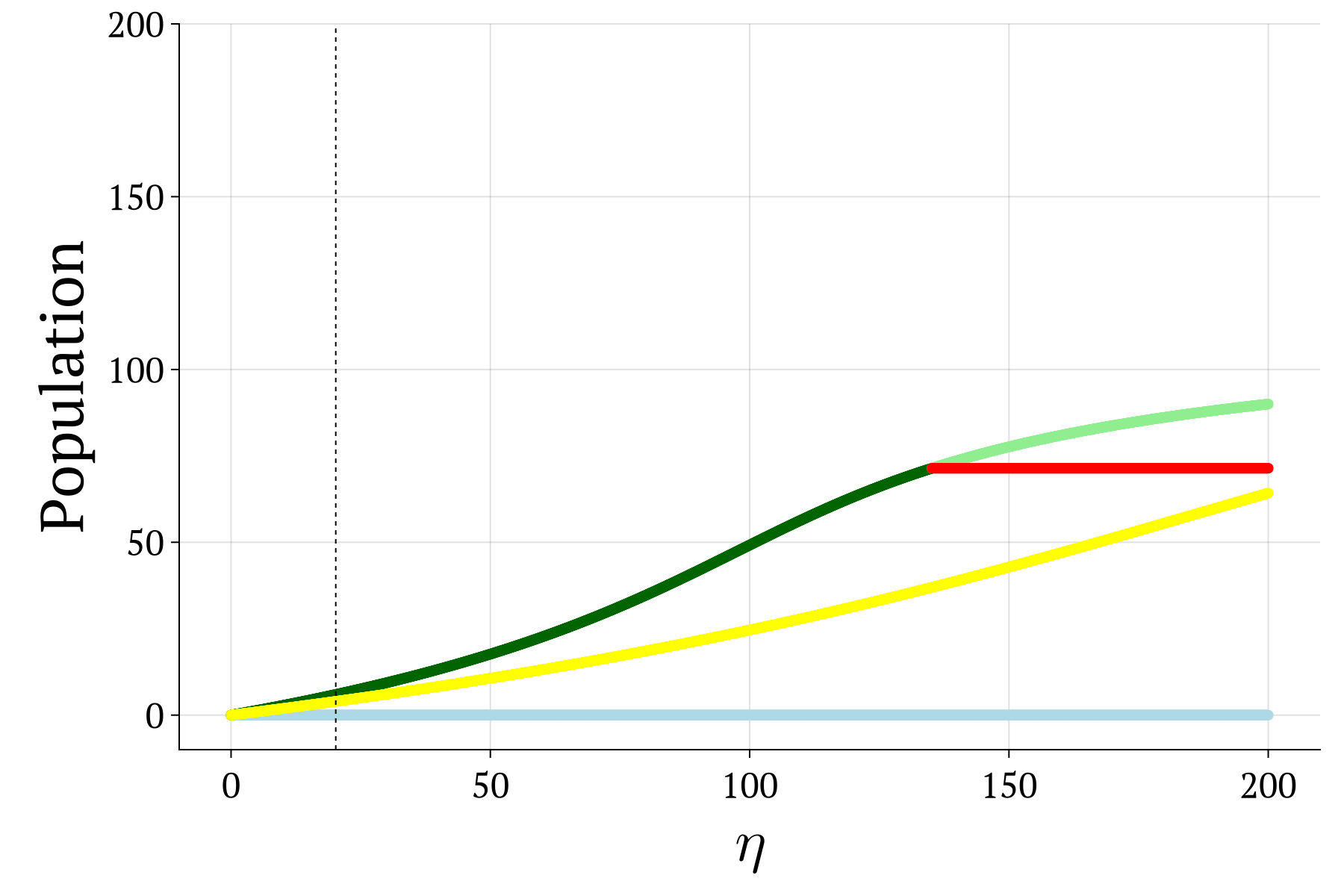}
\end{subfigure}\\
\begin{subfigure}{\columnwidth}
\includegraphics[width=\linewidth]{figures/Chemo_bifurcation_legend.png}
  \end{subfigure}
  \vspace{-1cm}
\caption{Bifurcation diagrams for $\omega$, $\phi$, and $\eta$ when $T_2$ is chemo-resistant and $T_1$ employs immune checkpoint evasion. $\theta_1=\theta_2=0.5$ and $\theta_1=0.5 < \theta_2=0.8$ in the left and right columns, respectively.}
\label{fig:bif_chemo_ICE_extera}
\end{figure}

\begin{figure}[!ht]
\centering
\captionsetup[subfigure]{justification=centering}
\begin{subfigure}{0.45\columnwidth}
 \caption{}
    \includegraphics[width=\textwidth]{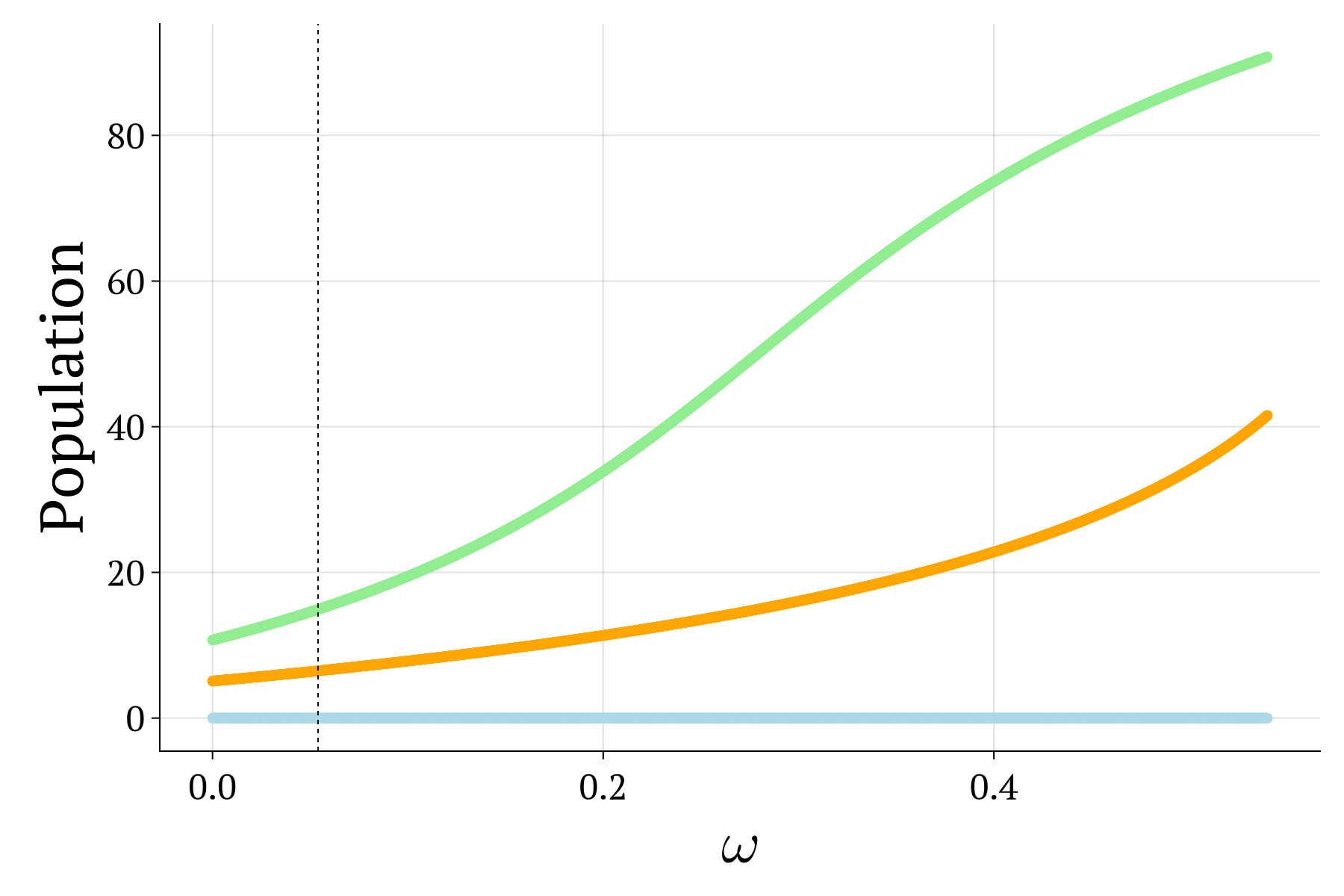}
\end{subfigure}
\begin{subfigure}{0.45\columnwidth}
 \caption{}
    \includegraphics[width=\textwidth]{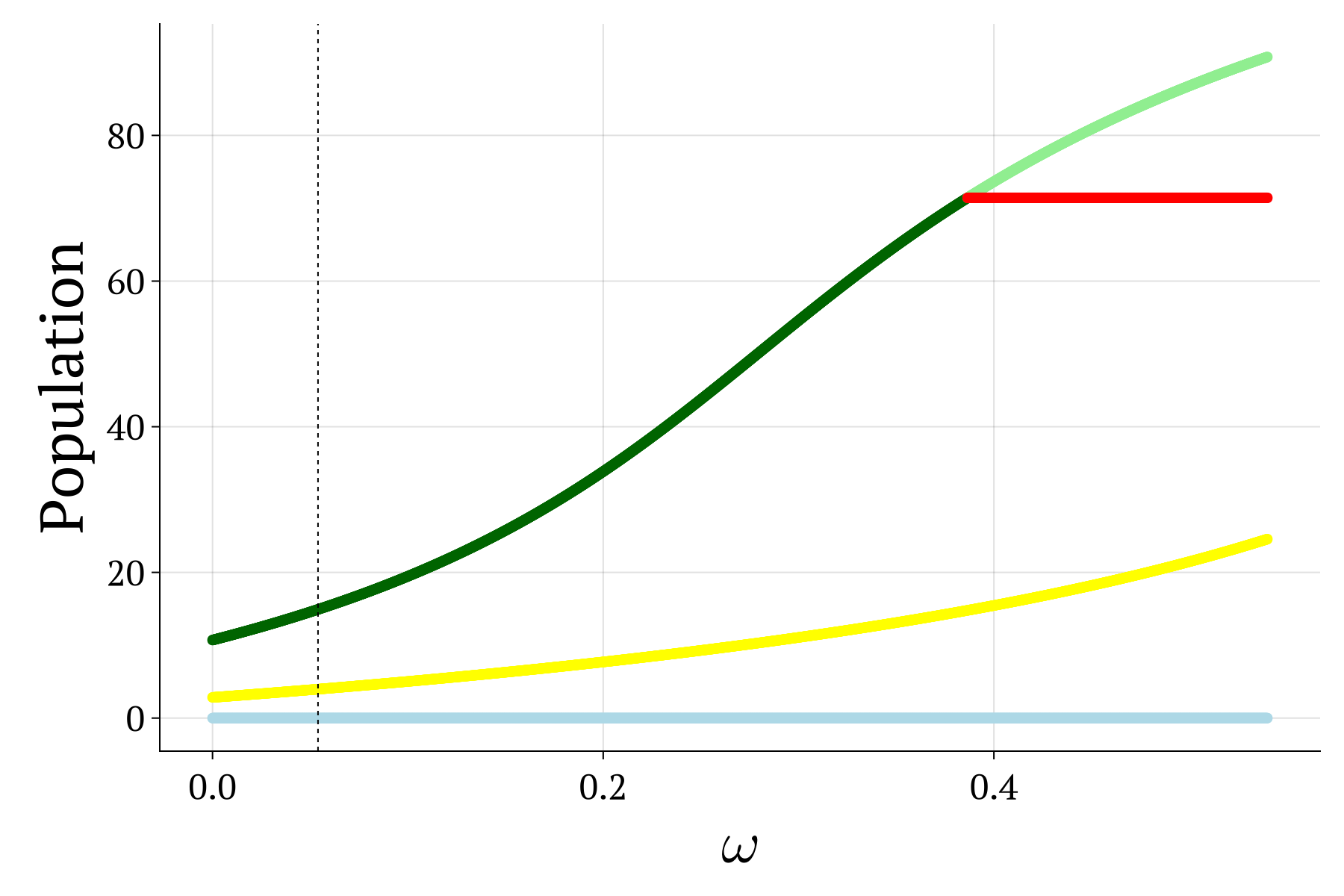}
\end{subfigure}
\begin{subfigure}{0.45\columnwidth}
 \caption{}
    \includegraphics[width=\textwidth]{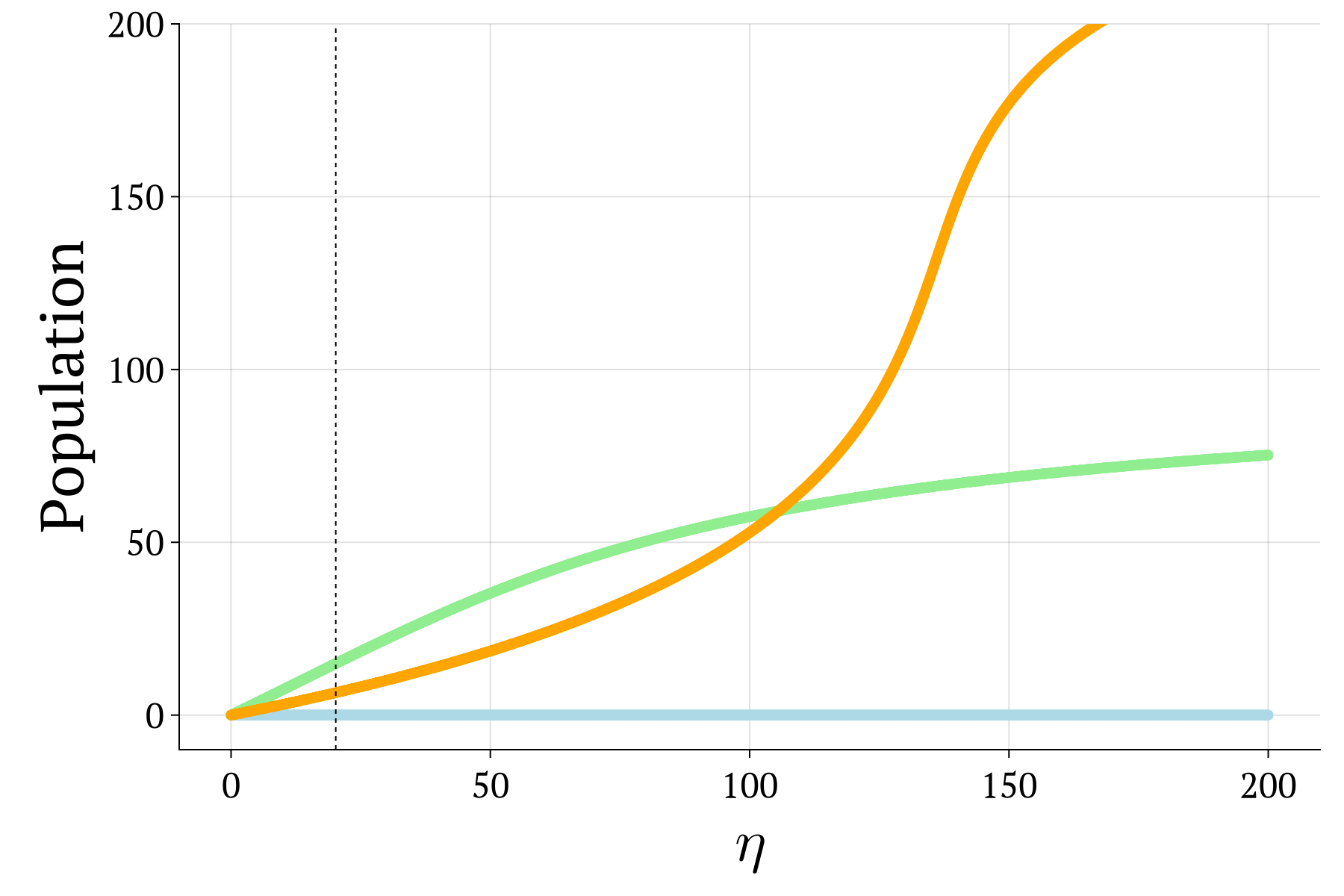}
\end{subfigure}
\begin{subfigure}{0.45\columnwidth}
 \caption{}
    \includegraphics[width=\textwidth]{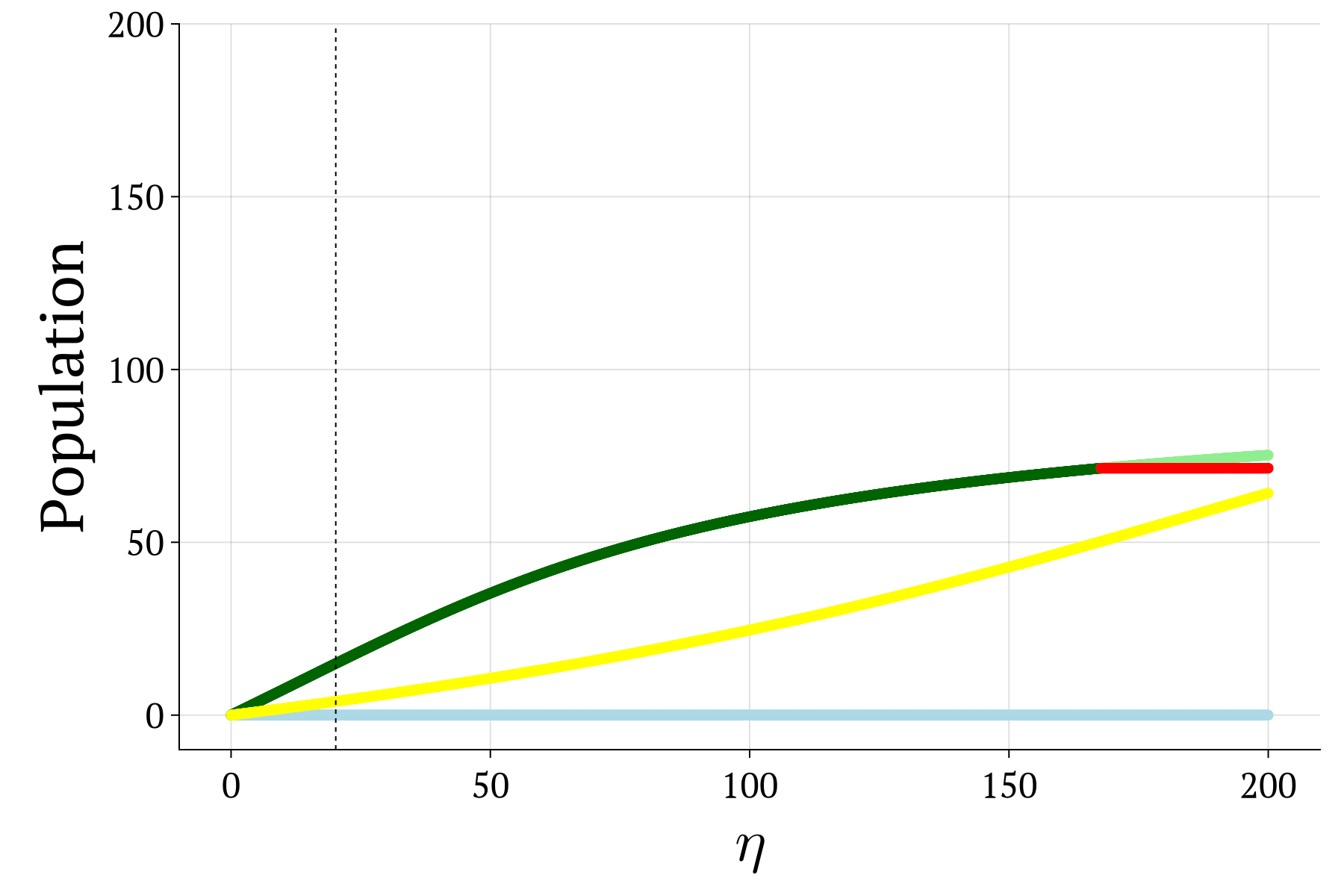}
\end{subfigure}\\
\begin{subfigure}{\columnwidth}
\includegraphics[width=\linewidth]{figures/Chemo_bifurcation_legend.png}
  \end{subfigure}
    \vspace{-1cm}
\caption{Bifurcation diagrams for $\omega$ and $\eta$ when $T_2$ is chemo-resistant and $T_1$ employs reduced antigen presentation. $\theta_1=\theta_2=0.5$ and $\theta_1=0.5 < \theta_2=0.8$ in the left and right columns, respectively.}
\label{fig:bif_chemo_RAP_extera}
\end{figure}

Figures \ref{fig:bif_chemo_ICE_extera} and \ref{fig:bif_chemo_RAP_extera} illustrate the remaining parameters. Although these parameters may be relevant for phenotype-guided therapies, they provide limited insight into combination therapy design. For example, increasing $\omega$ does not significantly alter the stability structure except when $\omega$ becomes sufficiently large to weaken the immune system. In this case, the immune response is unable to eliminate tumor cells, resulting in a stable coexistence equilibrium.

Overall, these supplementary bifurcation diagrams reinforce the conclusions of Section \ref{results:bifurcation_chemo_resistant} and demonstrate that optimal treatment strategies depend on both tumor phenotype and immune-evasion mechanism. Combination therapy consistently expands the parameter region corresponding to tumor control and delays resistance emergence.

\end{document}